\documentclass[%
 reprint,
superscriptaddress,
showpacs,
nofootinbib,
 amsmath,amssymb,
 aps,
showkeys
]{revtex4-1}

\usepackage{graphicx}
\usepackage{dcolumn}
\usepackage{bm}
\usepackage{lipsum}
\usepackage{amsmath,amssymb}
\usepackage{booktabs}
\usepackage[utf8]{inputenc}
\usepackage[T1]{fontenc}
\usepackage[normalem]{ulem} 
\usepackage{verbatim}
\usepackage[left=2cm, right=2cm, top=2cm]{geometry}

\usepackage{epigraph}
\usepackage{siunitx}
\usepackage[english]{babel}
\usepackage{listings}
\usepackage{graphicx}
\usepackage[utf8]{inputenc}
\usepackage{listings}
\usepackage{matlab-prettifier}
\usepackage{amsfonts}
\usepackage{amsmath}
\usepackage{amssymb}
\usepackage{amsthm}
\usepackage{array}
\newcolumntype{P}[1]{>{\centering\arraybackslash}p{#1}}
\newcolumntype{M}[1]{>{\centering\arraybackslash}m{#1}}
\usepackage{braket}
\usepackage{xcolor}
\usepackage{enumitem}   
\usepackage{tabularx}
\usepackage{booktabs}
\usepackage{mathtools}

\usepackage{multirow}

\usepackage{glossaries}

\usepackage[colorlinks = true,
            linkcolor = blue,
            urlcolor  = blue,
            citecolor = blue,
            anchorcolor = blue]{hyperref}
\usepackage{comment}
\usepackage{color}

\usepackage[capitalise]{cleveref}

\usepackage{orcidlink}

\makeatletter
\newcommand*{\rom}[1]{\expandafter\@slowromancap\romannumeral #1@}
\makeatother

\newcommand{\eps}{\epsilon}

\newcommand{\AMEND}[1]{#1}

\newacronym{lisa}{LISA}{Laser Interferometer Space Antenna}
\newacronym{emri}{EMRI}{extreme-mass-ratio inspiral}
\newacronym{imri}{IMRI}{intermediate-mass-ratio inspiral}
\newacronym{co}{CO}{compact object}
\newacronym{ode}{ODE}{ordinary differential equation}
\newacronym{gw}{GW}{gravitational wave}
\newacronym{few}{\textsc{few}}{\textsc{FastEMRIWaveforms}}
\newacronym{mbh}{MBH}{massive black hole}
\newacronym{imbh}{IMBH}{intermediate-mass black hole}
\newacronym{pn}{PN}{post-Newtonian}
\newacronym{gpu}{GPU}{graphics processing unit}
\newacronym{cpu}{CPU}{central processing unit}
\newacronym{gsf}{GSF}{gravitational self-force}
\newacronym{nr}{NR}{numerical relativity}
\newacronym{snr}{SNR}{signal-to-noise ratio}
\newacronym{bhpc}{BHPC}{Black Hole Perturbation Club}
\newacronym{bhpt}{BHPT}{Black Hole Perturbation Toolkit}
\newacronym{psd}{PSD}{power spectral density}
\newacronym{mcmc}{MCMC}{Markov chain Monte-Carlo}
\newacronym{aak}{\textsc{AAK}}{Augmented Analytic Kludge}
\newacronym{gpr}{GPR}{Gaussian process regression}
\newacronym{td}{TD}{time domain}
\newacronym{fd}{FD}{frequency domain}
\newacronym{tdi}{TDI}{time-delay interferometry}

\begin{document}

\title{The Fast and the Frame-Dragging: Efficient waveforms for asymmetric-mass eccentric equatorial inspirals into rapidly-spinning black holes}

\author{Christian E.~A.~Chapman-Bird\,\orcidlink{0000-0002-2728-9612}} 
\email[]{c.chapman-bird@bham.ac.uk}
\affiliation{Institute for Gravitational Wave Astronomy \& School of Physics and
Astronomy, University of Birmingham, Edgbaston, Birmingham B15 2TT, UK}
\affiliation{School of Physics and Astronomy, University of Glasgow, Glasgow G12 8QQ, UK}

\author{Lorenzo Speri\,\orcidlink{0000-0002-5442-7267}} 
\affiliation{European Space Agency (ESA), European Space Research and Technology Centre (ESTEC), Keplerlaan 1, 2201 AZ Noordwijk, the Netherlands}

\author{Zachary Nasipak\,\orcidlink{0000-0002-5109-9704}} 
\affiliation{School of Mathematical Sciences and STAG Research Centre, University of Southampton, Southampton, UK, SO17 1BJ}

\author{Ollie Burke\,\orcidlink{0000-0003-2393-209X}} 
\affiliation{School of Physics and Astronomy, University of Glasgow, Glasgow G12 8QQ, UK}
\affiliation{Laboratoire des 2 Infinis - Toulouse (L2IT-IN2P3), Université de Toulouse, CNRS, F-31062 Toulouse Cedex 9, France}

\author{Michael L.~Katz\,\orcidlink{0000-0002-7605-5767}} 
\affiliation{NASA Marshall Space Flight Center, Huntsville, Alabama 35811, USA}

\author{Alessandro Santini\,\orcidlink{0000-0001-6936-8581}}
\affiliation{Max Planck Institute for Gravitational Physics (Albert Einstein Institute), Am M\"{u}hlenberg 1, 14476 Potsdam, Germany}

\author{Shubham Kejriwal\,\orcidlink{0009-0004-5838-1886}}
\affiliation{Department of Physics, National University of Singapore, 21 Lower Kent Ridge Rd, Singapore 119077}

\author{Philip Lynch\,\orcidlink{0000-0003-4070-7150}}
\affiliation{Max Planck Institute for Gravitational Physics (Albert Einstein Institute), Am M\"{u}hlenberg 1, 14476 Potsdam, Germany}

\author{Josh Mathews\,\orcidlink{0000-0002-5477-8470}}
\affiliation{Department of Physics, National University of Singapore, 21 Lower Kent Ridge Rd, Singapore 119077}

\author{Hassan Khalvati\,\orcidlink{0000-0001-5313-9282}}
\affiliation{Perimeter Institute for Theoretical Physics, Ontario, N2L 2Y5, Canada}
\affiliation{University of Guelph, Guelph, Ontario N1G 2W1, Canada}

\author{Jonathan E.~Thompson\,\orcidlink{0000-0002-0419-5517}}
\affiliation{School of Mathematical Sciences and STAG Research Centre, University of Southampton, Southampton, UK, SO17 1BJ}
\affiliation{Theoretical Astrophysics Group, California Institute of Technology, Pasadena, CA 91125, U.S.A.}

\author{Soichiro Isoyama\,\orcidlink{0000-0001-6247-2642}}
\affiliation{Department of Physics, National University of Singapore, 21 Lower Kent Ridge Rd, Singapore 119077}

\author{Scott A.~Hughes\,\orcidlink{0000-0001-6211-1388}}
\affiliation{Department of Physics and MIT Kavli Institute, MIT, Cambridge, MA 02139 USA}

\author{Niels Warburton\,\orcidlink{0000-0003-0914-8645}}
\affiliation{School of Mathematics \& Statistics, University College Dublin, Belfield, Dublin 4, Ireland}

\author{Alvin J.~K.~Chua\,\orcidlink{0000-0001-5242-8269}}
\affiliation{Department of Physics, National University of Singapore, 21 Lower Kent Ridge Rd, Singapore 119077}
\affiliation{Department of Mathematics, National University of Singapore, Singapore 119076}

\author{Maxime Pigou\,\orcidlink{0000-0001-6997-2500}}
\affiliation{Laboratoire des 2 Infinis - Toulouse (L2IT-IN2P3), Université de Toulouse, CNRS, F-31062 Toulouse Cedex 9, France}

\begin{abstract}
Observations of gravitational-wave signals emitted by compact binary inspirals provide unique insights into their properties, but their analysis requires accurate and efficient waveform models.
Intermediate- and extreme-mass-ratio inspirals (I/EMRIs), with mass ratios $q \gtrsim 10^2$, are promising sources for future detectors such as the Laser Interferometer Space Antenna (LISA).
Modelling waveforms for these asymmetric-mass binaries is challenging, entailing the tracking of many harmonic modes over thousands to millions of cycles. 
The FastEMRIWaveforms (\textsc{few}) modelling framework addresses this need, leveraging precomputation of mode data and interpolation to rapidly compute adiabatic waveforms for eccentric inspirals into zero-spin black holes.
In this work, we extend \textsc{few} to model eccentric equatorial inspirals into black holes with spin magnitudes $|a| \leq 0.999$.
Our model supports eccentricities $e \leq 0.9$ and semi-latus recta $p \leq 200$, enabling the generation of long-duration IMRI waveforms, and produces waveforms in $\sim100\,\mathrm{ms}$ with hardware acceleration.
Characterising systematic errors, we estimate that our model attains mismatches of $\sim 10^{-5}$ (for LISA sensitivity) with respect to error-free adiabatic waveforms over the majority of the parameter space. 
We find that kludge models can introduce errors in signal-to-noise ratios (SNRs) as great as $^{+60\%}_{-40\%}$ and induce marginal biases of up to $\sim 1\,\sigma$ in parameter estimation.
We show that LISA's horizon redshift for I/EMRI signals varies significantly with $a$, reaching a redshift of $3$ ($15$) for EMRIs (IMRIs) with only minor $(\sim10\%)$ dependence on $e$ for an SNR threshold of 20.
For signals with $\rm{SNR} \sim 50$, spin and eccentricity-at-plunge are measured with uncertainties of $\delta a \sim 10^{-7}$ and $\delta e_{\rm f} \sim 10^{-5}$. 
This work advances the state-of-the-art in waveform generation for asymmetric-mass binaries, providing open-source tools for the investigation of I/EMRI astrophysics and data analysis.
\end{abstract}

\maketitle

\tableofcontents

\section{Introduction}

The advancement of \gls{gw} astronomy has significantly enhanced our understanding of astrophysical phenomena, particularly through the  analysis of compact binary mergers identified by ground-based LIGO-Virgo-KAGRA \gls{gw} detector network~\cite{KAGRA:2021vkt,LIGOScientific:2014pky,VIRGO:2014yos,KAGRA:2018plz,KAGRA:2021duu}.
To facilitate these efforts, models for the gravitational waveforms emitted by these coalescences have been developed and refined over time~\cite{Varma:2019csw,Ramos-Buades:2023ehm,Pratten:2020ceb}.
However, there exists a region in the parameter space that has not yet been explored by \gls{gw} detectors: systems where the masses $m_{1,2}$ of the binary components differ substantially (i.e., $m_1 \gg m_2$)~\cite{LISAConsortiumWaveformWorkingGroup:2023arg}. 
These asymmetric-mass binaries are not the primary focus of current ground-based detectors, which have identified sources with mass ratios of up to $q = m_1/m_2\sim 10$ (in the case of \textsc{GW190814}~\cite{LIGOScientific:2020zkf,LIGOScientific:2024elc,LIGOScientific:2020stg}), and even up to $q \sim 26$ in the case of the event GW191219\_163120~\cite{KAGRA:2021vkt} (although this mass ratio estimate may be unreliable due to the inaccuracy of the waveform models used in this region of parameter space).

Asymmetric-mass binaries are anticipated to be key sources for future observatories. 
When the smaller body is a \gls{imbh} with $m_1 \sim 10^3\,M_\odot$, their GWs are in the band of next-generation ground-based detectors such as Cosmic Explorer~\cite{Punturo:2010zz} or the Einstein Telescope \cite{Reitze:2019iox} and deci-Hertz detectors~\cite{Sedda:2019uro} such as (B-)DECICO~\cite{Seto:2001qf,Nakamura:2016hna,Kawamura:2020pcg} and lunar \gls{gw} detectors~\cite{Branchesi:2023sjl}).  
Binaries in which the larger body is a massive black hole, $m_1 \in [10^4, 10^8]\,M_\odot$, radiate in the milli-Hertz band targeted by detectors such as \gls{lisa}~\cite{Amaro-Seoane:2007osp,LISA:2024hlh}, TianQin and Taiji~\cite{TianQin:2020hid,2021PTEP.2021eA108L,Gong:2021gvw}.

One of \gls{lisa}'s primary scientific objectives is to study the properties and environments of black holes in the local Universe with observations of \glspl{imri} and \glspl{emri}, characterised by mass ratios of $q \in [10^2, 10^4]$ and $q \in [10^{4}, 10^{6}]$, respectively~\cite{Amaro-Seoane:2012lgq}.
Binaries with even larger mass ratio are also of interest \cite{Amaro-Seoane:2019umn,Gourgoulhon:2019iyu}.
While measurement prospects for \glspl{imri} are less well-explored due to a lack of accurate and efficient waveform models in this regime, the analysis of \gls{lisa} observations of \gls{emri} signals will enable measurements of binary parameters with sub-percent precision~\cite{Babak:2017tow, Berry:2019wgg}.
Such precision facilitates rigorous tests of general relativity~\cite{Barack:2003fp,Gair:2012nm,Maselli:2020zgv,Maselli:2021men,Barsanti:2022ana,Speri:2024qak} and offers insights into the environments surrounding massive black holes~\cite{Bonga:2019ycj,Yang:2019iqa,Yin:2024nyz,Dyson:2025dlj,Speri:2022upm}.
This, in turn, enhances our understanding of the mass function of \glspl{mbh}~\cite{Gair:2010yu,Chapman-Bird:2022tvu}, 
the dense stellar environments in galactic cores~\cite{Amaro-Seoane:2007osp} and the gas disks surrounding these black holes~\cite{Barausse:2006vt,Barausse:2007dy,Yunes:2011ws,Barausse:2014tra,Barausse:2014pra,Speri:2022upm,Khalvati:2024tzz,Duque:2024mfw,Copparoni:2025jhq}, providing insights into the many proposed formation channels of these systems~\cite{Pan:2021oob,Pan:2021ksp,LISA:2022yao,Naoz:2022rru}.
Moreover, \gls{emri} signals are effective probes of cosmic expansion \cite{MacLeod:2007jd,Laghi:2021pqk,Liu:2023onj,LISACosmologyWorkingGroup:2022jok, toscani2024strongly} and may even enable \gls{lisa} calibration errors to be constrained~\cite{Savalle:2022xpv}.

\begin{figure*}
\begin{center}
    \includegraphics[width=0.9\linewidth]{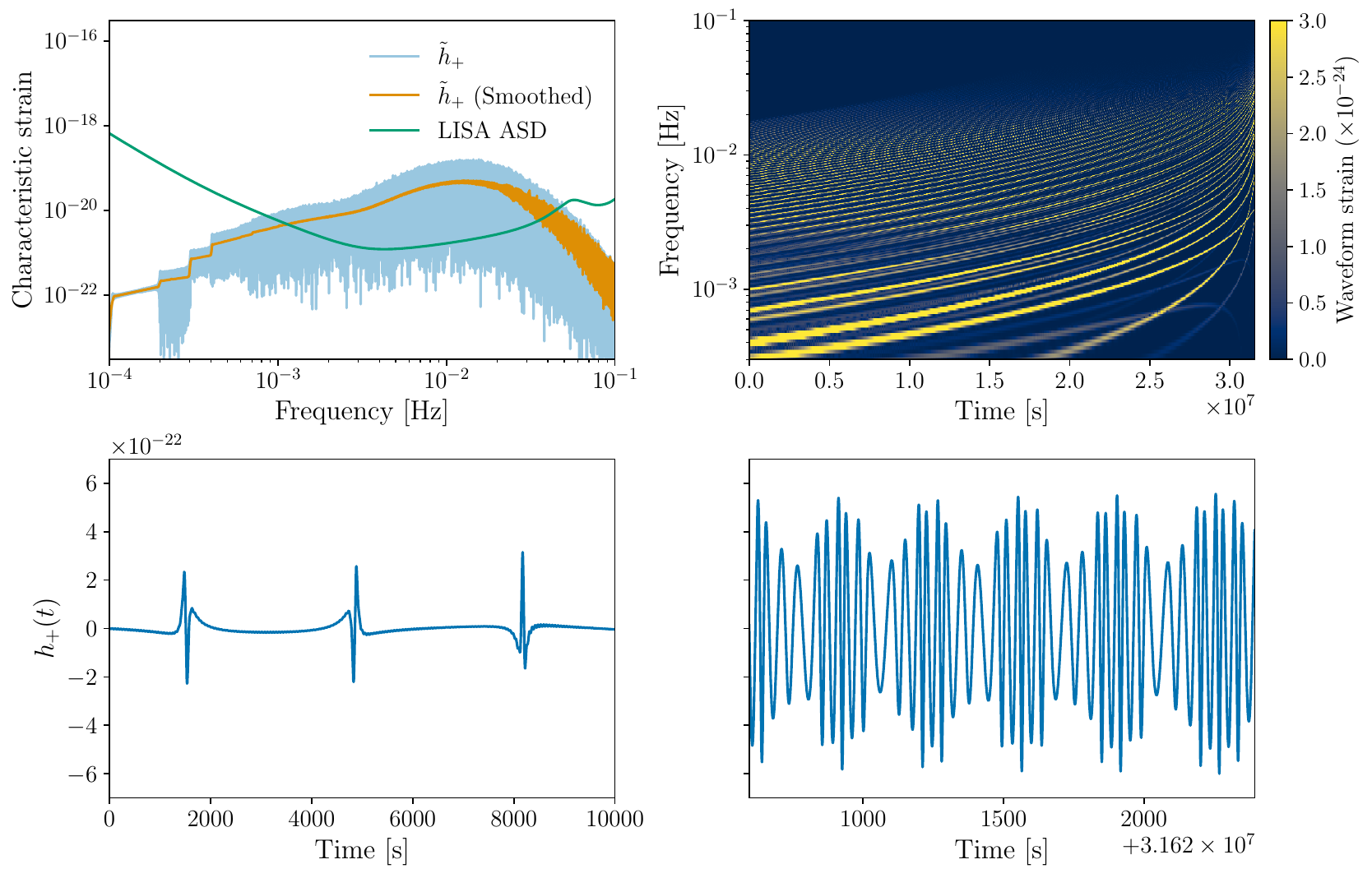}
    \caption{
    Waveform for a retrograde eccentric \gls{emri} into a spinning \gls{mbh} in the frequency (upper-left), time-frequency (upper-right) and time domain (early and late times in lower-left and lower-right respectively).
    This system has parameters $\{m_1, m_2, a, p_0,e_0, d_\mathrm{L}\} = \{10^5\,M_\odot, 30\,M_\odot, -0.998, 28.3, 0.85, 1\,\mathrm{Gpc}\}$, and plunges after one year.
    When observed with \gls{lisa} over this duration, this signal has an \gls{snr} of 41.
    The rich harmonic mode structure of the waveform evolves as the binary circularizes and the trajectory enters the strong-field region of the \gls{mbh} (see also \cref{sec:amplitudes}).
    Despite the waveform's complexity and size, it is generated by \gls{few} in less than $100\,\mathrm{ms}$ of wall-time for an A100 \gls{gpu}, sufficiently fast enough for full parameter estimation studies to be performed on a timescale of hours.
    }
    \label{fig:tf-waveform}
\end{center}
\end{figure*}

Achieving these scientific goals entails the unbiased analysis of these signals, a procedure reliant upon accurate waveform models.
Waveforms from \glspl{imri} and \glspl{emri} are among the most challenging to construct accurately of any compact binary system~\cite{pound2022black, barack2018self, barack2009gravitational}.
As the trajectory of the secondary object is (in general) expected to be eccentric and inclined with respect to the primary object's spin-momentum vector~\cite{Alexander:2017aln} (leading to Lense--Thirring precession of the orbital plane), the inspiral is tri-periodic~\cite{Schmidt:2002qk}, which manifests as a rich structure of tens of thousands of strong harmonic (sideband) modes in the gravitational waveform~\cite{Hughes:2021exa,Drasco:2005kz} (as shown in \cref{fig:tf-waveform}).
Models must accurately track the \gls{gw} phase evolution of these modes to sub-radian precision over the thousands to millions of orbital cycles that the secondary object typically completes in-band, which is an infeasible task for the \gls{nr} techniques that underpin existing models of comparable-mass binaries (despite ongoing efforts to address scale disparity in \gls{nr} simulations~\cite{Lousto:2020tnb,Rosato:2021jsq,Lousto:2022hoq,Wittek:2024gxn,Wittek:2024pis}).
Moreover, waveform models must also be highly efficient; computational wall-times of less than one second are vital for the analysis of \glspl{imri} and \glspl{emri} to be a feasible prospect, especially in the context of a global \gls{lisa} analysis framework (e.g., Refs.~\cite{Katz:2024oqg,Deng:2025wgk,Littenberg:2023xpl}).

The most promising approach that satisfies these stringent modelling constraints is the \gls{gsf} paradigm~\cite{Mino:1996nk,Quinn:1996am,gralla2008rigorous,Detweiler:2011tt,Gralla:2012db,pound2012second,Harte:2014wya}, where the metric of the binary is found by perturbatively solving the Einstein Field Equations in powers of the small mass ratio around the metric of the primary black hole; we refer readers to Refs.~\cite{Poisson:2011nh,barack2018self,pound2022black} for reviews.
In particular, the modern waveform generation scheme in the \gls{gsf} paradigm is built upon the multiscale framework~\cite{Hinderer:2008dm,Miller:2020bft,pound2022black,Mathews:2025nyb}, leveraging the quasi-periodic orbital dynamics of asymmetric-mass binaries 
(see, e.g, Refs.~\cite{Yunes:2010zj,Chiaramello:2020ehz,Zhang:2020rxy,Ramos-Buades:2021adz,Zhang:2021fgy,Albanesi:2021rby,Nagar:2022fep,Albertini:2022dmc,Albertini:2023aol,Albanesi:2023bgi,Albertini:2023aol,Cheung:2023lnj,Gamboa:2024hli,Loutrel:2024qxp,Paul:2024ujx,Cheung:2024byb,Akpinar:2025huz,Morras:2025nlp,Planas:2025feq,Strusberg:2025qfv} for other efforts towards \glspl{imri} and \glspl{emri} waveform generation). 
An important feature of this framework is that it enables waveform generation to be divided into an expensive ``offline'' step (in which many numerical data products are computed once and stored) and a fast ``online'' step in which these data products are interpolated and integrated to build waveforms quickly and accurately.
In the case of the inspiral dynamics\footnote{In this work, we exclusively focus on the inspiral stage of waveform modelling, as the merger-ringdown stage does not significantly contribute to the measurement precision of \gls{emri} parameters.
While our waveform model does not incorporate these stages of coalescence, they may become more impactful for \glspl{imri} and their inclusion is worth investigating in this regard. 
Recent advancements have been made in terms of merger-ringdown modelling in the \gls{gsf} framework (e.g. Refs.~\cite{Apte:2019txp,Kuchler:2024esj,Becker:2024xdi,Lhost:2024jmw,Kuchler:2025hwx}), highlighting the potential for the future extension of our model to include this stage; see also our concluding discussion in~\cref{sec:conclusions}.}, 
at leading order,
one can approximate the change of energy, angular momentum, and Carter constant~\cite{Carter:1968rr} of the binary due to the emission of \glspl{gw} using so-called ``flux-balance formulae'' \cite{Mino:2003yg,Drasco:2005is,Sago:2005fn,Tanaka:2005ue,Isoyama:2018sib,Grant:2024ivt} in what is known as the adiabatic approximation~\cite{Hughes:2001jr,Finn:2000sy,Mino:2006em,Sundararajan:2008zm,Hughes:2021exa,Isoyama:2021jjd,Kerachian:2023oiw,Nasipak:2023kuf}. 
These fluxes have been calculated using both numerical codes~\cite{Nakamura:1987zz,Shibata:1993uk,Shibata:1993yf,Shibata:1994xk,kennefick1998stability,Hughes:1999bq,Glampedakis:2002ya,Drasco:2005kz,Hughes:2005qb,Mino:2005an,Fujita:2009us,Harms:2014dqa,Fujita:2020zxe,Chen:2023lsa}, 
and in analytic \gls{pn} expansions~\cite{Mino:1997bx,Sasaki:2003xr,Ganz:2007rf,
Fujita:2012cm,Shah:2014tka,Fujita:2014eta,Munna:2020iju,Munna:2020som,Munna:2023vds,
Sago:2024mgh,Castillo:2024isq}. 
It has been determined through a multiscale analysis (and subsequently verified with data analysis simulations in Ref.~\cite{Burke:2023lno}) that one must go beyond leading order adiabatic waveforms and include the first post-adiabatic corrections to the dynamics of the binary to achieve the sub-radian accuracy required for \gls{lisa}'s scientific goals. 
This requires knowledge of not only the first order in $\epsilon = 1/q$ \gls{gsf}~\cite{Barack:2007tm,barack2009gravitational,Barack:2010tm,Osburn:2014hoa,Isoyama:2014mja,Wardell:2015kea,vandeMeent:2016pee,vandeMeent:2017bcc,Dolan:2023enf} but also the second order \gls{gsf}~\cite{Pound:2019lzj,Warburton:2021kwk,Durkan:2022fvm}\footnote{The effects of the secondary object's spin also enter at this first post-adiabatic order~\cite{Drummond:2023wqc,Skoupy:2023lih,Piovano:2024yks,Mathews:2025nyb}.}.
When included, \gls{gsf} waveforms show remarkable agreement with \gls{nr} even at $q \sim 10$~\cite{Wardell:2021fyy}.
These corrections can be included in an efficient way by leveraging averaging techniques~\cite{Tanaka:2005ue,Mino:2005an,Mino:2006em}, the multi-scale expansion~\cite{Hinderer:2008dm,Miller:2020bft,pound2022black,Mathews:2025nyb} or near-identity averaging transformations~\cite{VanDeMeent:2018cgn,Lynch:2021ogr,Lynch2022,Lynch:2023gpu,Drummond:2023wqc,Lynch:2024ohd}.
This leads to the expensive offline step mentioned above, where one densely tiles the intrinsic \gls{emri} parameter space such that it may be rapidly interpolated during waveform generation.
As the \gls{gsf} data varies smoothly over parameter space, techniques such as cubic spline interpolation can be applied at low computational cost, facilitating rapid modelling of the inspiral dynamics and trajectory evolution.
Unlike the dynamics, the amplitude evolution of each waveform mode needs only to be known to leading order for \glspl{emri} (see, e.g., Appendix C of Ref.~\cite{Burke:2023lno}).
These amplitudes also vary smoothly over parameter space and are readily interpolated.

Even once the vast majority of computational effort is shifted into these up-front data-generation procedures, interpolating many thousands of waveform modes at the density of the \gls{lisa} data stream and summing over them to construct the full waveform is still a computationally expensive and memory-intensive procedure.
The \gls{few} project and modeling framework is a pioneering effort in addressing these computational costs.
It is a software package capable of rapidly computing \gls{lisa}-scale, fully relativistic adiabatic waveforms in both the time and frequency domains in sub-second speeds~\cite{Chua:2020stf,Katz:2021yft,chapman_bird_2025_15630565,Speri:2023jte}
It integrates standalone modules to generate \gls{emri} waveforms on both \glspl{gpu} and \glspl{cpu}, with the \gls{gpu} version demonstrating a speedup of more than three orders of magnitude compared to its \gls{cpu} counterpart for eccentric inspirals into non-spinning black holes.
This significant acceleration rendered the Bayesian parameter estimation of high-fidelity \gls{emri} signals feasible on timescales of hours/days for the first time; \gls{few} was applied in Ref.~\cite{Katz:2021yft} to perform $\sim 100$ such analyses, and has since seen extensive application in the field of \gls{emri} data analysis (e.g., Refs.~\cite{Speri:2022upm,Copparoni:2025jhq,Duque:2024mfw,Khalvati:2024tzz,Speri:2024qak,Burke:2023lno,toscani2024strongly}).

A significant limitation of the original \gls{few} implementation is the restriction of its fully-relativistic waveform generation to zero-spin systems.
It is anticipated that \glspl{imbh} and \glspl{mbh} in nature will exhibit a wide variety of spin magnitudes that are closely linked to their formation and evolutionary history (see, e.g., the review in Ref.~\cite{Reynolds:2020jwt}).
A rapidly-spinning primary object significantly alters waveform morphology; in the case of prograde systems (where the primary's spin-vector is aligned with the orbital angular momentum vector), the inspiral extends deep into the strong-field region of the central object, enhancing signal detectability, harmonic mode content and the precision of parameter estimates~\cite{burke2020constraining,Babak:2017tow}.
The development of accurate and rapid waveform models that incorporate such features is therefore vital for both the development of analysis techniques and studies of scientific prospects for realistic \gls{emri} signals.

In this work, we address this need with the release of \gls{few} v2, extending its domain of validity to spinning systems and covering a large portion of the parameter space of Kerr eccentric equatorial inspirals at adiabatic order.
Specifically, we support dimensionless primary spin magnitudes of up to $0.999$ that exceeds the Thorne limit ($\sim 0.998$ \cite{Thorne:1974ve}) which represents the astrophysical ``limit" of Kerr-spin parameters accounting for spin-up (and spin-down) mechanisms arising from accretion\footnote{Methods for spinning up \glspl{mbh} beyond the Thorne limit are discussed in~\cite{bardeen1970kerr,S_dowski_2011,Arbey_2020}; further investigations regarding \glspl{emri} into these near-extremal black holes can be found in~\cite{gralla2016inspiral,Compere:2017hsi, burke2020transition, burke2020constraining}.}.
Similar to expanding coverage of parameter space in primary spin, we also extend our coverage of semi-latus rectum and eccentricity to $\sim 200$ and $0.9$, respectively (c.f., $\sim 16$ and $0.7$ in \gls{few} v1).
This provides the necessary support for the generation of long-duration \gls{imri} waveforms.
We also take this opportunity to thoroughly assess the systematics of waveform generation with \gls{few} (with specific focus on the impact of interpolation errors), ensuring that our model produces waveforms that meet the accuracy requirements imposed by \gls{lisa} (when neglecting post-adiabatic effects).
From the results of this assessment, we conclude that our model is most robust for spins of up to $\sim 0.998$ and initial eccentricities of up to $\sim 0.85$.
While our model does not include the post-adiabatic corrections necessary for the unbiased analysis of real signals (see Ref.~\cite{Burke:2023lno} for a discussion on this point), these additional contributions may be readily folded in as they become available, and it reproduces many features of these signals (such as detectability and parameter estimation precision) sufficiently well for use in studies of \gls{imri} and \gls{emri} science.
As our model also accurately produces the rich harmonic mode spectrum of these sources (\cref{fig:tf-waveform}), it will also serve as an effective test-bed for the development of search and identification algorithms for these signals (which is an area of active research~\cite{Cole:2025sqo,Strub:2025dfs,Ye:2023lok,chua2022one,Badger:2024rld,Zhang:2022xuq,Yun:2023vwa}) and will facilitate the construction of realistic test datasets as part of upcoming LISA Data Challenges~\cite{Baghi:2022ucj}.

This paper is organized as follows.
In \cref{sec:few-overview}, we provide an overview of the \gls{few} framework for fast waveform generation.
We then describe the extensions to this framework (both in the incorporation of primary spin and in the general improvement of the software) that define our waveform model in \cref{sec:implementation}.
To verify that our model is accurate and robust, we perform an extensive validation study of our systematic assumptions in~\cref{sec:validation}.
After assessing the computational cost of our model, we apply it in the examination of multiple facets of \gls{emri} science in~\cref{sec:results}.
Our conclusions and outlook of future work are then presented in~\cref{sec:conclusions}.

Some incidental matters are relegated to Appendices. 
We explain the new mass convention that \gls{few} v2 adopts in~\cref{app:mass-convention}. 
The data grid and coordinates that we use for the model implementation are explained in~\cref{app:grid_layout_appendix}.
The source parameter prior probability distribution for the Monte-Carlo studies in this work 
is summarized in~\cref{app:monte-carlo-parameters}, 
and we provide a concise description of our \gls{gw} data analysis methods (which we apply throughout this work) in~\cref{app:data_analysis_fundamentals}. 
In~\cref{app:model-comparison}, we provide additional validation of our waveform model 
by comparing it to existing models and data from a specific cases.
Finally, the marginal posterior distributions corresponding to the analyses performed in~\cref{sec:inference-subsection} are provided in~\cref{app:corner_plots}.

Access information for repositories containing both the \gls{few} package and the resources required to reproduce the results presented in this work can be found in the Data Availability statement.
Throughout this manuscript, we employ geometrized units unless otherwise stated, setting $G = 1 = c $.

\section{Waveforms for eccentric equatorial inspirals}
\label{sec:few-overview}

We consider a binary with masses $m_1 \gg m_2$ and spins $\vec{\chi}_1$ and $\vec{\chi}_2$. In our current treatment, we set $\vec{\chi}_2=\vec{0}$: in Refs.~\cite{Mathews:2025nyb,Drummond:2023loz, Drummond:2023wqc,Feng:2021sax,Rahman:2021eay,Xu:2022fyp} it is demonstrated that effects due to the higher multipole moments of the \gls{co}, including its spin, enter the inspiral dynamics at post-adiabatic orders, and can therefore be ignored in our adiabatic model.
We choose our source frame so that $m_1$ sits at the origin with its spin $\vec{\chi}_1$ aligned with the $z$-axis. The magnitude of $\vec{\chi}_1$ is then parametrized in terms of the (dimensionless) spin parameter $a = |\vec{\chi}_1|/m_1^2$. 
Following \cite{Hughes:2021exa,Isoyama:2021jjd}, we construct the waveform strain at null infinity $h = h_+ - ih_\times$ by summing over harmonic modes.
Each mode is defined by its phase $\Phi_{mkn}$ and amplitude $H_{\ell m k n}$, which evolve over the course of an inspiral.
The mode indices $(\ell, m, k, n)$ refer to the multipole of the gravitational radiation and the harmonic of the fundamental frequencies of the orbit (with $m$, $k$ and $n$ corresponding to azimuthal, polar, and radial motion, respectively).
For inspirals restricted to the equatorial plane, which have no spin-induced orbital precession, the modes with $k \neq 0$ vanish due to symmetry such that
\begin{equation}
    \label{eq:source-frame-waveform}
    h(t) = \frac{\mu}{d_\mathrm{L}}\sum_{\ell m n}H_{\ell mn}(t, \theta, \phi) e^{-i\Phi_{mn}(t)},
\end{equation}
where $\mu = m_1 m_2/M$ is the reduced mass of the binary, $M=m_1+m_2$ is its total mass, $d_\mathrm{L}$ is the luminosity distance to the source, and $(\theta, \phi)$ are, respectively the source-frame polar and azimuthal viewing angles of the observer.
Our notation for the masses in our waveform model deviates from the majority of previous adiabatic waveform models in the \gls{gsf} literature, e.g., \cite{Fujita:2020zxe, Hughes:2021exa, Isoyama:2021jjd, Katz:2021yft, Nasipak:2023kuf, Khalvati:2024tzz}, which define $\mu = m_2$ and $M=m_1$. Instead, our mass convention is chosen to align with what is commonly used when modelling comparable-mass binaries, e.g., \cite{Blanchet:2013haa, Blanchet:2023bwj, Pompili:2023tna, Albanesi:2025txj}. Further discussion of this change in convention is provided in~\cref{app:mass-convention}.

In the following section, we review how~\cref{eq:source-frame-waveform} is constructed in the \gls{few} framework, highlighting where assumptions made for non-spinning systems must be modified in the extension of this framework to spinning systems.

\subsection{Inspiral trajectories into spinning black holes}
\label{sec:trajectory-intro}

In order to compute the amplitude and phase of each mode in~\cref{eq:source-frame-waveform}, we must first obtain the orbital trajectory of the \gls{co} by integrating its equations of motion.
Operating in the multiscale framework at adiabatic order~\cite{Hinderer:2008dm,Hughes:2016xwf,pound2022black,Isoyama:2021jjd}, we parameterize the inspiral trajectory at each moment in time by three quasi-Keplerian orbital elements --- $p(t)$ (semi-latus rectum), $e(t)$ (eccentricity) and $x_I(t)$ (cosine of the orbital inclination)--- along with the fundamental orbital phases $\Phi_{A}(t)$ with $A=\{r,\theta,\phi\}$ (up to initial conditions). 
In this work, we specifically choose these elements such that they satisfy the geodesic relations between semi-latus rectum, eccentricity, inclination and the (dimensionless) orbital frequencies $\hat{\Omega}_A(a,p,e,x_I)$, as described in \cite{Lynch:2024hco} (built on works \cite{Glampedakis:2002ya,Schmidt:2002qk,Fujita:2009bp,KerrGeodesicsPackage}), so that it adheres to the convention set out in the LISA convention document~\cite{lisa_rosetta_stone_2025}.\footnote{Some works treat $p$ as dimensionless (e.g., \cite{Fujita:2009bp}), while others take $p$ to scale with $m_1$ (e.g., \cite{Lynch:2024hco}). In this work, we make all quantities, including $p$, dimensionless and explicitly introduce the dependence on mass through the equations of motion. We discuss our choice of mass convention in \Cref{app:mass-convention}.}
The equations of motion then describe how these quantities evolve with respect to time.
At adiabatic order, the evolution of the orbital elements is driven solely by the orbit-averaged back-reaction due to \gls{gw} emission, and the evolution of the fundamental phases is equivalent to that of the tangent geodesic (which changes with respect to time due to the slowly evolving orbital elements).
The adiabatic equations of motion are therefore a system of six \glspl{ode} that must be solved numerically.
Three of these \glspl{ode} are associated with the evolution of the (dimensionless) orbital elements $\alpha \in \{p, e, x_I\}$:
\begin{equation}
    \label{eq:elements-ode}
    \frac{\mathrm{d}\alpha}{\mathrm{d}t} = \frac{\nu}{M} \left[ \hat{f}^{(0)}_\alpha(a, p, e, x_I) + \mathcal{O}(\nu)\right],
\end{equation}
where $\nu = \mu / M$ is the symmetric mass ratio, and we define our time parameter $t$ to scale with $M$. Note that in most prior EMRI literature, including the first version of \gls{few}~\cite{Katz:2021yft}, the small parameter is taken to be the (small) mass ratio $\epsilon = m_2/m_1 \le 1$, and $t$ is defined in terms of $m_1$ not $M$. Here we opt to switch to the symmetric-mass convention, as doing so is known to improve agreement with intermediate and comparable mass binaries \cite{Wardell:2021fyy}. As $\epsilon = \nu + O(\nu^2)$, changing from the small to the symmetric mass ratio (and likewise scaling time from $m_1$ to $M$) only affects the waveform phase at post-adiabatic order, which we do not consider in this work. A detailed discussion of this change in mass convention can be found in Appendix \ref{app:mass-convention}.

At adiabatic order, the forcing functions $\hat{f}^{(0)}_\alpha$ can be related to the asymptotic fluxes of energy and angular momentum (as well as a rate of change of the Carter constant, which can be deduced from asymptotic radiation fields and locally-defined quantities along the orbit) via flux balance laws~\cite{Sago:2005fn,Drasco:2005is,Tanaka:2005ue,Isoyama:2018sib,Grant:2024ivt}.
Due to axial symmetry, $\hat{f}_{x_I}$ is always equal to zero for equatorial inspirals. 
The remaining three \glspl{ode} describe the evolution of the fundamental phases ($\Phi_\phi,\Phi_\theta,\Phi_r$), which are obtained by integrating their corresponding fundamental frequencies
\begin{equation}
    \label{eq:phases-ode}
    \frac{\mathrm{d}\Phi_A}{\mathrm{d}t} = \frac{1}{M}\hat{\Omega}_A(a, p, e, x_I),
\end{equation}
with $A=\{r,\theta,\phi\}$ and the dimensionless frequencies $\hat{\Omega}_A(a,p,e,x_I)$ found in, e.g.,  Refs.~\cite{Glampedakis:2002ya,Schmidt:2002qk,Fujita:2009bp,KerrGeodesicsPackage}. 

These phases can then be related to the $\{r,\theta,\phi\}$ Boyer-Lindquist coordinates of the secondary using the relations found in Ref.~\cite{Lynch:2024hco}.
The phase evolution of each mode in~\cref{eq:source-frame-waveform} is then
\begin{equation}
    \label{eq:mode-phase}
    \Phi_{mkn}(t) = m\Phi_\phi(t) + k\Phi_\theta(t) + n\Phi_r(t).
\end{equation}
However, as only modes with $k=0$ contribute to the waveform for equatorial inspirals, we may write $\Phi_{mn} = m\Phi_\phi + n\Phi_r$ and ignore $\Phi_\theta$.
By combining~\cref{eq:elements-ode,eq:phases-ode}, and integrating over a radiation reaction timescale ($\sim M/\nu$), we can see that the truncation in the evolution of the orbital elements leads to an error in the orbital phase of $O(\nu^0)$~\cite{Hinderer:2008dm}\footnote{In this work, we are ignoring the \gls{gsf} effects of transient resonances~\cite{Flanagan:2010cd}, which first enter the inspiral evolution at $\mathcal{O}(\nu^{1/2})$ past the adiabatic order~\cite{vandeMeent:2013sza,Berry:2016bit,Lynch:2024ohd}. 
Transient resonances for equatorial eccentric inspirals are even weaker, entering at $\mathcal{O}(\nu^1)$ (i.e. first post-adiabatic order)~\cite{vandeMeent:2014raa} and can therefore be neglected in our adiabatic model.}.
This is sufficient to demonstrate the \gls{few} framework for \glspl{imri} and \glspl{emri}, and to obtain the quantitive results of the studies performed in this work. 
Once post-adiabatic corrections to the evolution of the orbital elements are known, they can easily be incorporated as additional forcing terms into~\cref{eq:elements-ode,eq:phases-ode} and quickly evaluated via multiscale expansion \cite{pound2022black,Mathews:2025nyb} or near-identity transform algorithms \cite{Lynch:2021ogr,Lynch:2023gpu}.

We generate inspiral trajectories by numerically integrating~\cref{eq:elements-ode,eq:phases-ode} from a set of initial conditions $\{p_0, e_0, \Phi_{\phi0}, \Phi_{r0}\}$ with an adaptive \gls{ode} solver, returning a sparse trajectory consisting of $\sim 100$ points.
Formally, the adiabatic approximation breaks down as the trajectory approaches the separatrix $p_\mathrm{sep}(a, e, x_I)$, which represents the innermost stable orbit in Kerr spacetime~\cite{stein2020location}. As $p \rightarrow p_\mathrm{sep}(a, e, x_I)$, the \gls{co} ``transitions'' across the boundary of the separatrix and plunges into the \gls{mbh} on timescales where the adiabatic approximation of the inspiral trajectory is no longer valid~\cite{ori2000transition}. 
In practice, it is sensible to terminate the trajectory at an intermediate point $p_\mathrm{sep} + \Delta p_\mathrm{buf}$ prior to crossing the separatrix. 
Not only is this regime better modelled by a multiscale transition-to-plunge expansion (e.g., Refs.~\cite{Compere:2021iwh,Compere:2021zfj,Kuchler:2024esj,Becker:2024xdi,Lhost:2024jmw,Kuchler:2025hwx}), but in our parameterization the forcing functions $\hat{f}^{(0)}_{\alpha}$ diverge as $p \to p_{\rm sep}$~\cite{kennefick1998stability,Hughes:2024tja}\footnote{This divergence is due to the Jacobian between the parameters $(p, e, x_I)$ and the integrals of motion $(E, L_z, Q)$; see~\cref{eq:forcesJac}.  As discussed in Ref.\ \cite{Hughes:2024tja}, one can avoid this issue by changing the parameterization, though at the cost introducing other numerical complications.} which leads to numerical instabilities in \gls{ode} integration techniques such as Runge--Kutta methods.
The point at which one should switch from an inspiral to a transitional model (in terms of matching at first post-adiabatic order) is not yet fully understood for generic trajectories.  
We can get some guidance by considering quasi-circular orbits, for which it is well understood that the transition away from adiabatic inspiral happens when $p - p_{\rm sep} \approx K(\nu/10^{-5})^{2/5}$~\cite{ori2000transition, Apte:2019txp, Kuchler:2024esj}; the numerical coefficient $K$ is roughly $0.05$ for zero-spin systems, and varies with primary spin and orbital inclination by a factor of order several tens of percent.
As our inspiral model extends to $p - p_\mathrm{sep} = 2\times10^{-3}$ (see~\cref{sec:trajectory}), we expect it will be capable of incorporating such a transitional component in the future, though extending these transition models to eccentric orbits is still an area of active research~\cite{Becker:2024xdi,Lhost:2024jmw}.

In principle, this mathematical framework for obtaining inspiral trajectories is the same as that described in \cite{Katz:2021yft}, which focused on inspirals into Schwarzschild black holes.
However, extending this framework to incorporate spin on the primary component introduces additional complexity that raises computational costs and accentuates any systematic errors present (such as interpolation error) for the following reasons:
\begin{itemize}
    \item Introducing $a$ increases the dimensionality of the $\hat{f}^{(0)}_{p,e}$ data grids from two to three dimensions.
    This increases both the computational cost of interpolating these grids and the error incurred in this interpolation, the latter of which must be carefully controlled to ensure waveforms are accurate.
    \item 
    For $a \neq 0$, the separatrix no longer takes the simple form $p_\mathrm{sep} = 6+2e$ 
    and a more costly root-finding operation is often required~\cite{stein2020location,Ng:2025maa}\footnote{
    In principle, it is indeed possible to express the Kerr equatorial separatrix 
    in a closed-form analytical expression $p_\mathrm{sep}(a,\,e)$
    through the solution of degree 4 polynomials for $p$ (cf. Ref.~\cite{Compere:2021bkk}). However, the expression is relatively lengthy (involving nested radicals) and it is found that the computational efficiency is similar to that of the pre-existing numerical root finding method. For that reason, we shall not employ this analytical expression in this work.}.
    \item The separatrix moves to smaller values of $p$ as $a \rightarrow 1$.
    As $\hat{f}^{(0)}_{p,e}$ and $\hat{\Omega}_{\phi, r}$ vary increasingly rapidly as $p$ decreases, the \gls{ode} integrator must take smaller steps to maintain a given error tolerance, further increasing computational costs.
\end{itemize}

A description of how the \gls{few} inspiral generation framework must be modified to address these issues is presented in~\cref{sec:trajectory}. 

\subsection{Mode amplitudes for Kerr inspirals}
The amplitude of each mode may be written as
\begin{equation}\label{eq:Hlmn}
    H_{\ell m n}(t) = A_{\ell m n}(t){}_{-2}S_{\ell m n}(\theta; \hat{\omega}_{mn})e^{im\phi},
\end{equation}
where ${}_{-2}S_{\ell m n}$ are spheroidal harmonics with spin-weight $-2$, and 
\begin{equation}
\label{eq:amps}
    A_{\ell m n}(t) = -2\frac{\hat{Z}_{\ell m n}^\infty(a, p, e, x_I)}{\hat{\omega}_{mn}^2(a, p, e, x_I)}.
\end{equation}
In analogy to~\cref{eq:mode-phase}, $\hat{\omega}_{mn} = m\hat{\Omega}_\phi + n\hat{\Omega}_r$ is the frequency of the mode, and the amplitude $\hat{Z}_{\ell m n}^\infty$ is obtained by solving the Teukolsky equation \cite{Hughes:2021exa}.
As the orbital elements $(p, e, x_I)$ evolve during an inspiral, the dependent quantities $\hat{\omega}_{mn}$, $\hat{Z}_{\ell m n}^\infty$ and $S_{\ell m n}$ evolve as well; we indicate this by expressing $A_{\ell m n}$ and $H_{\ell m n}$ as functions of time.    

The evolution of the spheroidal harmonics has the potential to greatly increase the cost of computing each mode, since it appears that one must continually recompute these functions as $\hat{\omega}_{mn}$ evolves along the inspiral.  
To avoid this cost, we use the fact that the spin-weighted spheroidal harmonics $_{-2}S_{\ell m n}(\theta; \hat{\omega}_{mn})e^{im\phi}$ can be expanded into spin-weighted \textit{spherical} harmonics $_{-2}Y_{\ell m}(\theta,\phi)$ very efficiently (indeed, the spheroidal harmonics exactly reduce to spherical harmonics when $a = 0$~\cite{Teukolsky1973}).
These functions do not vary during an inspiral and are inexpensive to compute. 
We take advantage of this behavior by projecting the amplitudes $A_{\ell m n}$ onto the spherical harmonic basis, allowing us to build each wave mode using angular functions that do not vary along the trajectory. 
The mode spectrum which results is generally broader in the deep strong field accessible to prograde inspirals (i.e., higher multipoles contribute more significantly to the overall waveform), which places more stringent requirements on the framework used to generate these amplitudes. 
The remapping procedure and subsequent modifications to the \gls{few} amplitude module that are required in order to accommodate this increased complexity are described in~\cref{sec:amplitudes}.

\section{Model implementation}
\label{sec:implementation}
\subsection{Inspiral trajectory}
\label{sec:trajectory}

The accurate and efficient integration of~\cref{eq:elements-ode,eq:phases-ode} is an essential component of the \gls{few} framework.
However, as pointed out in~\cref{sec:trajectory-intro}, incorporating the effects of spin presents additional challenges that must be addressed if accuracy and efficiency are to be retained.
In this section, we will demonstrate how the existing \gls{few} framework for trajectory generation (which previously modelled zero-spin systems) can be modified for use as part of a Kerr equatorial eccentric waveform model.
We will also discuss general modifications to the trajectory module of \gls{few} we have made that improve the accuracy and efficiency of trajectories and waveforms, but are not limited in scope to the introduction of \gls{mbh} spin into the model.

\subsubsection{Computation of trajectory fluxes}
\label{sec:flux-calculations}
To evolve the trajectories, we must compute the forcing functions $\hat{f}^{(0)}_{p,e}$ 
on the righthand side of~\cref{eq:elements-ode}. 
At adiabatic order, $\hat{f}^{(0)}_{p,e}$ are related to the flux of energy and angular momentum lost by the system (both to null infinity and through the event horizon of the primary) due to \gls{gw} emission,
\begin{align}
\label{eq:forcesJac}
    \hat{f}^{(0)}_\alpha = -\frac{m_1^2}{m_2}\left[\frac{\partial \alpha}{\partial {E}}\left\langle\frac{dE}{d{t}} \right\rangle_\mathrm{GW} + \frac{\partial \alpha}{\partial {L}}\left\langle\frac{d{L}}{d{t}} \right\rangle_\mathrm{GW}\right],
\end{align}
where $\alpha = (p,e)$ and the Jacobian elements ${\partial \alpha}/{\partial E}$ and 
${\partial \alpha}/{\partial L}$ are analytic functions of $(a,p,e,x_I)$ as given in Appendix B of Ref.~\cite{Hughes:2021exa}.
The time-averaged \gls{gw} fluxes $\langle{dE}/{dt}\rangle_\mathrm{GW}$ and $\langle{dL}/{dt}\rangle_\mathrm{GW}$ are calculated from the Teukolsky equation using techniques in black hole perturbation theory, as described in, e.g., Refs.~\cite{Fujita:2020zxe, Skoupy:2021asz, Hughes:2021exa}. 
We omitted the Carter flux $\langle{dQ}/{dt}\rangle_\mathrm{GW}$ in~\cref{eq:forcesJac} as it is zero for equatorial orbits.
The fluxes are also functions of $(a,p,e,x_I)$, but must be determined numerically. 

In this work, we use the \texttt{Python} library \textsc{pybhpt} \cite{Nasipak:2023kuf, zachary_nasipak_2025_15627818} to compute $\left\langle{dE}/{d{t}} \right\rangle_\mathrm{GW}$ and $\left\langle{dL}/{d{t}} \right\rangle_\mathrm{GW}$. Fluxes are computed to a requested precision of $10^{-8}$. We chose this value as a balance between computational cost and model accuracy, since multiscale analyses suggest that errors in the fluxes induce $O(\nu^{-1})$ errors in the \gls{gw} phasing \cite{Hinderer:2008dm}.
The computational cost of each flux calculation ranges from $\sim 1$ to $\sim 1000$ seconds and is highly dependent on the values of $(a,p,e,x_I)$. As \textsc{pybhpt} performs computations in the frequency-domain, we find that it is most efficient and accurate for orbits where $p > p_\mathrm{sep} + 1$ and $e < 0.1$, since these orbits have a very narrow frequency spectrum. The computational cost of the flux calculations then grows as $p \rightarrow p_\mathrm{sep}$ and $e \rightarrow 1$. (See~\cref{app:flux-timing} for further details on the timing of flux calculations.)
Compared to the rest of the waveform generation, the fluxes are incredibly expensive to compute. In the following section, we describe how this computational cost is circumvented in the \gls{few} framework.

\subsubsection{Rapid computation of ODE derivatives}
In order to integrate~\cref{eq:elements-ode,eq:phases-ode}, we must first specify the form of $\hat{f}^{(0)}_{p,e}$ and $\hat{\Omega}_{\phi, r}$.
The latter terms are simple to incorporate; analytic expressions for the dimensionless orbital frequencies of geodesics in generic Kerr, which are equal to $\hat{\Omega}_{\phi, r}$ at adiabatic order, have existed in the literature for some time (see e.g. \cite{Fujita:2009bp,vandeMeent:2019cam} and references therein) and are inexpensive to evaluate.
We employ these expressions in our framework, simplifying terms involving $x_I$ (given $x_I = 1$) where possible for efficiency and numerical stability.

Obtaining the forcing functions is less straightforward, as the data generation procedure in~\cref{sec:flux-calculations} is orders of magnitude too expensive to be applied on-the-fly during trajectory integration.
However, these fluxes vary smoothly over the parameter space and are ideal targets for interpolation schemes.
We therefore opt to generate forcing function values en-masse, producing data grids in parameter space over which to perform this interpolation.
Producing these data grids is computationally expensive, but must only be performed once; as the methods described in~\cref{sec:flux-calculations} are straightforward to parallelise, this can be achieved relatively quickly on distributed computing resources.
Of particular note is the accuracy to which the fluxes must be estimated, which is $\mathcal{O}(\nu)$ for sub-radian accumulated phase error over the course of a typical trajectory evaluation.

Setting up the forcing function interpolants entails three main decisions that affect the interpolation accuracy and computational cost.
The first is to choose an interpolation scheme.
To avoid bottlenecking trajectory generation, the cost of evaluating the interpolants should be as low as possible while retaining accuracy (provided that the number of data grid points required to attain this performance is not prohibitively large).
We therefore choose to employ a tricubic spline interpolant over a uniform grid, as implemented in the \textsc{multispline} package~\cite{multispline}.
While uniform spline methods enforce more stringent requirements on the grid spacing parameters (requiring a uniform step-size in all three dimensions), they are also far more efficient as they eliminate costly index-solving operations that are required when spacing is non-uniform.
We select $E(3)$ boundary conditions for all parameters~\cite{BEHFOROOZ1989231}, as this has been shown to improve forcing function interpolation accuracy for quasi-circular inspirals in previous work~\cite{Nasipak:2023kuf}.
With this choice of interpolating function, each set of two flux values is computed in $<\SI{1}{\micro\second}$, which is sufficient to obtain year-duration trajectories in $\sim \SI{10}{\milli\second}$.

Next, we must specify our interpolation variables.
This is essential for our framework because a regular grid cannot be constructed in $(a, p, e)$ space (recalling that $x_I = 1$ for equatorial inspirals) due to the boundary imposed by $p_\mathrm{sep}$ that varies across the parameter space.
Choosing sensible interpolation variables can also spread out rapid variation in the fluxes across the parameter space (especially near the separatrix), improving the accuracy of interpolation in these regions.
To this end, we format our data grids in terms of three interpolation variables $(u, w, z)$ that are related to $(a, p, e)$ by bijective and analytically-invertible transformations.
For brevity, we will not state these transformations in the main text here (they are given in full in \cref{app:grid_layout_appendix}) but instead summarize the important features of the resulting grid structure below:
\begin{itemize}
    \item Our grids span $a \in [-0.999, 0.999]$ and $p \in [ p_\mathrm{sep} +10^{-3}, 200]$.
    For $p - p_\mathrm{sep} > 9.001$, $e \in [0, 0.9]$.
    \item For smaller values of $p$, we employ an eccentricity taper such that the grids span a smaller ranges of eccentricities as $p$ decreases.
    Data with high $e$ and low $p$ are extremely computationally expensive to compute accurately, but sources in this region of the parameter space are short-lived and therefore are less likely to be detectable by \gls{lisa}.
    In order to reduce computational costs, we exclude this region of parameter space from our model.
    \item We separate the parameter space into ``inner'' and ``outer'' regions (with some smaller overlap between them) such that this taper finishes at $p - p_\mathrm{sep} = 9.001$ without introducing a discontinuity into our grid coordinate transformations (which can lead to interpolation artifacts).
\end{itemize}
The shape and grid-point distribution of these grids is also shown in \cref{fig:data-domains} and \cref{fig:flux-timing} in \cref{app:grid_layout_appendix}, which clearly displays the structure of the eccentricity taper.
Functions for performing these mappings are implemented in \gls{few} and are readily applicable in the construction of user-specified grids.

Finally, we must choose the dimensions of the grid.
With all other aspects of the problem fixed, this is essentially a trade-off between the accuracy of the interpolant and the computational cost of producing the grids.
However, assessing how interpolation error in trajectory modelling propagates to the accuracy of the resulting waveform is not straightforward.
We therefore opt to identify empirically what flux grid dimensions are required to satisfy waveform accuracy requirements.
Based on this process, we settle on the dimensions $(N_u, N_w, N_z)$ = $(129, 65, 65), (65, 33, 33)$ for the inner and outer grid regions respectively.
As we will later demonstrate in \cref{sec:validation}, this ensures that grid point density (and therefore, the accuracy of the interpolated forcing functions) is sufficient for our model to be highly robust over the majority of the parameter space.

\subsubsection{ODE integration with continuous solution}
\label{sec:dense-output}
In the previous iteration of \gls{few}, trajectories were intregrated numerically with routines from the \gls{ode} module of the \textsc{gsl} software package~\cite{GSL}.
This provided the necessary tools to obtain trajectories in milliseconds without the need to develop and test \gls{ode} integrator codes.
However, this choice came with the significant drawback of inflexibility; it required that trajectory models be implemented in \texttt{C} (such that end users were forced to re-build the \gls{few} package for any non-trivial modifications to the inspiral model), and prohibits extension of the integration scheme to the \gls{gpu}.

In \gls{few} v2, we address these limitations with a bespoke implementation of the explicit embedded Dormand--Prince 8(5,3) Runge--Kutta method~\cite{hairer1993solving}.
The integrator logic is implemented entirely in \texttt{Python}, consisting of vectorised array operations in order to remain performant.
It accesses the \gls{ode} derivatives via a new class interface that the user can readily adapt in order to substitute these derivatives for any \texttt{Python} function they choose.
We anticipate this flexibility will greatly streamline the process of extending the \gls{few} framework to investigate fundamental physics or environmental effects, which typically entails the modification of the trajectory fluxes (see e.g., Refs.~\cite{Kocsis:2011dr,Barausse:2014tra,Speri:2022upm,Duque:2024mfw,Speri:2024qak} for examples of such modifications).

One key component of this new integrator framework is access to the \textit{continuous solution} of the \gls{ode} solver. 
This allows for the construction of a $C^1$-continuous interpolation of the evolution of the \gls{ode} variables as a 7-th order piecewise polynomial~\cite{hairer1993solving}.
As an example, the azimuthal phase between time $t_i$ and $t_{i+1}$, where $t_i$ is the time of the $i$-th node of the numerical ODE solution, can be interpolated via
\begin{align}
    {}^{i}\Phi_\phi(t) =& {}^ic_0 + s({}^ic_1 + \bar{s}({}^ic_2 + s ({}^ic_3 + \bar{s} ({}^ic_4 \nonumber \\
                        & +s ({}^ic_5 + \bar{s}({}^ic_6 + {}^ic_7 s))))))\,,
\end{align}
where $\bar{s} = 1-s$ and  $s = (t - t_i) / (t_{i+1} - t_i) \in [0,1)$.
The coefficients ${}^ic_{j}$ for $j \in [0,7]$ are functions of twelve intermediate evaluations of the \gls{ode} derivatives performed during an integration step (along with three further evaluations) and are therefore relatively inexpensive to compute. 
The explicit relation between the intermediate steps and the ${}^i c_j$'s can be found in Ref.~\cite{hairer1993solving} in terms of a constrained system of linear equations in terms of the Runge--Kutta coefficients.

The benefits greatly outweigh the additional computational cost of three additional derivative evaluations.
A continuous approximation of $p(t)$ and $e(t)$ greatly simplifies the final root-finding step of trajectory evaluation (which was previously performed with an iterative Euler-step method) and retains the desired error tolerance of the integrator, improving the stability of the trajectory with respect to small perturbations in initial conditions.
In \gls{few} v1, cubic splines of the orbital phases are constructed as this continuous solution is not available, resulting in unstable waveform derivatives (and therefore, information matrices~\cite{vallisneri2008use}) and (somewhat unexpectedly) led to relatively poor reconstruction of the more slowly-varying waveform phasing in the early inspiral where trajectory points are typically very sparsely distributed.
Access to an accurate interpolation of the orbital phases remedies this behaviour, ensuring that waveform numerical derivatives remain smooth and phasing is consistent for the entire waveform.

\begin{figure}[t]
    \centering
    \includegraphics[width=0.99\columnwidth]{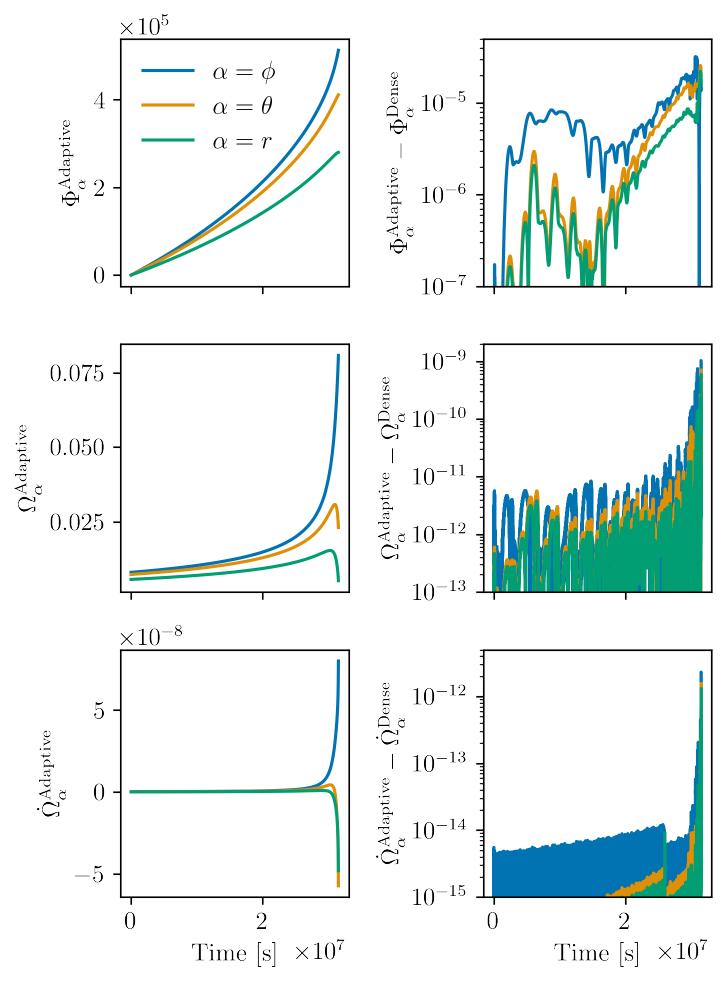}
    \caption{Continuous solutions for the orbital phase, frequency and frequency derivative (left) of the trajectory obtained with an adaptive stepping integrator, and their absolute difference (right) with respect to a densely-stepped (fixed step size) integration. The considered binary system has component masses $(10^6, 10 M_\odot)$ and $(a, e_0)=(0.998, 0.5)$. The polar phase and its derivatives (denoted here by $\theta$) do not enter waveform generation for equatorial inspirals, but are included here for completeness.}
    \label{fig:dense-output-phasing}
\end{figure}

A continuous approximation for the phase also yields similar piecewise polynomials for the orbital frequencies and their derivatives with respect to time; as shown in~\cref{fig:dense-output-phasing}, this approximation is very accurate (exceeding 9 decimal places in the frequencies for the majority of the inspiral).
This is particularly useful in the construction of frequency-domain waveforms (which require this information as an input) and will enable resonance effects (which require careful monitoring of these frequencies, as well as accurate frequency derivatives) to be seamlessly folded into the \gls{few} framework in the future.

\subsubsection{Numerical integration of ODE system}
\begin{figure}[t]
    \centering
    \includegraphics[width=0.95\columnwidth]{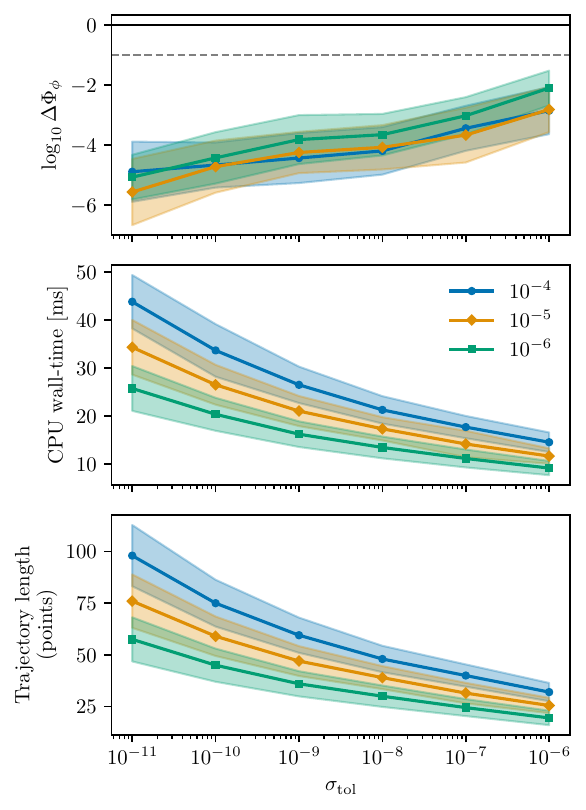}
    \caption{Trajectory characteristics as a function of \gls{ode} integrator absolute tolerance $\sigma_\mathrm{tol}$ for $100$ draws of four-year duration inspirals with $\epsilon$ of $10^{-4}$ (blue circles), $10^{-5}$ (orange diamonds) and $10^{-6}$ (green squares) respectively.
    \textbf{Top panel:} Deviation of the phase $\Phi_\phi$ at end of inspiral with respect to a trajectory with $\sigma_
    \mathrm{tol} = 10^{-13}$. \textbf{Middle panel:} Wall-time per trajectory evaluation. \textbf{Bottom panel:} Number of adaptive points in the trajectory solution.}
    \label{fig:traj_ode_error}
\end{figure}

With our framework for obtaining the derivatives in \cref{eq:elements-ode,eq:phases-ode} and integrating the resulting \gls{ode} system in place, we are ready to evolve a trajectory from a set of initial conditions $\{p_0, e_0, \Phi_{\phi0}, \Phi_{r0}\}$ given specific values for $M$, $\nu$ and $a$.
To improve the numerical stability of the integrator, we first rescale the problem such that the integration variables vary by $\mathcal{O}(1)$ across an inspiral.
We achieve this by integrating the system with respect to the radiation-reaction timescale $t_\mathrm{rr} = \nu t$ and eliminating the pre-factor $\nu$ from \cref{eq:elements-ode} accordingly.
Rather than scaling \cref{eq:phases-ode} similarly, we leave it unchanged (in order to achieve the desired property of $\mathcal{O}(1)$ scaling for all parameters) and instead rescale the orbital phase evolution post-integration by $\nu^{-1}$.
This approach is valid in the adiabatic approximation because (by definition) the $\hat{f}^{(0)}_{p,e}$ are independent of the orbital phases.
A consequence of this choice is that, for an integrator absolute error tolerance $\sigma_\mathrm{tol}$, the effective error tolerance on the orbital phases is $\sigma_\mathrm{tol} /\nu$.
This is in fact a desirable outcome: we are mainly interested in tuning the accuracy with which $p(t)$ and $e(t)$ are evolved (as small errors in these parameters will gradually accumulate over an inspiral, coupling in turn to the orbital phase evolution), whereas the error in the reconstructed phase will remain of order $\sigma_\mathrm{tol} /\nu$, which is small and has negligible impact on waveform phase accuracy. 
This scaling is demonstrated in \cref{fig:traj_ode_error}, where phase error grows in proportion to (but is several orders of magnitude larger than) the error tolerance of the \gls{ode} integrator.
Trajectory wall-time varies significantly with respect to \gls{ode} error; to ensure trajectories are accurate to within $10^{-3}$ radians, we set $\sigma_\mathrm{tol} = 10^{-11}$ by default, but this can be relaxed by the end user according to their accuracy requirements.
We also note that trajectory wall-time will be hardware-dependent; the timings presented in \cref{fig:traj_ode_error} were obtained with an Apple M3 processor.

\AMEND{Trajectory integration can be performed either forwards or backwards with respect to time (the latter achieved by negating \cref{eq:elements-ode,eq:phases-ode})}.
The size of the first integration step from the starting point is chosen automatically with standard techniques~\cite{GLADWELL1987175}.
The integrator then steps adaptively, producing sparse trajectories of $\sim 100$ points in length.
\AMEND{During forwards integration,} we terminate adaptive time-stepping once the integrator either attempts to step to a point with $p < p_\mathrm{stop}$, where $p_\mathrm{stop} = p_\mathrm{sep} + \Delta p_\mathrm{buf}$ with $\Delta p_\mathrm{buf} = 2\times10^{-3}$, or exceeds the maximum duration $T$ set by the user\footnote{Recall that we terminate our grid at $p_\mathrm{sep} + 10^{-3}$, both to avoid numerical instabilities in the adiabatic equations of motion near the separatrix and because the dynamics in this regime are better modelled in a transition-to-plunge (rather than inspiral) framework. 
We then choose $\Delta p_\mathrm{buf} = 2\times 10^{-3}$ to prevent trajectories from reaching this grid boundary.}. 
In the former case, we must then place a point at $p = p_\mathrm{stop}$ to avoid truncation errors in the results of data analysis.
As the adaptive stepping procedure first attempted to step past $p_\mathrm{stop}$, we have access to a continous solution in this interval; we therefore perform a numerical root-finding operation via Brent's method to find the value of $t$ where
\begin{equation}
    \delta p_\mathrm{stop}(t) = p(t) - p_\mathrm{stop}(p(t), e(t))
\end{equation}
is equal to zero.
This typically converges to within $\sigma_\mathrm{tol}$ of the root in $\sim 5$ iterations and contributes little to the overall cost of a trajectory evaluation.
Once the root $t_\mathrm{root}$ has been found, the final trajectory point is obtained by evaluating the continuous solution at $t_\mathrm{root}$.
In the latter case of a trajectory duration exceeding $T$, this root-finding operation is not required and the 
continuous solution is simply evaluated at $t=T$ in order to obtain the last trajectory point.
\AMEND{Backwards integration proceeds similarly, except that $p_\mathrm{stop}$ is now determined by the boundaries of the forcing function data grids (see \cref{app:grid_layout_appendix} for these definitions).
If backwards integration reaches this grid boundary before the duration exceeds $T$, a final point is placed at the boundary and integration is terminated.}

The outputs of the trajectory integration are then fed along to later stages of waveform generation. 
The intermediate evaluation coefficients for the orbital phases are passed directly to the waveform summation stage (in a similar manner to which cubic spline coefficients for these quantities were inserted in the original \gls{few} implementation).
As the mode amplitudes are functions of $p(t)$ and $e(t)$, they cannot be interpolated with the continuous \gls{ode} solution; we instead retain the existing behaviour, passing the sparsely-distributed $p(t)$ and $e(t)$ points to the amplitude module to be evaluated (and subsequently interpolated with cubic splines).
Improvements made to this amplitude module are discussed in the next subsection.

\subsection{Waveform amplitudes}
\label{sec:amplitudes}

\begin{figure*}
\begin{center}
    \includegraphics[width=0.8\linewidth]{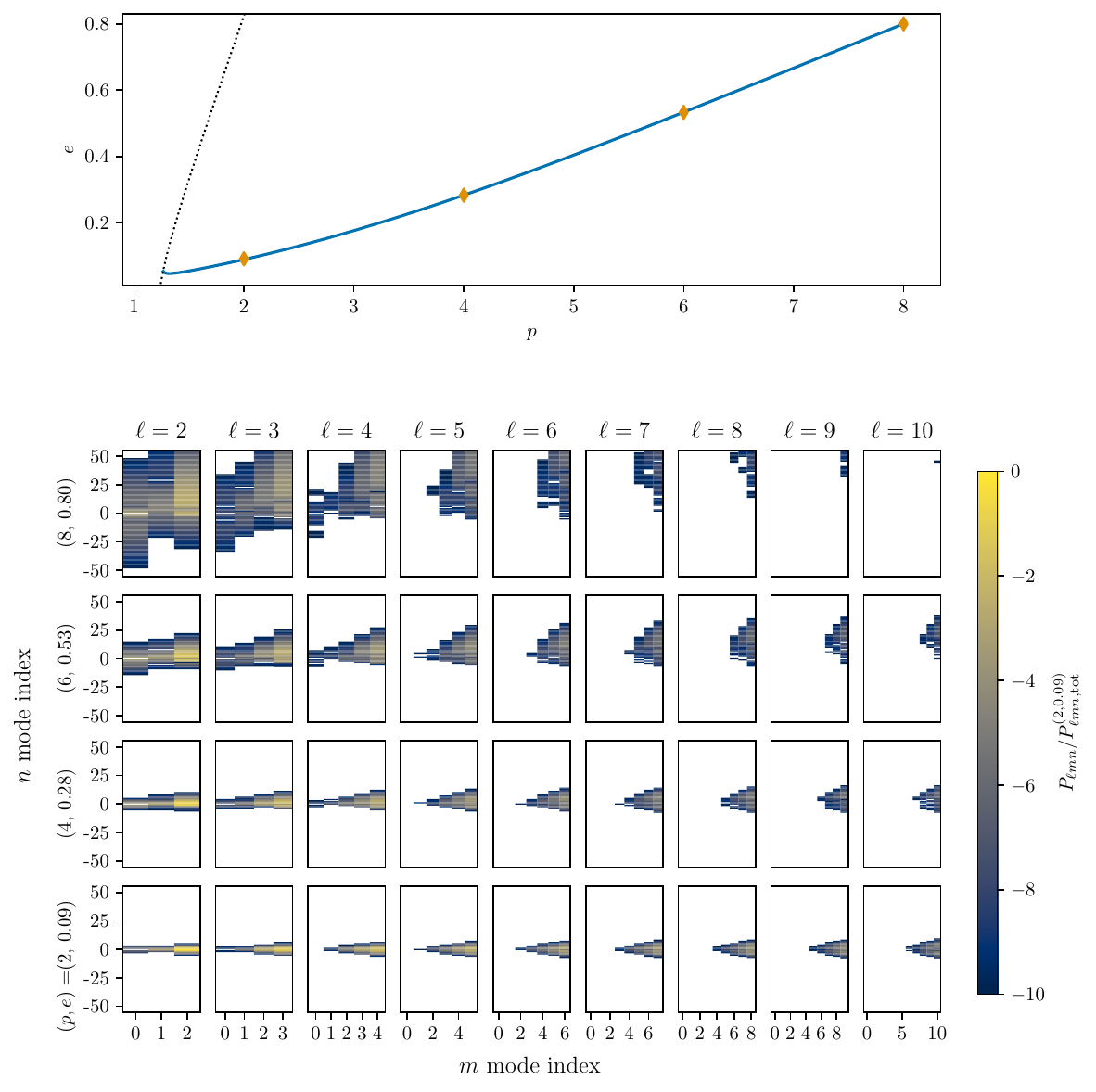}
    \caption{\textbf{Top panel:} The blue curve represents a highly eccentric and equatorial inspiral trajectory for $\epsilon = 10^{-5}$ and $a = 0.998$. 
    The black dotted line describes separatrix $p_{\text{sep}}$, which evolves as a function of eccentricity. The orange diamonds represent individual snapshots of the trajectory. \textbf{Bottom panel:} A plot of the mode spectrum normalised by the total mode power $P_{\ell mn,\rm tot}^{(p = 2, e = 0.09)}$ at the points in the trajectory indicated by the orange diamonds. 
    Modes with a normalised power below $10^{-10}$ are not shown.
    As the orbital parameters evolve, the distribution of the mode power with respect to mode index changes significantly, presenting the need for wide mode index coverage in a typical inspiral.
    The number of modes per row that account for $99\%$ of the total power in that row are $(129, 40, 24, 26)$; the union of these mode sets consists of $160$ elements (due to the evolving shape of the mode spectrum).
    For $99.999\%$ of the total power (our default for waveform generation) these counts become $(718, 345, 199, 174)$ and $988$ respectively.
    }
    \label{fig:heatmap_mode_amplitudes} 
\end{center}
\end{figure*}

\subsubsection{Teukolsky amplitude data grids}
\label{sec:teukolsky-amplitudes}
As with the orbital-element forcing functions, the mode amplitudes $A_{\ell m n}$ are too expensive to compute as part of a rapid waveform generation framework.
Instead, we may obtain $A_{\ell m n}$ via interpolation over pre-computed data grids. 
Extending the amplitude data from Schwarzschild spacetimes to Kerr spacetimes presents two key complications:
\begin{enumerate}
    \item Kerr trajectories evolve more ``deeply'' in the strong field of the \gls{mbh} (i.e, they can have smaller values of $p$), leading to more $(\ell,m,n)$ harmonics contributing to the \gls{gw} strain.
    \item The \gls{gw} strain naturally decomposes onto a basis of \emph{spheroidal} harmonics that, like the amplitudes, also evolve in time -- note the dependence on $\hat{\omega}_{mn}$ in the spheroidal harmonic in \cref{eq:Hlmn}.
\end{enumerate}
To address the former issue, we simply precompute more waveform harmonics for the Kerr model, specifically all $(\ell,m,n)$ modes within the limits $2\leq \ell \leq \ell_\mathrm{max}=10$, $|m| \leq \ell$, and $|n| \leq n_\mathrm{max} = 55$.
For computational efficiency, we only compute mode amplitudes with $m \geq 0$ and infer amplitudes for $m < 0$ with a conjugate mode symmetry~\cite{Hughes:1999bq,Fujita:2020zxe}, as was performed in \gls{few} v1~\cite{Katz:2021yft}.

To circumvent the second complication, we expand the spheroidal harmonics in terms of the spherical harmonics, as proposed in \cite{Katz:2021yft} and implemented in other Kerr models \cite{Nasipak:2023kuf, Khalvati:2024tzz}. 
This is expressed by the relation
\begin{equation} \label{eq:YlmToSlm}
    S_{\ell m n}(\theta, \hat{\omega}_{mn})e^{i m \phi} = \sum_{j=\ell_\mathrm{min}}^{\infty}b_{\ell m n}^j (t) _{-2} Y_{j m}(\theta, \phi),
\end{equation}
where $\ell_\mathrm{min} = \mathrm{max}(2, |m|)$ and $b_{\ell m n}^j(t)$ are spherical-spheroidal mixing coefficients.
Further discussion regarding this expansion [and the computation of $b_{\ell m n}^j(t)$] can be found in, e.g., Appendix A of \cite{Hughes:1999bq}.
The time dependence of the spheroidal harmonics is then absorbed by $b_{\ell m n}^j(t)$. 
Inserting \cref{eq:YlmToSlm} into \cref{eq:source-frame-waveform}, we then sum over the spheroidal harmonic mode amplitudes $A_{\ell m n}$ to obtain their corresponding \emph{spherical} harmonic amplitudes,\footnote{See also, e.g., Appendix A of \cite{Yunes:2010zj},  
Sec.~IIIB of \cite{vandeMeent:2016pee} 
and Sec.~IIB of \cite{Nasipak:2023kuf} for explicit derivations of this procedure.}
\begin{align}
\label{eq:sphericalproj}
    \mathcal{A}_{\ell m n} = \sum_{j = \ell_\mathrm{min}}^\infty b^{\ell}_{jmn} A_{j m n}.
\end{align}
After this remapping, the waveform strain is then written as
\begin{equation}
    \label{eq:remapped-source-frame-waveform}
    h(t) = \frac{\mu}{d_\mathrm{L}}\sum_{\ell m n}\mathcal{A}_{\ell mn}(t) _{-2}Y_{\ell m}(\theta, \phi) e^{-i\Phi_{mn}(t)}.
\end{equation}
As \cref{eq:remapped-source-frame-waveform} expresses the waveform strain for eccentric equatorial Kerr inspirals in the same form as that of Schwarzschild inspirals, we are free to apply \gls{few}'s rapid mode summation framework without modification.
This approach is readily extendable to eccentric and inclined Kerr inspirals once flux and mode amplitude data for these systems are available~\cite{Fujita:2009us,Fujita:2025pc,Katz:2021yft,Hughes:2021exa,Chen:2023lsa, Kerachian:2023oiw}.

\subsubsection{Interpolation of mode amplitudes}
\label{sec:amplitude-interpolation}
Due to the large number of harmonic modes that contribute significantly to a typical \gls{emri} waveform, obtaining their amplitudes efficiently via interpolation is a vital component of the \gls{few} waveform generation framework.
\AMEND{
As was performed in \gls{few} v1, we opt to interpolate the complex amplitudes in terms of their real and imaginary components.
These quantities vary smoothly over the parameter space, and this allows for both the leading-order term $|\mathcal{A}_{\ell m n}|$ and the post-adiabatic contribution $\psi_{\ell m n} = \arg(\mathcal{A}_{\ell m n})$ to be obtained directly, at the cost of the additional memory needed to store data grids for both components at full resolution.
While the latter term is not strictly-speaking necessary for an adiabatic model, we seek to demonstrate that all effects at first post-adiabatic order can readily be included in the FEW framework, and so include this for completeness.
While it is tempting to interpolate $|\mathcal{A}_{\ell m n}|$ and $\psi_{\ell m n}$ directly (with the accuracy requirements on the latter being relaxed due to its post-adiabatic nature), the former is non-smooth (due to zero-crossings in the amplitude surfaces) and the latter must be unwrapped to avoid discontinuities (which is challenging in multiple dimensions).
We plan to investigate whether a form of amplitudes exists that retains the benefits of both conventions in more detail in future work.
}

An important feature of this interpolation problem is that it is highly parallelisable with respect to both the mode indices $(\ell, m, n)$ and the parameters $(a, p, e)$.
It is therefore sensible to design an interpolation framework that exploits this by computing mode amplitudes over many mode index combinations and parameter values simultaneously.
For \gls{few} v1, this was achieved by fitting a reduced-order model (in the form of a neural network) that rapidly evaluated a set of basis coefficients for multiple sets of input parameters.
These coefficients were then converted into a complete set of mode amplitudes (i.e. over all sets of mode indices) via a matrix dot product; see Section 4 of \cite{Katz:2021yft} for further discussion.
This approach is computationally efficient and well-suited to acceleration via \gls{gpu} vectorisation techniques.
However, it suffers two significant drawbacks.
First, the model struggles to accurately estimate the amplitudes of weaker (but still significant) modes.
This was identified in previous work and was observed to limit the accuracy of the waveforms obtained to mismatches $\sim 10^{-5}$~\cite{Katz:2021yft}, which is on the order of the accuracy required for unbiased analysis of the loudest expected astrophysical \gls{emri} signals (which may have \glspl{snr} in the high hundreds~\cite{Babak:2017tow}; see \cref{sec:waveform-validation} for further discussion).
Note that for quieter sources, larger mismatches than this will not necessarily lead to parameter biases~\cite{Katz:2021yft}; such a requirement is only enforced by the (potential) existence of particularly loud (and therefore 
informative) signals. 

Even if this limitation can be overcome (e.g. by tuning the loss function used to train the neural network), a second drawback to this approach is that it is completely inflexible with respect to the requested mode content of the waveform.
For instance, if the user wishes to generate a waveform containing only a handful of modes (e.g. as part of the early stage of a signal extraction algorithm), the model must first generate every mode amplitude before discarding the majority of them.
Potential improvements to the mode selection process that reduce the number of mode amplitudes to be generated, such as a parameterized ``mask'' in mode index space, will similarly be affected (albeit to a lesser extent).
We therefore seek an alternative approach that addresses these limitations, accurately and adaptively constructing the mode amplitude spectrum without compromising efficiency.

As each $\mathcal{A}_{\ell m n}$ varies smoothly over the parameter space, in a similar manner to the fluxes, they can be accurately modelled with interpolation methods.
However, the accuracy with which each mode amplitude must be recovered is far less stringent than that of the forcing functions, as systematic errors in these amplitudes do not accumulate over the course of inspiral, whereas inaccurate \gls{ode} derivatives lead to secularly growing errors in the inspiral trajectory (and hence phasing).
Estimating each mode amplitude with a relative error of $\sim10^{-2}$ is sufficient for systematic biases to remain small with respect to measurement precision for astrophysically-relevant systems with \glspl{snr} of $\sim 100$~\cite{lindblom2008model}, with some dependence on how \gls{snr} is distributed between modes (which varies as a function of spin, eccentricity and orbital separation).
What is important is to attain this relative error across the spectrum of mode amplitudes (which spans several orders of magnitude). Basis-compression approaches like reduced-order modelling can struggle to achieve this global relative error.

We instead opt to perform spline interpolation, which is particularly well-suited to this task and addresses the limitations outlined in the previous paragraph.
Interpolating each mode amplitude separately ensures that the weaker modes in the spectrum are still estimated accurately.
Mode amplitudes are readily interpolated in parallel, but as each interpolation operation is easily separable from the others, it is straightforward to evaluate only a requested sub-set of the full mode spectrum.
The relationship between $\mathcal{A}_{\ell m n}$ and $a$ is sufficiently weak (at least for non-extremal systems~\cite{burke2020constraining,gralla2016inspiral,Nasipak:2023kuf}) that linear interpolation with respect to this parameter performs adequately over the majority of the parameter space, further reducing computational costs.
We therefore interpolate mode amplitude data with a ``bicubic+linear'' spline interpolation framework, striking a balance between accuracy and efficiency.
Similarly to the case of the fluxes, we define data grid coordinates $(u, w, z)$ connected to $(p, e, a)$ by invertible transformations that are described in \cref{app:grid_layout_appendix}. 
These transformations are of similar form to those used for the forcing function grid setup, but with adjusted coefficients to ensure the strong-field region is sufficiently densely sampled to capture the rapid amplitude variations that are characteristic of this region of the parameter space.
The form of the eccentricity tapering is correspondingly altered to ensure it retains the same shape as the flux grids, such that the forcing function and amplitude interpolation domains (and therefore, the domain over which the waveform is defined) are consistent.
The value of the mode amplitude $\mathcal{A}_{\ell m n}$ given grid coordinates $(u, w, z)$ is then written as
\begin{equation}
    \mathcal{A}_{\ell m n} =  (\mathcal{A}_{\ell m n}^\mathrm{+} - \mathcal{A}_{\ell m n}^\mathrm{-})\frac{z -\hat{Z}_\mathrm{-}}{\hat{Z}_\mathrm{+} - \hat{Z}_\mathrm{-}} + \mathcal{A}_{\ell m n}^\mathrm{-},
\end{equation}
where $\mathcal{A}_{\ell m n}^\mathrm{-,+} \equiv \mathcal{A}_{\ell m n}(\hat{Z}_\mathrm{-,+})$ are obtained via a bicubic spline interpolant over $(u, w)$, and $(\hat{Z}_\mathrm{-}, \hat{Z}_\mathrm{+})$ are the values of the grid points immediately (below, above) $z$.

To justify the wide range of modes we include in our waveform model, in \cref{fig:heatmap_mode_amplitudes} we apply our amplitude interpolant to demonstrate how the morphology of the waveform mode spectrum changes over an eccentric $(e_0 = 0.8)$ inspiral into a rapidly-spinning $(a=0.998$) \gls{mbh}, plotting the mode power $P_{\ell m n} = |A_{\ell m n}|^2$ as a function of $(\ell, m , n)$.
At early times, the waveform contains rich $n$-mode content due to the high initial eccentricity, but the mode spectrum rapidly tails off with respect to $\ell$ as the \gls{co} is orbiting in the relatively weak field.
As the inspiral circularizes due to \gls{gw} emission, the higher-order $n$-mode contribution gradually weakens;  meanwhile, the high spin of the \gls{mbh} allows the \gls{co} trajectory to extend deep into the strong field of the \gls{mbh}, significantly accentuating the higher $\ell$-mode content of the waveform.
Typically, for systems with higher eccentricities at lower semi-latus rectum values than what is shown in \cref{fig:heatmap_mode_amplitudes}, more power is contained in the higher $n$-mode content of the spectrum (particularly for higher values of $\ell$)~\cite{Drasco:2005kz,Fujita:2009us}.

One significant drawback of interpolating each mode amplitude with cubic splines (as opposed to a reduced-basis representation) is that the size of the spline coefficient arrays that must be stored in memory grows linearly with the number of modes supported in the model.
For the 6993 modes that we interpolate in our waveform model, these coefficients arrays total $\sim \SI{5}{\giga\byte}$ in size, which is sufficiently compact to fit in the memory of most \glspl{gpu}. 
However, some care will be required once \gls{few} is extended to inclined and eccentric inspirals in future work, as the number of waveform modes (and therefore the memory requirements) will increase significantly (with typically as many as $\sim 10^5 - 10^6$ waveform modes~\cite{Katz:2021yft,Hughes:2021exa,Isoyama:2021jjd}).

\subsection{Mode selection and waveform summation}
\label{sec:selection-summation-intro}
While the inclusion of spin on the primary introduces additional complexity to the generation of inspiral trajectories and mode amplitudes, once these have been addressed, the waveform may be constructed in much the same way as for spinless systems.
We review this procedure below, which is largely unchanged except for the incorporation of the continuous \gls{ode} solution described in \cref{sec:dense-output}; see \cite{Katz:2021yft} and \cite{Speri:2023jte} for more information regarding these stages of waveform generation in the time domain and frequency domain, respectively.

Before performing the summation in \cref{eq:source-frame-waveform}, we apply a mode selection process to reduce the number of modes that must be summed without sacrificing accuracy, which proceeds as follows.
First, at each sparse trajectory point, we compute the mode power $|A_{\ell m n}|^2$ for each waveform mode in our amplitude data grids.
We then sort the list of mode power values at each individual point, perform a cumulative summation over this list and truncate the sum once it exceeds a fraction $(1 - \kappa)$ of the total power.
We assume that modes that do not pass this cut-off do not strongly contribute to the overall waveform.
However, to ensure that the waveform is smooth with respect to time, we take the union of the sets of modes that pass the individual cut-offs and retain only these modes for waveform summation.
The tuning parameter $\kappa$ controls the accuracy of the waveform with respect to this procedure; in line with previous work, we set $\kappa = 10^{-5}$ by default. 
We have verified that this value is a sensible choice for a variety of \glspl{emri} and \glspl{imri}, as discussed in~\cref{sec:inference-subsection}.

After mode selection, we perform the summation in \cref{eq:source-frame-waveform} over the remaining sets of mode indices.
For time-domain waveforms, we interpolate the sparse quantities $\mathcal{A}_{\ell m n}$ and $\Phi_{m n}$ (with cubic splines and the continuous \gls{ode} solution respectively) on a uniform grid of time values of spacing $\mathrm{d}t$, summing the interpolated waveform modes together according to \cref{eq:source-frame-waveform}.
Frequency-domain waveforms are obtained via the approach described in \cite{Speri:2023jte}, making use of the stationary phase approximation.
As this approach assumes that the frequency evolution of each mode is described by cubic splines (such that the stationary phase relation has analytic roots), we retain this structure rather than incorporating the continuous solution depicted in \cref{fig:dense-output-phasing}.
The frequency-domain waveform also requires higher-order frequency derivatives to approximate the Fourier phase in each bin; as the derivatives of this cubic spline are less accurate than those obtained with the continuous solution, we rewrite this part of the frequency-domain summation to use the coefficients output by the \gls{ode} solver.
We found that this significantly improved the consistency of the time-domain and frequency-domain waveforms.
Modifying the root-finding procedure of the stationary phase approximation to incorporate our more accurate solution for the frequency evolution (e.g. by taking the analytic root as an input to Newton's method, which will converge in 1-2 iterations) will further improve this consistency at minimal computational cost, and is left to future work.

Once either time- or frequency-domain waveforms have been obtained in the source-frame, they are transformed to the detector frame (as described in \cite{katz2021fast}) for use in data analysis. 

\section{Validation of model accuracy}
\label{sec:validation}
With our waveform generation framework fully specified, we will now explore the accuracy of our waveform model at adiabatic order.
This is made difficult by lack of accurate waveform templates to compare against: numerical relativity simulations have not been computed (however, see recent efforts in Refs.~\cite{Lousto:2022hoq,Wittek:2024gxn,Wittek:2024pis}) in this region of the parameter space (especially for the number of orbital cycles [$\sim \mathcal{O}(\nu^{-1})$] required for useful comparisons to be made), and \gls{pn} models are not trustworthy in the strong-field region of our domain where confirming the accuracy of our model is most pertinent.
We therefore opt for a different approach in which we quantify significant sources of systematic error in our model (which in combination are responsible for any differences between our model and ``exact'' adiabatic \gls{emri} waveforms), with the aim of understanding how these errors propagate to the accuracy of our constructed waveforms.
In this way, we can roughly quantify the faithfulness of our model with respect to an ``error-free'' adiabatic model without the need for independent waveforms to compare against.
Throughout this section, we will examine each component of our waveform generation in turn: we first consider the trajectory module (\cref{sec:trajectory-validation}), followed by the amplitude module (\cref{sec:amplitude-validation}) and conclude with an examination of how these systematics impact accuracy at the waveform template level (\cref{sec:waveform-validation}).

In addition to this self-contained examination of our model, we also verify that it correctly matches other models in limiting cases in~\cref{app:model-comparison}.
In combination with the systematic validation that we perform throughout this section, the results of these comparisons confirm that our model produces accurate waveforms for spinning and/or equatorial \glspl{imri} and \glspl{emri} across its domain of validity of $|a| \leq 0.999$ and $e_0 \leq 0.9$.
In this process, we identify that our model is least accurate for: $a > 0.99$ and $p - p_\mathrm{sep} \lesssim 1$; and $e \gtrsim 0.85$. 
These are regions where self-force data vary rapidly and are most challenging to interpolate.

\subsection{Trajectory validation}
\label{sec:trajectory-validation}
In this subsection, we verify that our model for the inspiral trajectory of the \gls{co} accurately traces the orbital phasing of the system to adiabatic order over the long timescales typical of the asymmetric-mass binary systems observable by \gls{lisa}.
As our model only incorporates adiabatic contributions to the inspiral evolution, one could argue that we need only construct the phase evolution to an accuracy of $\mathcal{O}(1)$ radian, as we cannot hope to match trajectories that include post-adiabatic contributions
to better than this scale (and typically, ignoring post-adiabatic terms regularly leads to $10$-$100$ radians of phase error~\cite{Isoyama:2021jjd, Wardell:2021fyy, Burke:2023lno, Lynch:2024ohd}, at least for the quasi-circular \glspl{emri}).
However, we can assume that post-adiabatic effects will be implemented in our framework once they are readily obtainable (as they are necessary to achieve sufficient accuracy for data analysis purposes), so this is not a useful requirement for us to satisfy.
In this work, our focus remains on the accuracy of our model for the adiabatic contribution to the inspiral. 
Quantitatively assessing how phase errors translate into systematic biases in EMRI parameter estimation is not straightforward. However, in Ref.~\cite{Burke:2023lno}, which examined quasi-circular inspirals into non-spinning \glspl{mbh}, a phase error of $\sim 10^{-1}$ radians was sufficiently small for any resulting biases to remain subdominant to statistical uncertainties. Our aim therefore is to ensure that our trajectory model reproduces inspiral phasing with at least this level of accuracy.

The faithfulness of the phase evolution of our trajectory model with respect to error-free adiabatic trajectories is wholly dependent on the accuracy to which the forcing functions $\hat{f}^{(0)}_{p,e}$ are estimated.
In our framework, there are two potential sources of systematic error at this stage:
\begin{enumerate}
    \item The accuracy with which the $\hat{f}^{(0)}_{p,e}$ data grids are computed; as $\hat{f}^{(0)}_{p,e}$ are obtained from infinite mode-sums of oscillatory integrals, they are expected to contain errors due to limitations in numerically resolving the integrands and from the finite truncation of the mode-sum.
    \item The spline interpolation of these data grids, which is closely tied to the grid-point density and choice of parameter conventions.
\end{enumerate}
We begin with the first of these error sources by comparing our computed data grids with two independent data-sets.
In order for this comparison to be made, it is still necessary for us to interpolate our data --- the results we present here are therefore a combination of interpolation error and data computation errors.
The purpose of these comparisons is to ascertain whether any regions have been computed with significant errors compared to the errors incurred in this interpolation (we will examine interpolation error separately later in this subsection).

\begin{figure}[t]
    \centering
    \includegraphics[width=0.95\columnwidth]{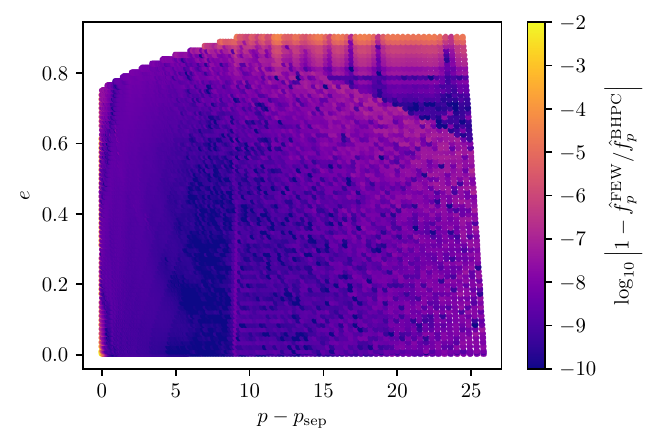}
    \caption{
    Relative error in $\hat{f}_{p}^{(0)}$ between an interpolation of our data grid and the corresponding data of Ref.~\cite{Fujita:2020zxe}, available from the BHPC data lake~\cite{BHPClub}.
    The two datasets agree to better than $8$ digits over the majority of the parameter space. 
    Here we examine prograde orbits with $a = 0.5$; we observe similar agreement for other spin values $a \in [-0.9, 0.9]$.
    }
    \label{fig:bhpc_flux_comparison}
\end{figure}

The first of these datasets was computed by the authors of Ref.~\cite{Fujita:2020zxe} 
(built on the \gls{bhpc} code developed in Refs.~\cite{Fujita:2004rb,Fujita:2005kng,Fujita:2009us})
and is publicly available on the \gls{bhpc} website~\cite{BHPClub}. It broadly and densely covers a large fraction of our parameter space.
The authors accompany their data with an error estimate informed by their truncation of mode-summations.
For $p \lesssim p_\mathrm{sep} + 1$, the estimated error becomes prohibitively large for validation purposes --- in this region, the discrepancy between our data and theirs grows proportionally to the predicted error.
We therefore excise this region of the parameter space from this comparison (and will examine it separately below with a separate dataset).
The region of the \gls{bhpc} data we retain extends to $p \sim 25$ and $e = 0.925$, providing good coverage of a large fraction of our grid.
Beyond $p \sim 25$, we validate our data separately with  \gls{bhpc}'s \gls{pn} results (see \cref{app:weakfieldappendix}).
The relative error between an interpolation of our data and the \gls{bhpc} data for $\hat{f}_p^{(0)}$ is shown in \cref{fig:bhpc_flux_comparison} --- for brevity, we do not include the comparison with $\hat{f}_e^{(0)}$, which follows similar behaviour.
We attain excellent agreement with this dataset across the parameter space, with relative error remaining below $\sim 10^{-8}$ over the vast majority of the grid.
Above $e \sim 0.8$, errors steadily rise to $\sim 10^{-4}$.
This is expected: mode-sums with respect to $n$ converge slowly in this region, requiring one to compute and sum over hundreds of $n$-modes. The large $n$-harmonics become increasingly more difficult to resolve at floating point precision and introduce their own source of error if they are under-resolved (similar to how it is incredibly difficult to calculate the 100th harmonic in a Fourier series expansion). For a single $n$-mode, this error is typically subdominant to the total error, but if the mode-sum is not truncated early enough, the accumulation of these small errors can saturate the overall relative accuracy of the results.
As both our data and that of the \gls{bhpc} are impacted by this limitation, it is not clear which of the two have been computed more accurately in this high eccentricity region, only that the two datasets are consistent to $\sim 4$ digits of precision.
This comparison is performed for prograde orbits with $a=0.5$ --- we find similarly good agreement across all spin values $(a=0.1,0.3,0.5,0.7,0.9)$ for both prograde and retrograde configurations, confirming that we have accurately computed and interpolated our $\hat{f}_{p,e}^{(0)}$ data grids over this part of the parameter space.

\begin{figure}[t]
    \centering
    \includegraphics[width=0.95\columnwidth]{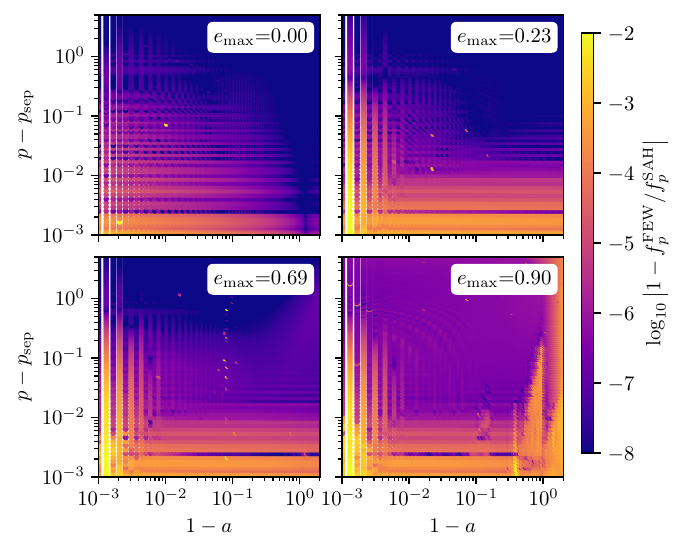}
    \caption{
    Relative error in $\hat{f}_{p}^{(0)}$ between an interpolation of our data grid and a dataset computed with the \textsc{GREMLIN} code.
    The four panels shown correspond to four slices in eccentricity that follow the tapering scheme described in \cref{app:grid_layout_appendix}.
    The two datasets generally agree to$\sim 6$ digits for $p - p_\mathrm{sep} \gtrsim 10^{-2}$, increasing to $\sim 4$ digits below this value.
    Some data quality issues are also evident in the lower right panel; see main text for discussion.
    }
    \label{fig:SAH_nearISO_comparison}
\end{figure}

To inspect the accuracy of our data grids in the strong-field region $p \rightarrow p_\mathrm{sep}$, we also compare against data computed with the \textsc{GREMLIN} code, independently developed by author Hughes and previous collaborators.  
(See Ref.\ \cite{Drasco:2005kz} for an overview of this code's foundations, and Refs.\ \cite{ThroweThesis, OSullivan:2014ywd} for discussion of updates to notation and methods that this code uses.  
Additional recent optimizations and algorithmic improvements, including a port to the Open Science Grid \cite{osg07}, will be described in a future publication, accompanied by a public release after removing proprietary code currently used by this package.)  
This data is more densely sampled in this region than our grids in $a$ and $p$; comparisons are done on $4$ slices in eccentricity that follow a taper similar to the one we include in our grid reparametrization.  
We distinguish forcing functions computed with \gls{few} and \textsc{GREMLIN} with superscript ``FEW'' and ``SAH'' respectively.
The relative error between an interpolation of our data and each of these slices is shown in \cref{fig:SAH_nearISO_comparison} for the case of $\hat{f}_{p}^{(0)}$.
We observe excellent agreement (with relative errors less than one part in a million) for $p - p_\mathrm{sep} \gtrsim 10^{-2}$.
Closer to the separatrix, we observe a larger discrepancy between the datasets, with relative errors reaching $\sim 10^{-3}$; this is expected as (especially at larger eccentricities) it is computationally challenging to compute $\hat{f}_{p,e}^{(0)}$ in this region, and our interpolation error begins to dominate at high values of $a$ (as indicated by the alternating bands of larger error with respect to $a$), as we will examine further below.
However, \glspl{emri} typically complete relatively few orbital cycles in this region of the parameter space (compared to their total number of in-band cycles) before crossing the separatrix, so even relatively large errors will not significantly impact the phase accuracy of the trajectory.

While the deviations observed are due in part to errors in the interpolation of our data grids, they are also impacted by differences in the convergence criteria associated with each software.
The \textsc{gremlin} computations truncate at either a relative precision of $10^{-7}$ or up to $n = 250$ closest to the separatrix, whereas \textsc{pybhpt} cuts off at either a relative precision of $10^{-8}$ or $n = 300$; as these higher $n$-modes can contain significant errors due to limited numerical precision, truncating the mode-sum at different values of $n$ may yield results that differ substantially beyond the first few digits.
Moreover, relative precision cut-offs may not be directly comparable between datasets, as this quantity may be estimated in different numerical frameworks.
Likewise, the \gls{bhpc} data were computed with mode-sums truncated at up to $n = 1000$ or $\ell = 25$ for the (highly-eccentric) strongest-field orbits to ensure a relative precision at a level of $10^{-6}$~\cite{Fujita:2020zxe}, which may again yield significant differences with our data (such as those observed for the highest eccentricities in~\cref{fig:bhpc_flux_comparison}).
Errors due to these differences in methodology as a function of parameter space are not well understood at present; further investigation of these systematics is beyond the scope 
of this work and reported in a separate work~\cite{Khalvati:2025znb} 
(but is worthy of future study, particularly for eccentric and inclined orbits).

Two other significant features are present in the heatmaps shown in \cref{fig:SAH_nearISO_comparison}.
The first is that for $a \rightarrow 0.999$, we observe a significant increase in relative error in alternating bands of constant $a$.
These bands correspond to $a$-values lying on grid nodes (low interpolation error) and exactly between grid nodes (high interpolation error) respectively, and their existence suggests that our grid density with respect to $a$ is insufficient near $p_\mathrm{sep}$ (even despite our reparameterization, which concentrates grid points in this region).
As we will see later in this subsection, this has little impact on the accuracy of our trajectories.
The second feature is observed for the largest eccentricity slice (bottom-right panel of \cref{fig:SAH_nearISO_comparison}), $a \in [0, 0.5]$ and $p - p_\mathrm{sep} \lesssim 1$, where a patch of larger relative error is observed.
These points correspond to $e \sim 0.5$ near the separatrix (due to the eccentricity taper with respect to $e_\mathrm{max} = 0.9$, see \cref{app:grid_layout_appendix}) and the larger errors are a consequence of ill-converged mode-sums typical of more eccentric orbits near $p = p_\mathrm{sep}$.
As this eccentricity slice is at the edge of our grid, trajectories typically do not perform many orbital cycles in this region before their eccentricities decay to lower values where this numerical precision issue is absent; we therefore do not anticipate this will significant impact the accuracy of our model.
In this eccentricity slice, a steadily-growing relative error is also observed for $a \rightarrow -0.999$, which corresponds to $e \rightarrow 0.9$ and is explained by the differences in convergence criteria between implementations (as described above).

\begin{figure}[t]
    \centering
    \includegraphics[width=0.95\columnwidth]{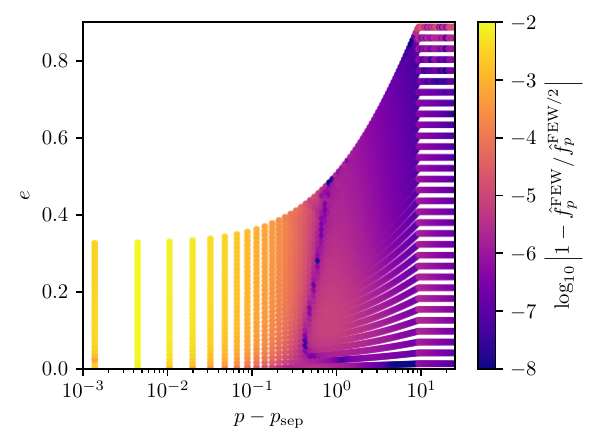}
    \caption{
    Relative error in $\hat{f}_{p}^{(0)}$ between data grid values and those obtained via interpolation over this grid down-sampled by a factor of 2 (denoted ``FEW'' and ``FEW/2'').
    The relative errors shown provide an upper bound on the interpolation error incurred by the full-resolution interpolant. 
    Shown is a slice in $(p, e)$ with $a = 0.9985$, which is the spin value for which our interpolant is the most inaccurate.
    }
    \label{fig:downsample-flux-comparison}
\end{figure}

Taking these comparisons with independent datasets as a reference point, we assess the impact of interpolation error on our estimation of $f_{p,e}^{(0)}$ as a function of the inspiral parameters.
As our data grids have dimensions of the form $2^n +1$ for some positive integer $n$, we can construct a grid of dimension $2^{n-1} + 1$ (halving the point density) by omitting every second point in the grid, which we will refer to as ``down-sampling'' the grid by a factor of $2$.
We can then evaluate this down-sampled grid at the omitted points, placing the upper bound on the interpolation error for the full grid across the parameter space.
The comparison is performed in \cref{fig:downsample-flux-comparison} for $a \sim 0.998$, the spin value at which we find the interpolation errors to be most significant (in line with what is observed in \cref{fig:SAH_nearISO_comparison}); again, we omit our comparison with $\hat{f}_e^{(0)}$ for brevity, which yields similar error estimates.
Over the majority of the parameter space, this interpolation error estimate exceeds those shown in \cref{fig:SAH_nearISO_comparison,fig:bhpc_flux_comparison} where we compare to independent datasets.
This means that for a trajectory model built with data grids down-sampled by a factor of 2 (or higher), interpolation error due to insufficient point density will be the dominant contribution to orbital dephasing.

\begin{figure}[t]
    \centering
    \includegraphics[width=0.95\columnwidth]{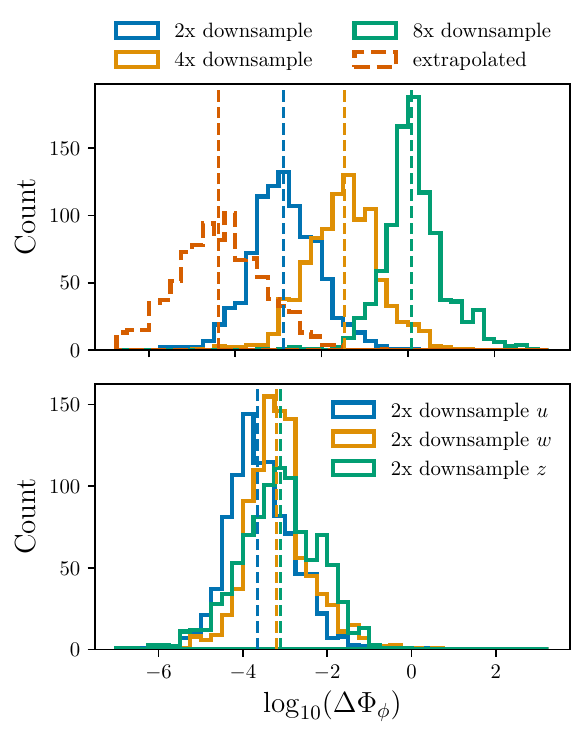} 
    \caption{Dephasing due to down-sampling of $\hat{f}_{p,e}^{(0)}$ interpolant data grids for a sample of $10^3$ four-year inspirals. In all cases, dashed vertical lines indicate median values. \textbf{Top panel:} Convergence of dephasing when one down-samples by a factor of either $2^1$ (blue), $2^2$ (yellow), or $2^3$ (green) in all grid dimensions; the red dashed histogram is obtained by extrapolating down-sampled phase errors to the full grid density ($2^0$). \textbf{Bottom panel:} Dephasing when halving the sampling density in either the $u$ (blue), $w$ (yellow), or $z$ (green) grid dimension only (with $\{u, w, z\}$ corresponding to $\{p, e, a\}$ respectively). The histograms overlap significantly, indicating appropriately-proportioned grid dimensions, and lie between the extrapolated and lowest-order down-sampled histograms in the top panel.}
    \label{fig:Flux_Downsampling}
\end{figure}

We will now use this result to roughly quantify the accuracy of our model.
We consider models built from data-grids down-sampled by factors of $2$, $4$ and $8$, and compare phase trajectories constructed with these down-sampled models to those produced by our full-scale model (which has the full grid density).
In the top panel of \cref{fig:Flux_Downsampling}, we plot histograms for dephasings between each down-sampling scenario and our full-resolution model for a random sample of $10^3$ inspirals with $p_0$ tuned such that they inspiral after 4 years (see \cref{app:monte-carlo-parameters} for details of the random-draw procedure).
We observe that these dephasings display power-law convergence (i.e., in line with the power of $2$ of the down-sampling factor) as a function of grid density.
By fitting this convergence trend between grid density and dephasing for each data point, we then extrapolate to obtain an estimate of the phase error at the full grid density, resulting in the dashed histogram in the top panel of \cref{fig:Flux_Downsampling}.
Taking this extrapolation as a rough estimate of our model's dephasing with respect to a ``perfect'' adiabatic inspiral model, we predict that our model attains a phase accuracy of $\sim 10^{-4}\mathrm{~rad}$ over the majority of the parameter space, with a higher tail extending to $\sim 10^{-2}\mathrm{~rad}$.
This is significantly smaller than our original target of $10^{-1}$~rad, and should be sufficiently accurate for the analysis of \gls{emri} signals that \gls{lisa} is expected to observe.
We emphasise that this extrapolation assumes that our interpolation errors are significantly larger than the numerical precision errors in our computed data.
In general, the worst-behaved systems have very high spins ($a>0.99$) and/or very high initial eccentricities ($e_0>0.8$). 
This highlights that these systems may require a more careful treatment (both in the generation and interpolation of forcing function data), an exploration of which we leave for future work.

It is also useful to understand whether insufficient sampling density in one of $u$, $w$ or $z$ (corresponding to $p$, $e$ and $a$) contributes significantly to interpolation error in $\hat{f}_{p,e}^{(0)}$.
To examine this more closely, we also construct trajectory models from grids down-sampled by a factor of $2$ in a single dimension at at time.
Histograms of dephasings between trajectories from these models and our full-resolution model are shown in the bottom panel of \cref{fig:Flux_Downsampling}. 
As these histograms overlap significantly, we conclude that the point density of our grids is well-proportioned with respect to grid dimension.
As expected, these histograms also lie between those corresponding to the extrapolated dephasings and the model constructed with grids down-sampled by $2$ in all dimensions.

\subsection{Mode amplitude validation}
\label{sec:amplitude-validation}
We now turn to examine the faithfulness of the amplitude module of our waveform generation framework.
In general, any systematic errors in the mode amplitudes $A_{\ell m n}$ should be significantly less impactful than those present in the forcing functions.
Contributions from the latter grow secularly over inspiral, whereas amplitude errors remain of fixed order.
For systematic biases in \gls{emri} parameter estimation to be sufficiently small (assuming no orbital phase errors), the relative error in each mode amplitude must be smaller (in an order-of-magnitude sense) than the inverse of the \gls{snr} of the mode~\cite{lindblom2008model}.
It is therefore important to verify that the scale of these errors is small.
In this subsection, we will compare the outputs of our amplitude module with an independent dataset to quantify the impact of errors in our interpolated amplitudes.
We will also explore where our choice to perform linear interpolation with respect to $a$ (as described in \cref{sec:amplitude-interpolation}) limits the accuracy our mode amplitude values when compared to spline interpolation along this dimension.

\begin{figure}[t]
    \includegraphics[width=0.95\columnwidth]{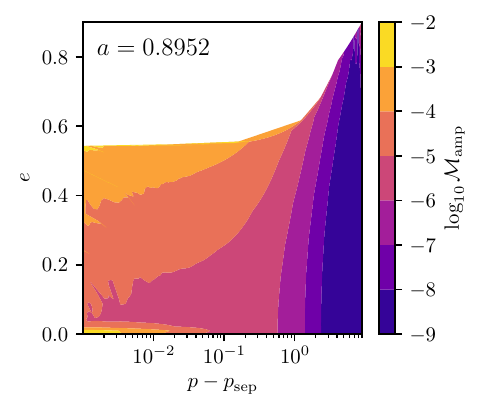}
    \caption{
    Mode amplitude (approximate) mismatch $\mathcal{M}_\mathrm{amp}$ (\cref{eq:mode-amp-mismatch}) between interpolated amplitudes from our model ($\mathcal{A}_{\ell m n}^{\mathrm{FEW}}$) and those produced with the \textsc{GREMLIN} code ($\mathcal{A}_{\ell m n}^{\mathrm{SAH}}$).
    Mismatch due to amplitude interpolation error increases significantly as $p$ decreases below $\sim p_\mathrm{sep} + 1$, and is slightly worse for higher eccentricities.
    This behaviour is due to our linear interpolation of amplitudes with respect to $a$ (explored in \cref{fig:bicubic-tricubic-amp-comparison}).
    }
    \label{fig:amp-mism-contour}
\end{figure}

As in \cref{sec:trajectory-validation}, we will validate our model by comparing interpolated mode amplitudes ($\mathcal{A}_{\ell m n}^{\mathrm{FEW}}$) against those produced with the \textsc{GREMLIN} code ($\mathcal{A}_{\ell m n}^{\mathrm{SAH}}$).
As our amplitude module produces $6993$ modes at each point in parameter space, examining any particular mode in isolation will not be particularly informative with regards to the accuracy of our amplitude model as a whole (especially as the relative importance of each mode is also a function of the parameter space).
Instead, in a similar vein to Ref.~\cite{Isoyama:2021jjd}, we will define a suitable figure of merit that encapsulates all mode amplitudes.
It is similar in form to that of the waveform mismatch (which we will apply in \cref{sec:waveform-validation}; see \cref{app:data_analysis_fundamentals} for details), and is written as 
\begin{equation}
    \label{eq:mode-amp-mismatch}
    \mathcal{M}_\mathrm{amp} = \frac{\sum_{\ell m n} \left(\mathcal{A}_{\ell m n}^{\mathrm{FEW}}\right)^{\star} \mathcal{A}^\mathrm{SAH}_{\ell m n}}{\left|\mathcal{A}^{\mathrm{FEW}}_{\ell m n}\right| \left|\mathcal{A}^{\mathrm{SAH}}_{\ell m n}\right|}\,,
\end{equation}
where $\star$ denotes complex conjugation and $|\cdot|$ is the amplitude vector norm (over the indices $\ell$, $m$ and $n$).
This quantity corresponds to a mismatch between two snapshot waveforms (i.e., constant in amplitude and frequency) with amplitudes $\mathcal{A}^{\mathrm{FEW}}_{\ell m n}$ and $\mathcal{A}^{\mathrm{SAH}}_{\ell m n}$, assuming that overlap terms between modes is small.
While this is not generally true, the inclusion of these terms is not required for us to explore how the accuracy of our modes varies over parameter space, and their exclusion enables us to avoid performing waveform computations at this stage.
This also ignores the spin-weighted spherical harmonics, as we wish to investigate differences independent of the viewing angle.
The $\mathcal{A}^{\mathrm{SAH}}_{\ell m n}$ data we compare against only extends to $p \sim 10$; see \cref{app:weakfieldappendix} for a comparison between our amplitude interpolation and \gls{pn} results in the weak-field limit.
In \cref{fig:amp-mism-contour}, we show how $\mathcal{M}_\mathrm{amp}$ varies over $(p, e)$ for $a=0.8952$ (which lies between two nodes in our amplitude data grids, maximising interpolation errors).
For $p \gtrsim p_\mathrm{sep} + 1$, our interpolated amplitudes agree well with the \textsc{GREMLIN} data for all eccentricities, with estimated mismatches $\leq 10^{-5}$.
Below this value, $\mathcal{M}_\mathrm{amp}$ grows rapidly with respect to $p$, reaching $\sim 10^{-3}$ at higher eccentricities for $p \gtrsim p_\mathrm{sep} + 10^{-2}$.
This suggests that our amplitude interpolation errors are significant near the separatrix.
In \cref{sec:waveform-validation}, we investigate the impact of amplitude interpolation on the waveform level and obtain results that agree quantitatively with those presented in \cref{fig:amp-mism-contour}.

\begin{figure}[b]
    \includegraphics[width=0.95\columnwidth]{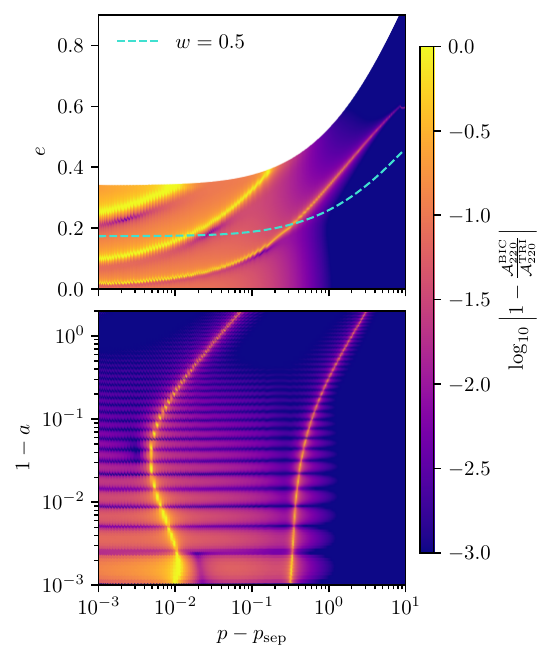}
    \caption{
    Relative difference between bicubic+linear (BIC) and tricubic interpolation (TRI) of the $A_{220}$ mode amplitude with respect to: $(p, e)$ with $a = 0.998$ (top panel); $(p, a)$ with grid coordinate $w=0.5$ (bottom panel).
    The eccentricity corresponding to this value of $w$ for $a=0.998$ is shown in the top panel as a cyan dashed line; for lower spins, $e$ at $p=p_\mathrm{sep}+10^{-3}$ will be higher than this line (see \cref{app:grid_layout_appendix}).
    The outputs of the two interpolants are in close agreement (better than $3$ decimal places), with the exception of the strongest-field region and spins greater than $0.9$ (regardless of eccentricity).
    Note that the real and imaginary components of ${\cal A}_{220}$ are interpolated separately; here we have plotted the relative difference between the complex amplitudes for visual clarity, as the components pass through zero more frequently.
    }
    \label{fig:bicubic-tricubic-amp-comparison}
\end{figure}

Recall that, in our amplitude interpolation framework (\cref{sec:amplitudes}), we opt to perform linear interpolation with respect to $a$ for the purposes of efficiency and to reduce memory usage.
This is motivated by the weak relationship between mode amplitude and spin (given constant $p$ and $e$).
However, our grid coordinate parametrization (which maps the curved inner surface of both the separatrix and the eccentricity taper to rectilinear coordinates) is highly non-linear near the separatrix, especially as $a\rightarrow 1$.
This effect is the main contributor to the larger amplitude mismatches we observe in \cref{fig:amp-mism-contour}.
To demonstrate this point and explore the impact of such behaviour on linear interpolation accuracy near the separatrix, we construct tricubic spline interpolants (identical in form to those applied to the forcing functions in \cref{sec:flux-calculations}) and compare their outputs to those of the bicubic+linear scheme.
For regions where the linear interpolation is a good approximation, we expect the outputs of these two interpolation methods to be in strong agreement.
We consider the ${\mathcal A}_{220}$ mode, which is typically the most significant mode in terms of \gls{snr} contribution for most of the inspiral: we found that comparisons performed for other strongly-contributing modes yielded similar results to this case.
Relative differences between the (complex) outputs of the bicubic+linear and tricubic interpolation configurations for this mode are shown in \cref{fig:bicubic-tricubic-amp-comparison}.
We find that (except for regions of the parameter space where $|{\mathcal A}_{220}| \rightarrow 0$) a linear interpolation in $a$ is sufficient for mode amplitudes to be computed accurately (better than 3 digits of precision) for $p-p_\mathrm{sep} \gtrsim 1$.
Larger relative differences occur where $|{\mathcal A}_{220}| \sim 0$, but as the amplitude is small here regardless of the interpolant used we do not expect this to significantly impact waveform accuracy.
For lower values of $p$ and $a \gtrsim 0.9$, the relative difference grows, reaching a few tens of percent at the highest spins and lowest orbital separations.
However, it is reasonable to expect that the impact of these errors on waveform accuracy will be subdominant with respect to errors of similar scale at larger orbital separations, as \glspl{emri} typically spend only a small fraction of their number of orbital cycles in this region.

For systems with particularly small mass ratios $\epsilon \lesssim \sim 10^{-6}$, which inspiral sufficiently slowly to accumulate hundreds of orbital cycles in the region $p-p_\mathrm{sep} < 1$, the impact of these amplitude interpolation errors is likely to be more significant.
As the primary focus of our waveform model is mass ratios $\epsilon \gtrsim 10^{-6}$, this is an acceptable limitation of our model (but will examine a candidate source with $\epsilon = 10^{-6}$ to understand this in more detail).
In \cref{sec:waveform-validation}, we will demonstrate that this reasoning holds for such systems by performing comparisons between amplitude models on the waveform level.

\subsection{Waveform validation}
\label{sec:waveform-validation}
Now that we have characterised the dominant sources of error in our trajectory and mode amplitude generation frameworks, we will now explore how these errors manifest in waveform templates produced with our model.
The gold standard for understanding the impact of systematic errors is to examine them through the lens of Bayesian parameter estimation, identifying how these errors propagate to the structure of the posterior distribution (\cref{eq:bayes}).
However, this procedure requires millions of likelihood evaluations (and therefore hours of \gls{gpu} wall-time) per comparison, and is therefore prohibitively computationally expensive for a global examination of the parameter space.
Techniques for approximating linear biases in parameter estimates~\cite{Miller2005,Flanagan:1997kp,Cutler:2007mi} (such as the approach we apply in \cref{sec:higher-mode-importance} for this purpose) reduce this to $\sim 10^2$ waveform evaluations per point in parameter space (which are required for computing waveform derivatives).
However, the resulting bias vectors in 13-dimensional parameter space are challenging to interpret in the context of a random draw of sample points over the \gls{emri} parameter space.

We instead opt to quantify the fidelity of our model in terms of the waveform mismatch $\mathcal{M}$, which is defined in terms of the noise-weighted inner product and therefore incorporates the sensitivity curve of the detector (in this case, \gls{lisa}). 
Definitions for the inner product and mismatch are given in \cref{app:data_analysis_fundamentals}. 
A mismatch of zero implies that two templates are identical, whereas a mismatch of one indicates that the two waveforms are orthogonal.
The accuracy requirements for a waveform model to be applied in data analysis without introducing bias must necessarily be expressed relative to the \gls{snr} $\rho$ of the signal being analysed (with louder signals requiring more faithful models --- see~\cite{Cutler:2007mi} for more discussion).
A useful (albeit typically strongly conservative) criterion derived in Refs.~\cite{lindblom2008model,Chatziioannou_2017,purrer2020gravitational,Toubiana:2024car} relates the \gls{snr} of the template signal $s$ ($\rho_s$) defined in \cref{eq:SNR_eq} to the minimum mismatch between this signal and a waveform template $h$ such that the two are indistinguishable.
It states that if
\begin{equation}
    \label{eq:lindblom}
    \mathcal{M}_\mathrm{min} \coloneqq \mathcal{M}(s, h) \leq \frac{D}{2\rho_s^2},
\end{equation}
with $D=12$ being the number of parameters in the waveform model that are estimated during inference (cf.~\cref{tab:emri_params}), then $s$ and $h$ are identical for the purposes of data analysis (at a confidence of $1\,\sigma$).
Choosing what constitutes an acceptable value of $\mathcal{M}_\mathrm{min}$ for our model therefore requires us to specify $\rho_s$.
For \glspl{emri} with $\epsilon \lesssim 10^{-4}$ situated at astrophysical distances ($d_\mathrm{L} \gtrsim 1$~Gpc), \gls{lisa} accumulates an \gls{snr} of up to $\sim 10^3$ over a few years of observation, given source parameters that maximise detectability (such as high \gls{mbh} spin).
See \cref{tab:emri_params} for some example systems, which we will study in more detail in~\cref{sec:inference-subsection}.
Informed by this order-of-magnitude estimate, we choose the value $\mathcal{M}_\mathrm{min} = 10^{-5}$; this is compatible with signal \glspl{snr} of several hundreds and (considering the conservative nature of \cref{eq:lindblom}) should represent sufficient accuracy for the purposes of \gls{emri} data analysis with \gls{lisa}. 
In what follows, we stress that we do not maximise our mismatch calculations over phase, coalescence time or other extrinsic parameters (such as source orientation); as this would always improve the mismatches we would observe, our results are therefore slightly conservative.
However, the impact of this choice is small for \glspl{emri}, as these mismatches are the result of small differences between many sets of harmonic modes that (in most cases) cannot all be significantly reduced by modifying such parameters.

\begin{figure}[t]
    \includegraphics[width=0.95\columnwidth]{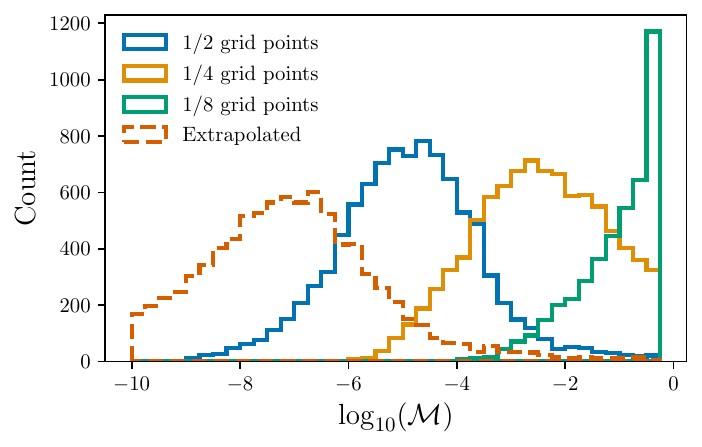}
    \caption{
    Similar to the top panel of \cref{fig:Flux_Downsampling}, but in terms of waveform mismatches instead of orbital dephasings.
    We consider four waveform models built with forcing function data grids of successively coarser resolutions.
    All mismatches are computed with respect to waveforms generated with the finest-resolution model.
    The majority of mismatches decrease exponentially (in proportion to the data grid density); the red dashed histogram is obtained by extrapolating this exponential scaling to the actual resolution of our data grid.
    A long upper tail of mismatches is evident, which corresponds to systems with high eccentricities (see \cref{fig:downsample-mismatch-scatters} and main text for discussion).
    }
    \label{fig:downsample-mismatch-histograms}
\end{figure}

\begin{figure}[t]
    \includegraphics[width=0.95\columnwidth]{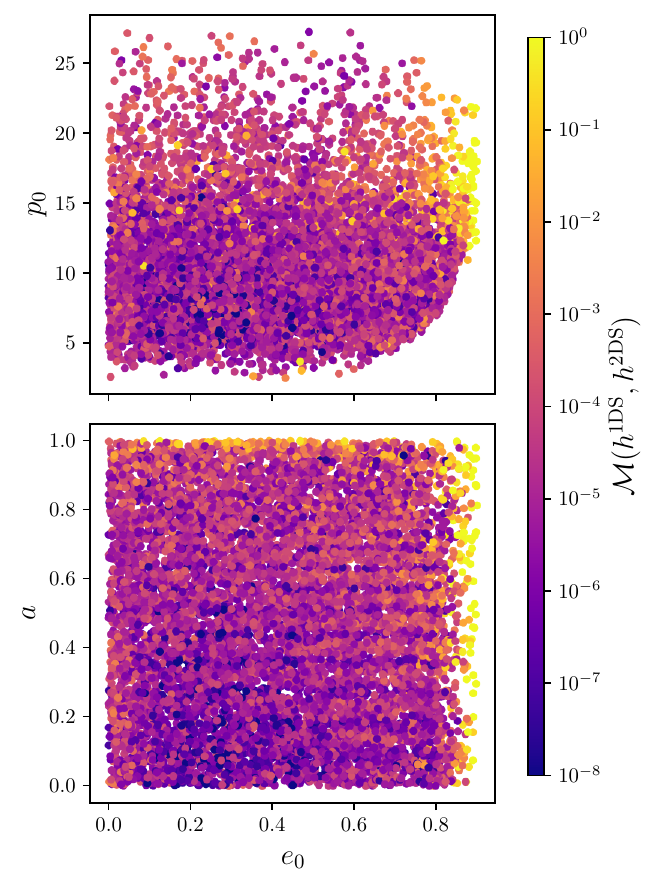}
    \caption{
    Mismatches between waveforms computed with models using full-resolution $(h^\mathrm{1DS})$ and half-resolution $(h^\mathrm{2DS})$ forcing function data grids (blue histogram in \cref{fig:downsample-mismatch-histograms}), with respect to $(e_0, p_0)$ (top panel) and $(e_0, a)$ (bottom panel).
    The mismatches shown can be interpreted as upper bounds on the accuracy of our full-resolution model at each point in parameter space.
    The accuracy of our model decreases significantly for $e_0 \gtrsim 0.85$.
    }
    \label{fig:downsample-mismatch-scatters}
\end{figure}

\subsubsection{Trajectory dephasing}

Based on the results presented in \cref{sec:amplitude-validation,sec:trajectory-validation}, the most significant source of systematic error in the constituent components of our waveform model (i.e., the trajectory and mode amplitude modules) is orbital dephasing due to interpolation errors in the forcing functions.
We will now quantify how this dephasing propagates to the accuracy of the computed waveforms, following a similar approach as was applied in \cref{sec:trajectory-validation} but with a focus on waveform mismatch (rather than orbital dephasing).
Randomly drawing $10^4$ points in parameter space (obtained according to the procedure detailed in \cref{app:monte-carlo-parameters}), we compute waveforms with both our (full-resolution) model and lower-resolution models with down-sampled forcing function data grids.
In \cref{fig:downsample-mismatch-histograms}, we summarize the mismatches between the full-resolution and lower-resolution models at these points in parameter space.
As in \cref{fig:Flux_Downsampling}, a clear trend is observed, with mismatches decreasing in proportion to grid resolution.
Fitting this exponential trend between grid resolution and mismatch for each data point (similarly to in \cref{sec:trajectory-validation}) and extrapolating this fit to our full grid density yields the red dashed histogram in \cref{fig:downsample-mismatch-histograms}.
Due to the poor phase accuracy of the lowest-resolution model (which regularly exhibits in excess of one radian of orbital dephasing) we do not include it when constructing this extrapolation.
This should be considered a rough, order-of-magnitude estimate of the accuracy of our full-resolution model; it is not completely representative of the mismatch between our model and an error-free adiabatic model (especially for the smaller mismatches $\lesssim 10^{-8}$, as we expect other minor sources of systematic error to become important at this level).
In general, we observe mismatches between waveforms from the full- and half-resolution models ($h^\mathrm{1DS}$ and $h^\mathrm{2DS}$ respectively) of less than $10^{-5}$ over the majority of the parameter space, with extrapolated mismatches improving upon this by roughly two orders of magnitude.
When compared to our mismatch requirement of $10^{-5}$, these results suggest that our waveform model is highly robust over a large fraction of the \gls{emri} parameter space.

We also find a small tail of higher mismatches $\mathcal{M} \rightarrow 1$ that are easily identified when $\mathcal{M}$ is examined as a function of $(a, p, e)$, as shown in \cref{fig:downsample-mismatch-scatters}.
Mismatch increases rapidly with respect to eccentricity beyond $e \sim 0.85$, which is not unexpected given the larger errors in the forcing functions in this region identified in \cref{fig:bhpc_flux_comparison}.
It is not clear whether the relatively poor performance of our model in this region is the result of data quality issues or insufficient interpolation density (or more likely, some combination of the two); regardless of the cause, these results suggest that users of our waveform model should exercise caution when choosing $e_0 \gtrsim 0.85$.
Some poorer mismatches are also observed for very high spins ($a \gtrsim 0.998$), which we attribute to the impact of insufficient grid density with respect to $a$ nearer to the separatrix (as was identified in \cref{fig:SAH_nearISO_comparison}).
Both of these problematic regions are near the edges of our domain of validity, where the forcing functions vary rapidly, are computationally expensive and are challenging to obtain accurately.
Given the resulting challenge in modeling systems with such high eccentricities and/or spins, treating these areas of the parameter space separately (e.g., by constructing separate data grids bespoke to these regions) is an idea worthy of investigation in future work.

\subsubsection{Amplitude interpolation errors}
As discussed in \cref{sec:amplitude-validation}, we do not expect that systematic errors in our mode amplitude interpolation will significantly impact the accuracy of our waveform model.
We will now test the validity of this assertion by comparing waveforms generated with two models that differ only in the generation of their amplitudes at each point output by the \gls{ode} solver: one uses amplitudes computed with our bicubic+linear interpolant (\cref{sec:amplitude-interpolation}), whereas the other directly solves the Teukolsky equation to obtain the mode amplitudes (\cref{sec:teukolsky-amplitudes}) using the \textsc{GREMLIN} code.
In what follows, we will refer to waveforms from these models as ``FEW'' and ``SAH'' respectively.
As the latter model is orders of magnitude more computationally expensive (despite amplitudes only being computed at $\sim 100$ points along the trajectory), we will consider three \gls{emri} systems in lieu of a more comprehensive study over the \gls{emri} parameter space.

\begin{figure}
    \includegraphics[width=0.98\columnwidth]{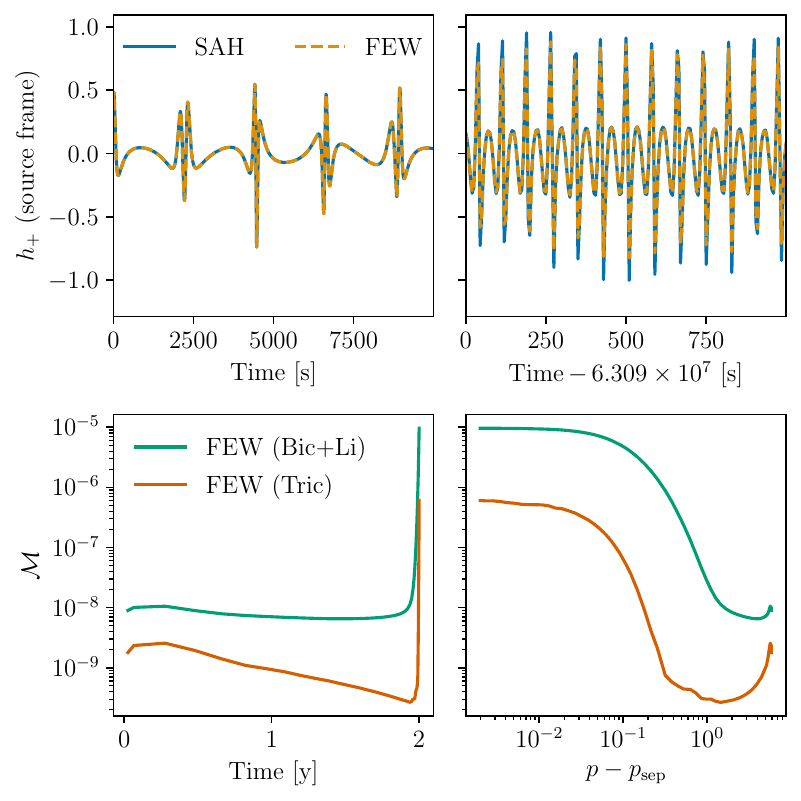}
    \caption{
    Comparisons between waveforms constructed with either interpolated (``FEW'') or directly-computed amplitudes (``SAH'') at each sparse trajectory point.
    This \gls{emri} has parameters $(a, p_0, e_0) = (0.998, 7.81, 0.7)$.
    \textbf{Top panels:} Time-domain strains at early and late times for waveforms built with exact Teukolsky mode amplitudes (blue line) and our bicubic+linear interpolation (yellow dashed line).
    Amplitude differences are evident near the end of inspiral.
    \textbf{Bottom panels:} Cumulative mismatches with respect to either time or semi-latus rectum between ``FEW'' and ``SAH'' waveforms.
    Here we examine both our fiducial bicubic+linear interpolation scheme (green line) and a tricubic interpolation scheme (red line).
    Mismatch increases as the inspiral approaches the separatrix; the larger mismatch in the green line suggests that this is a result of the linear interpolation in $a$ becoming increasingly inaccurate near the separatrix.
    }
    \label{fig:a0.998_amp_mismatch}
\end{figure}

The first system has parameters $\{m_1, \epsilon, a, p_0, e_0\} = \{10^6\,M_\odot, 10^{-5}, 0.998, 7.81, 0.7\}$ (with $p_0$ chosen such that the \gls{co} plunges after two years), and is representative of the region of our parameter space for which our mode amplitude interpolant is least accurate (i.e., high spin and eccentricity) for inspirals of this mass ratio and duration.
From \cref{fig:bicubic-tricubic-amp-comparison}, we do not expect weak-field amplitude interpolation errors to contribute significantly to waveform mismatch, so our choice of a two-year inspiral (over e.g., a four-year inspiral) will have little bearing on the results of this comparison.
In \cref{fig:a0.998_amp_mismatch}, we show the cumulative mismatch (i.e., the mismatch computed over increasingly large intervals of time) between the corresponding waveforms from these two models as a function of both $t$ and $p$.
Also shown in \cref{fig:a0.998_amp_mismatch} is the time-domain waveform strain near the beginning and end of inspiral, with differences in mode amplitudes visible by eye in the latter case.
At earlier times (in the weaker field), $\mathcal{M} \sim 10^{-8}$; this rises to $\mathcal{M} \sim 10^{-5}$ as $p \rightarrow p_\mathrm{sep}$.
This behaviour is expected based on the poorer performance of our interpolant close to the separatrix at high spins (\cref{fig:bicubic-tricubic-amp-comparison}).
The majority of the mismatch accumulates in the interval $p - p_\mathrm{sep} \in [1, 0.1]$, after which the inspiral completes the few remaining orbital cycles and the cumulative mismatch levels off.
To confirm that the primary cause of waveform inaccuracy near the separatrix is a result of our interpolation scheme (rather than issues of data quality), we also construct the mode amplitudes for this system with a tricubic interpolant and compare the resulting waveform with that of the ``SAH'' model.
The cumulative mismatch we obtain (shown in the bottom panel of \cref{fig:a0.998_amp_mismatch} in red) is lower than that obtained with our fiducial interpolation scheme by more than an order of magnitude, confirming that linear interpolation with respect to $a$ is indeed the limiting factor for the accuracy of our amplitude module near the separatrix.

While this sharp increase in mismatch may be alarming, it is important to note that the parameters of this \gls{emri} were chosen to probe the impact of the largest amplitude interpolation errors our model exhibits (given this mass ratio and duration). 
Additionally, as previously highlighted in \cref{sec:trajectory-intro}, the multi-scale expansion underpinning our model begins to break down in these final stages of inspiral; this region may therefore be more accurately captured by a transition-to-plunge framework~\cite{Kuchler:2024esj}.
As mismatch only accumulates significantly near the separatrix, the deviation should be larger for more extreme mass ratios.
To confirm this, we perform a similar analysis for an \gls{emri} of mass ratio $10^{-6}$, with parameters given by the second row of \cref{tab:emri_params}.
Note that $p \lesssim p_{\rm sep} + 0.5$ for this entire inspiral, so the two-year duration of this observation uses wave amplitudes from the region of parameter space where our amplitude interpolation is least accurate.
We do not show a visualisation of the results of this comparison for brevity, but observe a mismatch of $10^{-3}$ that steadily grows by a factor of $\sim 2$ over the course of the inspiral.
As expected, this is significantly larger than what was observed for the $\epsilon = 10^{-5}$ system.
This highlights the inherent difficulty in modelling these \glspl{emri} that spend many years in the neighbourhood of the separatrix; as self-force data varies rapidly in this region, it presents a significant challenge for interpolation schemes.
However, for \gls{lisa}, these sources will not have particularly high \glspl{snr} at typical astrophysical distances --- the \gls{snr} of this source is $30$ --- meaning that (according to \cref{eq:lindblom}) even given the poorer performance of our model in this region of the parameter space, we still do not expect to observe signficiant systematic biases during parameter estimation as a result.
While good performance over the entire parameter space is obviously desirable, there is also scope in future work to optimise waveform models based on the expected \glspl{snr} of astrophysical sources, which may improve the efficiency of these models without introducing significant biases into the results of data analysis.

To demonstrate the accuracy of our amplitude model away from the edges of our domain validity, we also consider a third \gls{emri} with arbitrarily-chosen parameters $\{m_1, \epsilon, a, p_0, e_0\} = \{10^5\,M_\odot, 10^{-4}, 0.5, 27.86, 0.4\}$.
We find that the waveforms from each model are in extremely close agreement: the mismatch between them remains at $\sim 10^{-10}$ throughout the entire inspiral.
This confirms that for inspirals with more typical spins and/or eccentricities, the amplitudes produced by our model are sufficiently accurate so as not to limit waveform accuracy (compared to other sources of error, such as forcing function interpolation).

Based on the results of these comparisons, we are confident that our waveform mode amplitude framework is highly robust over the majority of the parameter space.
Future development efforts in this area should be focused on the accurate interpolation of mode amplitudes near the separatrix, as well as the investigation of how close to the separatrix our multi-scale framework must extend to before the switch to a transitional framework is necessary.
These efforts will improve the accuracy of our framework for systems with more extreme mass-ratios $(\epsilon \lesssim 10^{-6})$.

\section{Results and discussion}
\label{sec:results}

In previous sections, we have described the implementation of our waveform model and demonstrated it is robust over its domain of validity.
Here, we apply our model to explore scientific and data analysis prospects for asymmetric-mass binary observations with \gls{lisa}.
We first characterise the computational cost of our waveform model (\cref{sec:timing}), which is closely tied to the feasibility of data analysis and parameter estimation.
We then examine the impact of semi-relativistic models for waveform mode amplitudes (which have been applied extensively in the literature) on the \gls{snr} of \gls{emri} sources (\cref{subsec:rel-amp-SNR}), and explore how these approximate models can induce biases during parameter recovery in \cref{sec:higher-mode-importance}.
Finding that \glspl{snr} computed with approximate models are subject to significant errors, in \cref{subsec:horizon_redshift} we apply our model to the exploration of the sky-averaged detection horizon for eccentric and rapidly-spinning \gls{emri} and \gls{imri} systems, examining its dependence on the component masses of the system.
We then examine the detectability of small eccentricities for a representative, rapidly-spinning \gls{emri} system in \cref{sec:marginal-eccentricity}, which is an area of great astrophysical interest with respect to the formation of \glspl{emri} in gas-dominated environments such as accretion disks.
We conclude this section with an exploration of parameter recovery for \gls{emri} and \gls{imri} sources (with parameters listed in ~\cref{tab:emri_params}) in the Bayesian inference context \AMEND{in \cref{sec:inference-subsection}}, highlighting similarities and differences in the results obtained between these two cases.
\AMEND{Unless otherwise stated, our analysis procedures throughout this section follow the conventions outlined in \Cref{app:data_analysis_fundamentals}.}

\subsection{Waveform computational cost analysis}
\label{sec:timing}

The principal objective of the \gls{few} package is to provide tools for the rapid generation of adiabatic waveforms by leveraging vectorised operations on \gls{gpu} hardware.
In this subsection, we will examine the computational cost of waveform generation in both the \gls{td} and the \gls{fd}. 
The procedures for constructing waveforms in these two domains differ only at the stage of the waveform summation modules, which have not changed significantly since the \gls{td} (Refs.~\cite{Katz:2021yft,chua2021rapid}) and \gls{fd} (Ref.~\cite{Speri:2023jte}) releases of the package. 
The differences in timing with respect to these previous works can therefore be attributed to: the expansion of the parameter-space coverage of the model, particularly the inclusion of spin and the evolution of the inspiral closer to the separatrix; and changes made to the trajectory and amplitude generation modules.
We warn the reader that the computational cost of the waveform can vary depending on the computing resources used; all computational wall-times reported in this section were obtained using an NVIDIA A100 \gls{gpu} and a $2\,\mathrm{GHz}$ AMD EPYC-7713 \gls{cpu}.
Furthermore, the timing distribution will depend on our procedure for sampling source parameters --- this is described in \cref{app:monte-carlo-parameters}.

\begin{figure}
    \includegraphics[width=0.95\columnwidth]{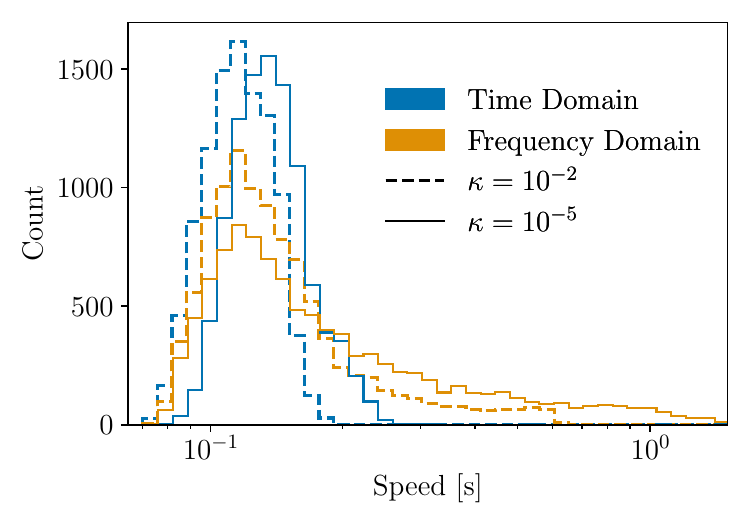}
    \caption{Wall-time of waveform generation with \gls{few} in either the \gls{td} (blue) or \gls{fd} (orange) output domains, for $10^4$ randomly-sampled sets of source parameters. 
    We show wall-times for $\kappa = 10^{-5}$ (dashed line) and $\kappa = 10^{-2}$ (solid line) to explore the impact of waveform mode content on computational costs. 
    The sampling interval is chosen to be $\mathrm{d} t=5\,\mathrm{s}$, with all inspirals plunging after four years.
    }
    \label{fig:timing_vs_sample_rate}
\end{figure}

We will examine the cost of our model for four-year inspirals with sampling cadence $\mathrm{d}t = 5\,\mathrm{s}$.
\AMEND{This sampling cadence is based roughly on what is required for systems with $M \sim 10^6\,M_\odot$; the computational cost for other choices of cadence can be roughly estimated by scaling the wall-time with respect to this value of $\mathrm{d}t$.}
We consider two mode selection thresholds, $\kappa =(10^{-2}, 10^{-5})$, to probe how this cost depends on the mode content requested in waveform generation.
Drawing $10^4$ sets of parameters, we show the corresponding distribution of computational wall-times per waveform evaluation in \cref{fig:timing_vs_sample_rate}. 
We obtain median wall-times of $0.11\,\mathrm{s}$ ($0.13\,\mathrm{s}$) per \gls{td} evaluation and $0.13\,\mathrm{s}$ ($0.15\,\mathrm{s}$) per \gls{fd} evaluation, for $\kappa = 10^{-2}\,(10^{-5})$.
The \gls{fd} wall-time distribution presents a long upper tail that extends to $\sim 1\,\mathrm{s}$; this feature corresponds to primary masses $m_1 \lesssim 10^6\,M_\odot$ with $e_0 \gtrsim 0.7$.
The reason for this discrepancy in timings between output domains is as follows (based on arguments presented in Ref.~\cite{Speri:2023jte}). 
The computational cost of the waveform scales with the number of data points at which each waveform mode must be evaluated. 
For \gls{td} generation, each array element receives contributions from all the selected harmonics (such that the overall summation cost scales linearly with the number of waveform modes). 
In contrast, the cost-per-mode of the \gls{fd} generation depends on the frequency evolution of that mode, as each mode is only evaluated in frequency bins it evolves through.
Therefore, the \gls{fd} waveform is most efficient when the modes of the signal span a compact range of frequencies, which corresponds to higher primary mass values; conversely, at lower primary masses and higher eccentricities many modes span a wide range of frequencies, leading to the tail observed in~\cref{fig:timing_vs_sample_rate}.
In terms of performance with respect to eccentricity at higher primary masses, it is argued in \cite{Speri:2023jte} that \gls{fd} waveform generation is more efficient than its \gls{td} counterpart for $e_0 \sim 0.7$ when $m_1 \sim 10^7$, and this remains true in our results.
\AMEND{
It is worth noting that frequency-domain waveforms generated via the stationary phase approximation can be downsampled to reduce computational cost, potentially improving efficiency for certain analyses~\cite{Speri:2023jte}. 
However, such downsampling reduces the effective signal-to-noise ratio, leading to biased parameter posteriors in noisy data, limiting its applicability for realistic EMRI inference.
}

The inclusion of primary spin increases the azimuthal frequency $\hat{\Omega}_\phi$ close to the separatrix by up to a factor of six (for $a = 0.999, e = 0)$, which in turn leads to many modes evolving over a much larger range of frequencies in a typical waveform than for the zero-spin case. 
This causes the \gls{fd} waveform generation to exceed that of \gls{td} generation for the majority of spins $a > 0$ (which is why the lower ends of the \gls{fd} timing distributions do not extend below their \gls{td} counterparts). 
We remind the reader that~\cref{fig:timing_vs_sample_rate} does not include retrograde orbits ($a < 0$), where we expect the opposite behaviour to occur and the relative cost of \gls{fd} waveform generation to decrease. 

We also examine the breakdown in wall-time between different waveform generation modules for the two output domains.
For a four-year inspiral with source parameters $\{m_1, \epsilon, a, e_0\} = \{10^6\,M_\odot, 10^{-5}, 0.9, 0.1 \}$, we measure wall-times of $0.10\,\mathrm{s}$ for the trajectory module, $0.02\,\mathrm{s}$ for the amplitude module and $0.15\,\mathrm{s}$ ($0.60\,\mathrm{s}$) for the \gls{td} (\gls{fd}) summation module (for $\kappa = 10^{-5}$).
The trajectory and summation modules therefore constitute the bulk of waveform generation computational cost.

To verify that the \gls{td} and \gls{fd} waveforms are consistent with each other, we conduct a mismatch analysis over the $10^4$ sets of drawn parameters. 
We find the median mismatch between the two to be $\sim 10^{-3}$, suggesting that the waveforms are in good agreement. 
We note that computing mismatches across output domains can be misleading due to the windowing / leakage effects, and that different window choices can significantly impact the mismatches obtained.
This was explored in more detail in Ref.~\cite{Speri:2023jte}; note that in that work, a Hann window was used for these comparisons, whereas here we instead adopt a Tukey window with shape parameter $0.005$.
Our choice of window does not affect any of the conclusions drawn in this subsection.

\subsection{Impact of relativistic amplitudes on waveform SNR}
\label{subsec:rel-amp-SNR}

Semi-relativistic waveform models known as ``kludges''~\cite{barack2004lisa, babak2007kludge, chua2017augmented} have been commonly-used tools to make qualitative statements describing detection rates, to perform large-scale parameter measurement studies, and to examine the impact of stochastic backgrounds composed of unresolved \gls{emri} signals on \gls{lisa} data analysis~\cite{Babak:2017tow,bonetti2020gravitational,piarulli2024test,Pozzoli:2023kxy,Chapman-Bird:2022tvu}.
These studies perform \gls{snr} calculations with kludge models that only incorporate a quadrupolar approximation (i.e., $\ell = 2 = m$) to the mode structure of the waveform.
As was identified in Refs.~\cite{Fujita:2020zxe,Khalvati:2024tzz}, these approximate quadrupolar amplitudes may be insufficiently accurate to adequately describe the \gls{emri} mode spectrum, particularly in the strong-field regime of gravity accessible to eccentric inspirals into rapidly-spinning black holes.
As we have already shown in~\cref{fig:heatmap_mode_amplitudes}, a large number of $(\ell, m, n)$ modes are required to faithfully reconstruct the true \gls{emri} signal. 
In this subsection, we will investigate through \gls{snr} comparisons (which are critical for determining detection rates) the impact of approximating the mode amplitudes of adiabatic \gls{emri} waveforms with those of the \gls{aak}, a kludge model that has seen extensive application in the literature. 
Understanding the regions of parameter space where the \gls{snr} is inaccurately estimated, and to what extent, is crucial for avoiding misrepresentations of the detection rate of \glspl{emri}.

Exploiting the modular nature of \gls{few}, we combine our relativistic trajectory model (\cref{sec:trajectory}) with two different amplitude models.
This results in two waveform models: the first uses our adiabatic mode amplitudes (\cref{sec:amplitudes}) and is the model we introduce in this work; the second uses the \gls{aak} mode amplitudes.
We will refer to these models as \textsc{Kerr} and \gls{aak} respectively throughout this subsection. 
The \textsc{Kerr} amplitudes, described in ~\cref{sec:amplitudes}, consider the harmonics $(\ell,m,n)$ for $2\leq \ell \leq 10$, $|m| \leq \ell$, and $|n| \leq 55$.
For the \gls{aak} model, which is quadrupolar~\cite{Blanchet:2013haa} (i.e., $\ell = 2 = m$), we set the number of Fourier modes to $n_{\mathrm{max}} = 50$; note that these modes do not have a one-to-one correspondence with those of our model (see Ref.~\cite{chua2017augmented} for definitions).
We confirmed that this value of $n_{\mathrm{max}}$ is sufficiently high to represent the \gls{aak} for the purposes of this analysis (i.e., our results are convergent with respect to $n_{\mathrm{max}}$).
For the \textsc{Kerr} model, we use the default mode-selection threshold $\kappa = 10^{-5}$; by construction, the \glspl{snr} computed with the \textsc{Kerr} model for $\kappa = 0$ and $\kappa = 10^{-5}$ will differ by $\sim 10^{-5}$, which is negligible on the scale of the \gls{snr} variations we identify in our results. 

For each set of source parameters, we choose $p_0$ such that inspirals plunge after \AMEND{four} years and fix extrinsic parameters (such as orientation angles and initial phases) to those given in the caption of~\cref{tab:emri_params}.
Our results are largely independent of this choice of extrinsic parameters.

\begin{figure}
    \includegraphics[width=\columnwidth]{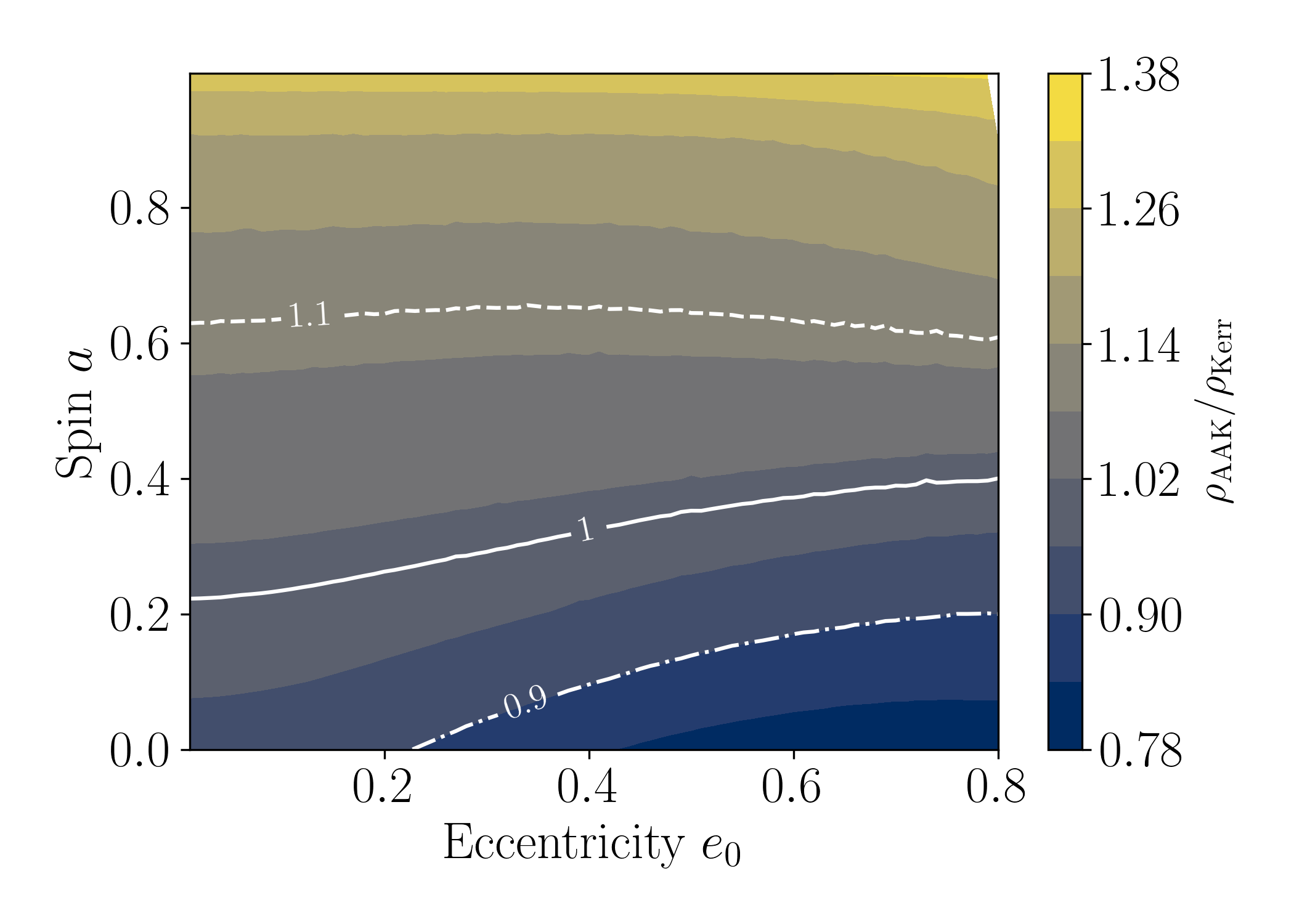}
    \caption{
    Ratio of the optimal \glspl{snr} computed with the semi-relativistic \gls{aak} mode amplitude model ($\rho_\mathrm{AAK}$) and the adiabatic model presented in this work ($\rho_\mathrm{Kerr}$), as a function of initial eccentricity $e_{0}\in[0,0.7]$ and primary spin $a\in[0,0.998]$.
    For all systems, we consider four-year inspirals with masses $(m_1, m_2)=(10^6, 10)\,M_\odot$. 
    Waveforms for both models are built with the same adiabatic trajectory model (\cref{sec:trajectory}). 
    The (dashed, solid, dot-dashed) white lines indicate the ratio of the \glspl{snr} for three reference values $\rho_{\rm AAK}/\rho_{\rm Kerr} \in \{1.1, 1, 0.9\}$.
    }
    \label{fig:AAK_Kerr_snr_comparison}
\end{figure}
In \cref{fig:AAK_Kerr_snr_comparison}, we plot the \gls{snr} $\rho$ as a function of $e_{0}$ and $a$ for \glspl{emri} with $(m_1, m_2) = (10^6,10)\, M_{\odot}$. 
The solid white line indicates the boundary where the two waveform models provide a consistent prediction of the signal-to-noise ratio $\rho_{\text{AAK}} / \rho_{\text{Kerr}} = 1$. 
Above this line (at higher spins), the \gls{aak} overestimates the \gls{snr} by up to $\sim$30\%, whereas below it underestimates it by up to $\sim$20\%.
This discrepancy between the models increases with eccentricity, particularly at lower values of $a$.
Two main factors contribute to the differences observed.
First, as the \gls{aak} is quadrupolar, it will become increasingly inaccurate as the orbital separation decreases and missing modes with $\ell > 2$ become more significant.
Second, as the mode amplitudes of the \gls{aak} are based upon a low-velocity approximation~\cite{Peters:1963ux}, they will also become increasingly inaccurate as the orbital separation decreases.
What we observe in \cref{fig:AAK_Kerr_snr_comparison} is a combination of these two effects.

The results of such a comparison will change depending on the masses of the system, due to the shape of the \gls{lisa} \gls{psd} and the change in $p_0$ required to hold the inspiral duration fixed.
To investigate this behaviour, we also perform this analysis for other values of $m_1$ and $m_2$ (keeping the mass ratio $\epsilon = 10^{-5}$ fixed), but do not show the corresponding figures for brevity.
For $m_1 = 10^{5}$, the faster rate of inspiral requires us to set $p_0$ higher (for a two-year inspiral) than if were to have larger primary masses. 
As most of the inspiral takes place in this weaker-field region (where the assumptions of the \gls{aak} fare better), and the strong-field emission is shifted to higher frequencies due to the lower total mass (where \gls{lisa} is less sensitive), we find only moderate deviations in the \gls{snr} of $\sim 9\%$ at most. 
Conversely, discrepancies between the models become more pronounced for $m_1 > 10^6$, with a lower $p_{0}$ leading to inspirals in the stronger-field regime where the amplitudes of the \gls{aak} fare worse. 
Indeed, for $m_1 = 10^{7} M_{\odot}$, we see an overestimation of the \gls{snr} of $\sim 60\%$ for higher $a$ and lower $e_0$, and an underestimation of $\sim 40\%$ for lower $a$ and higher $e_0$, deviations roughly twice as large as in the $m_1=10^6\,M_\odot$ case.
For all three cases, the differences between the \glspl{snr} computed with each model increased for larger eccentricities, with a larger overall difference for larger values of $m_1$.

From these results, we can conclude that one should be careful when employing the \gls{aak} (or other similar kludge waveform models) in prospective studies of \gls{lisa}'s scientific capabilities with \glspl{emri}.
This conclusion is qualitatively similar to that of Ref.~\cite{Khalvati:2024tzz}, which performed similar investigations in the quasi-circular case.
Our work reinforces this conclusion in the presence of eccentricity, and identifies that the \gls{aak} typically fares worse for eccentric systems than for quasi-circular ones.
As mentioned earlier in this subsection, the systematic errors in the \glspl{snr} computed with such models will impact the quantitative conclusions of \gls{emri} rate estimation studies.
In Ref.~\cite{Babak:2017tow}, detection rates and catalogues (that have since been applied extensively in the literature) were obtained using a kludge-based waveform model~\cite{Barack:2003fp}.
A similar analysis conducted using relativistic waveform models, such as ours, would likely yield lower detection rates, since many of these catalogues consisted solely of systems containing rapidly-rotating \glspl{mbh} with $a \in (0.9,0.998]$, and our results indicates that kludge models significantly overestimate the \glspl{snr} of these systems.
A more detailed study characterising these systematics will be explored in future work.

\subsection{Parameter recovery with approximate mode amplitude models}
\label{sec:higher-mode-importance}

\begin{figure}
    \centering
    \includegraphics[width=\columnwidth]{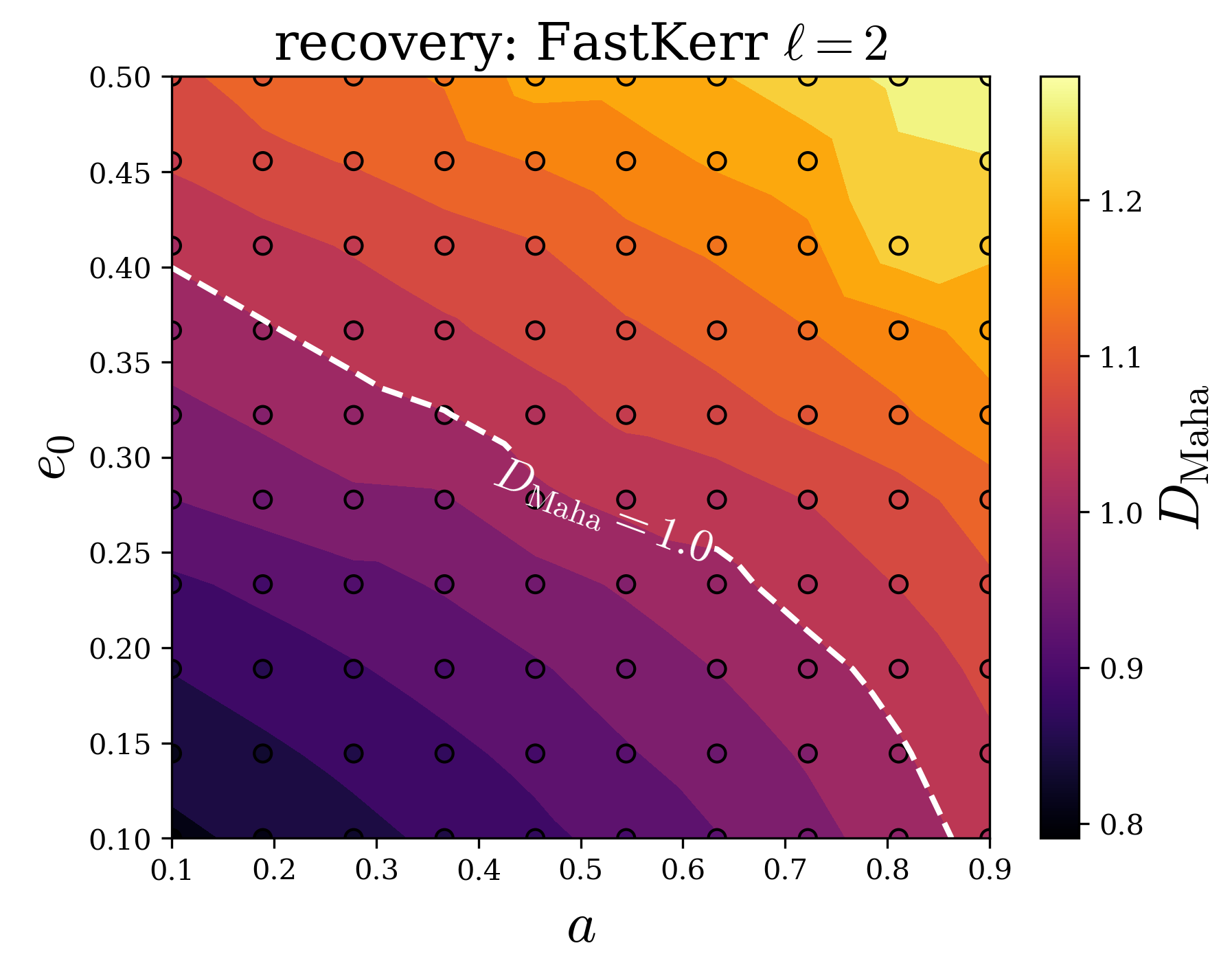}
    \includegraphics[width=\columnwidth]{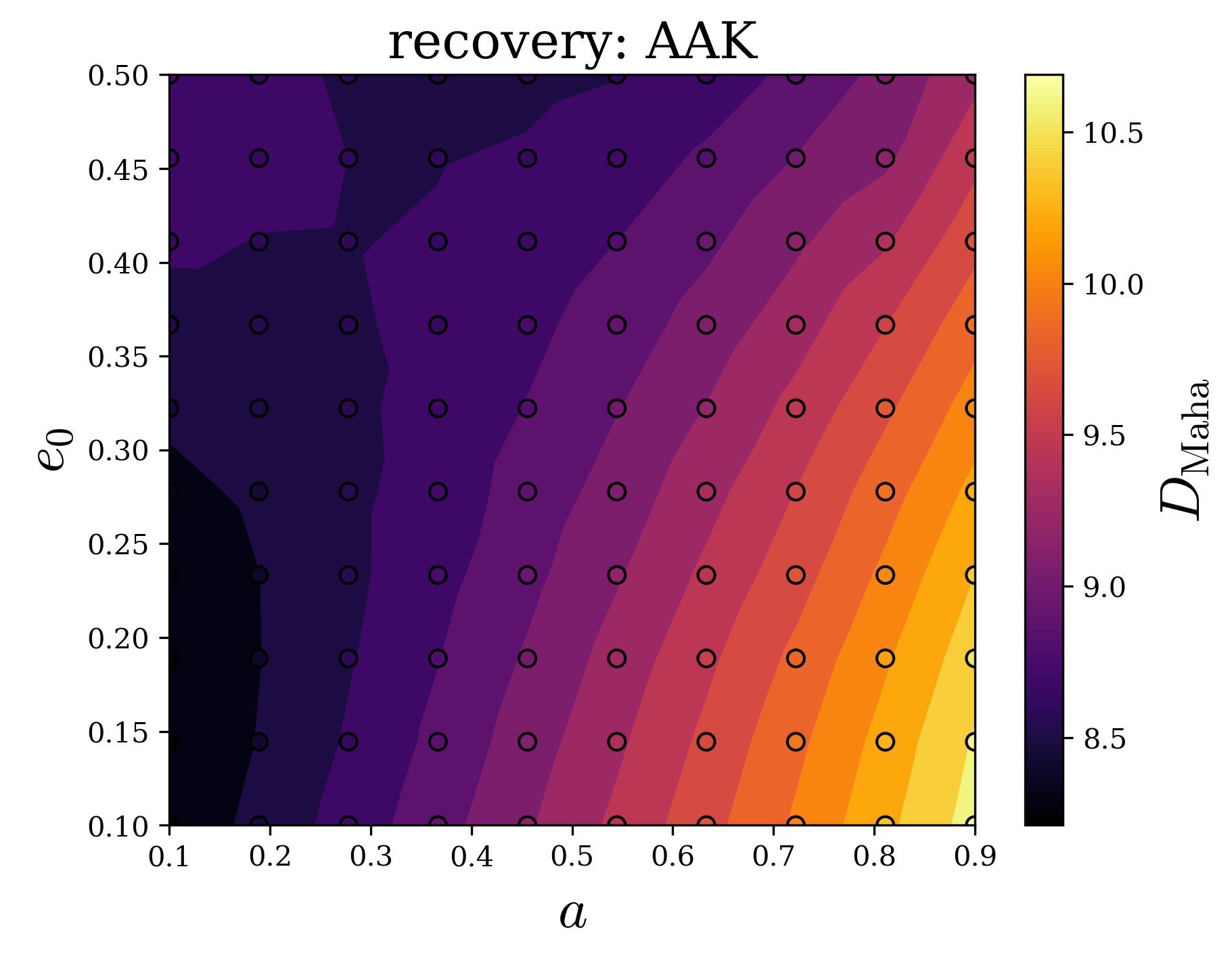}
    \caption{Distribution of sigma contour levels $D_{\rm Maha}$ of the best-fit $\boldsymbol{\theta}_{\rm bf}$ parameters with respect to the truths $\boldsymbol{\theta}_{\rm tr}$ visualized on a grid of $a$ and $e_0$ values. At each gridpoint (black circles), the injected signal is generated using the default  model and inferred with the approximate \textsc{Kerr$\ell2$} (top panel) and \textsc{AAK} (bottom panel) waveforms as described in the text, showing the impact of higher-multipole ($\ell > 2$) and relativistic mode amplitudes on parameter recovery, respectively. The plot corresponds to injected \gls{mbh} and \gls{co} (redshifted) masses $(m_1, m_2) = (10^6, 10) M_\odot$. The white line in the top panel represents $D_{\rm Maha} = 1.0$. Also note the difference in colour scale between the two panels.}
    \label{fig:approxwaveforms}
\end{figure}

In addition to inaccurate \gls{snr} calculations, approximate waveform models can also induce systematic biases in the inference of \gls{emri} parameters, e.g., due to missing physics~\cite{Kejriwal:2023djc, Speri:2024qak} or insufficient model accuracy~\cite{Cutler:2007mi,chua2017augmented,Katz:2021yft}. 
In this subsection, we explore the latter by studying the impact of incomplete or approximate waveform mode amplitude models on biases in parameter estimation.
We consider our fully-relativistic waveform model constructed in~\cref{sec:implementation} (and with the default mode-selection criterion as described in~\cref{sec:selection-summation-intro}) as the representative ``true'' model, and denote it as \textsc{Kerr}.
Similarly to \cref{subsec:rel-amp-SNR}, we consider two approximate models with different mode amplitudes but the same relativistic trajectory as this model (cf.~\cref{sec:trajectory}). 
Our first approximate model, \textsc{Kerr$\ell2$}, is composed of only the dominant $\ell=2$ modes (but accounts for all corresponding $m, n$ modes, where $|m| \leq \ell$ and $|n| \leq n_\mathrm{max} = 55$; cf.~\cref{sec:amplitudes}) while the second model, the \textsc{AAK}, uses semi-relativistic (``kludge'') amplitudes formulated in Ref.~\cite{chua2017augmented}, which are based on the quadrupole moment formalism (cf.~\cite{Blanchet:2013haa}). Unlike~\cref{subsec:rel-amp-SNR}, we do not control the mode content of the \textsc{AAK} waveform model. The modes are chosen instead according to the default mode-selection criteria in \textsc{FEW}, i.e., $n_\mathrm{max} = \mathrm{max}(4, 
\lfloor 30e_0 \rfloor)$.
We then infer the injected \textsc{Kerr} signal with these approximate models and assess their impact on parameter recovery.

Our analysis is set up as follows. We consider signals in a 2D grid of parameter points on the surface defined by $(a, e_0)$.
We select $N = 10$ points each along prograde inspirals $a \in [0.1, 0.9]$ and $e_0 \in [0.1, 0.5]$ for a total of 100 gridpoints. We choose only prograde inspirals ($a > 0.0$) because the effect of higher multipole modes of $A_{\ell m n}$ is less pronounced in retrograde inspirals; The mode amplitude $A_{\ell m n}$ of a given $\ell$ mode is $O(p^{-\ell/2})$ (or higher) for equatorial inspirals~\cite{Tagoshi:1995sh,Fujita:2010xj,Isoyama:2021jjd}, and the retrograde trajectory is completed within the weak field $p \gg 1$ in the majority of cases.
We also restrict the analysis to moderate eccentricity values for computational feasibility, as waveforms with $e_0 > 0.5$ are significantly more expensive due to their larger mode content.

We choose three detector-frame mass combinations, with $m_1 = \{0.5, 1.0, 1.5\}\times 10^6\, M_\odot$ paired with $m_2 = \{5.0, 10.0, 15.0\}\,M_\odot$ 
respectively, to gauge the impact of approximate models across different detector frequencies.
All other parameters are fixed to the following values: the sky location parameters $(\theta_S,\phi_S) = (\pi/5, \pi/6)$, the spin orientation parameters $(\theta_K, \phi_K) = (\pi/3,\pi/4)$, and the initial azimuthal and radial phases $\Phi_{\phi_0} = 0.0 = \Phi_{r_0}$. The initial semi-latus rectum $p_0$ is fixed to $p_{\rm plunge} + 0.5$ where $p_{\rm plunge}$ is the initial semi-latus rectum value that leads to plunge after one year of inspiral. Finally, the luminosity distance $d_\mathrm{L}$ is fixed such that each source has an \gls{snr} of exactly $100$ when measured using the true waveform. Note that this distance rescaling introduces a gradient in the source-frame masses $(m_1^{\mathrm{(s)}}, m_2^{\mathrm{(s)}})$ across the grid, which are related to the redshifted (detector-frame) masses as $(m_1^{\mathrm{(s)}}, m_2^{\mathrm{(s)}}) = (m_1, m_2)/(1+z_\mathrm{r})$ where $z_\mathrm{r}$ is the redshift which varies with $d_\mathrm{L}$. However, this does not influence our analysis, since we exclusively work in terms of detector-frame masses. 
We perform the analysis in the log-mass parametrization and vary all model parameters such that the parameter vector is $\boldsymbol{\theta} := (\ln{(m_1)},\ln{(m_2}), a, p_0, e_0, d_\mathrm{L}, \theta_S, \phi_S, \theta_K, \phi_K,\Phi_{\phi_0},\Phi_{r_0})$ with dimension $D = 12$. 

At each gridpoint, inference of the true signal with the approximate model generically introduces systematic biases in the best-fit parameter estimate that can be approximated in the high-SNR limit by the linear signal approximation given by Cutler and Vallisneri~\cite{Cutler:2007mi} (~\cref{eq:CVbias}). To quantify these biases, we calculate the sigma-contour level between the injected and the best-fit points, given by the Mahalanobis distance~\cite{mahalanobis}, notated as $D_{\rm Maha}$ and defined explicitly in~\cref{eq:Mahalanobis}. $D_{\rm Maha}$ quantifies how many sigma-levels away the biased parameter recovery point is from the injected parameters in the full $D$-dimensional space. It scales proportionately with the induced biases on the recovered parameters, such that $D_{\rm Maha} \in [0.0, \infty)$ and the null value is recovered at the injected parameter point. See discussion below~\cref{eq:Mahalanobis} for more details. In the rest of this section, we use ``Mahalanobis distance'' and ``sigma contours'' interchangeably.
We present our main results in the following two subsections. The first subsection (results \rom{1}) quantifies the sigma contour levels in the full-dimensional and the marginalized 1-dimensional spaces across the $(a, e_0)$ grid for a fixed total mass binary, while the second subsection (results \rom{2}) assesses the impact that different total EMRI masses ($m_1 + m_2$) have on the biases across the grid. We provide qualitative reasoning for the observed trends in both subsections and draw our conclusions in the final paragraph. 

\subsubsection{Results~\rom{1}: Sigma contour levels}
Our results are presented in~\cref{fig:approxwaveforms} for the $(m_1, m_2) = (10^6, 10)M_\odot$ case. For the \textsc{Kerr$\ell$2} case (top panel), we find that the sigma contours categorically scale with both $a$ and $e_0$. On the contrary, in the case of \textsc{AAK} (bottom panel), we observe an inverse scaling with $e_0$, with the typical sigma-contour levels $\sim 8-10$ times higher than \textsc{Kerr$\ell$2}. We now qualitatively explain these scalings.

\textit{Scaling with $a$}---At larger $a$'s, the separatrix of the orbit is closer to the horizon of the MBH, which will increase the number of highly-relativistic orbits. In this region, the importance of higher-order and fully relativistic mode amplitudes increases, and consequently, parameter recovery with the \textsc{Kerr$\ell$2} and \textsc{AAK} models is worsened. 

\textit{Scaling with $e_0$}--- For the \textsc{Kerr$\ell$2} model, as the eccentricity increases, higher $\ell$-mode amplitudes are no longer suppressed in general, and are more similar in magnitude to the $\ell = 2$ mode (cf. Refs.~\cite{Drasco:2005kz,Hughes:2021exa,Fujita:2009us}.) 
This leads to worsened parameter recovery with the \textsc{Kerr$\ell$2} model.
On the other hand, the \textsc{AAK} model does not scale as strongly with eccentricities, neither is the scaling monotonic across the grid. While this trend is qualitatively consistent with the findings in~\cref{fig:AAK_Kerr_snr_comparison} (see also~\cref{subsec:rel-amp-SNR}), our results warrant a detailed analysis in the future. 
We argue that at smaller initial eccentricities where the separatrix is closer to the \gls{mbh}, the \gls{co} completes more strong-field orbits, and thus the \textsc{AAK} model incurs larger biases. 

We also calculated the 1-dimensional marginalized sigma-biases ($z$-scores) on each parameter explicitly and found that the luminosity distance $d_\mathrm{L}$ incurs the largest biases among all parameters with a median value $\sim 0.3\sigma$ for \textsc{Kerr$\ell$2} and $\sim 1.5\sigma$ for \textsc{AAK} when $(m_1, m_2) = (10^6, 10)\,M_\odot$.
In both cases, the shift in $d_\mathrm{L}$ can be understood to make up for the loss of \gls{snr} compared to the true signal when it is recovered using the approximate waveforms (also see~\cref{subsec:rel-amp-SNR}). In the \textsc{Kerr$\ell$2} model, all other parameters are typically biased to $\lesssim 0.1\,\sigma$, showing robust recovery. However, the biases in the \textsc{AAK} model are a factor $\sim 3-5$ times higher (but still within $1\,\sigma$), consistent with the trend in~\cref{fig:approxwaveforms}. Two notable exceptions are the recovered phases $(\Phi_{\phi_0}, \Phi_{r_0})$, with biases a factor $\sim 10-12$ higher compared to the \textsc{Kerr$\ell$2} model, hinting at severe dephasing of the \textsc{AAK} waveform compared to the injected signal. 

\subsubsection{Results~\rom{2}: Other mass pairs}

Finally, we analysed two additional \gls{emri} sources with (i) $(m_1, m_2) = (0.5\times 10^6, 5)\,M_\odot$ (lower-mass), and (ii) $(m_1, m_2) = (1.5\times 10^6, 15)\,M_\odot$ (higher-mass).
In the \textsc{Kerr$\ell$2} case, we found the qualitative scaling of sigma contours with $a$ and $e_0$ to be the same. However, for \textsc{AAK}, the scaling with $e_0$ significantly depended on the total mass of the system: biases incurred by the lower-mass \gls{emri} were largely insensitive to $e_0$ and scaled with $a$, while the higher-mass \gls{emri} showed stronger dependence on both $a$ and $e_0$. While a detailed analysis is beyond the scope of this work, these scalings can be qualitatively attributed to the relation between the number of relativistic orbits completed, the total mass of the system, and the LISA sensitivity curve, similarly to the discussion in~\cref{subsec:rel-amp-SNR}.

Assessing the scaling of the average incurred biases with the total \gls{emri} masses (and consequently the detector's frequency band), we found opposing trends for the two approximate models, as described below.

\textsc{Kerr$\ell$2}\textit{ case---}On average, the lower-mass \gls{emri} incurred smaller biases while the higher-mass system was more biased. This behaviour is consistent with the results of Section \rom{5} D of Ref.~\cite{Katz:2021yft}. They argue that as the total mass of the binary pair increases, the \gls{gw} signal slides to lower detector frequencies where the instrument's noise suppresses contributions from lower multipole modes. Consequently, higher modes become important in the signal and inference biases are introduced in parameter recovery with the \textsc{Kerr$\ell2$} model.

\textsc{AAK}\textit{ case---}The typical biases incurred by the lower-mass \gls{emri} were the largest in this case, opposite to the trend observed in \textsc{Kerr$\ell$2}. This may be attributed to the larger number of strong-field orbits completed by the lower-mass \gls{emri}, leading to a larger accumulation of phase errors. The marginalized 1-dimensional sigma biases showed that, while other parameters typically incurred similar biases across all three systems, the lower-mass binary incurred $\approx 40\%$ larger biases in the initial phases $(\Phi_{\phi_0}, \Phi_{r_0})$ than the higher-mass system. This is consistent with our interpretation.\\

\noindent Overall, we find that the biases incurred using approximate models are small across a broad grid of $a$ and $e_0$ values. Even though the \textsc{AAK} waveforms led to sigma contour level biases of factor $8-10$ larger than \textsc{Kerr$\ell$2}, both models were able to recover all inferred parameters with 1-dimensional biases at $\lesssim 1\,\sigma$. This is useful, e.g., for coarse-grain recovery of the parameters during the search stage of \gls{emri} inference, for which faster and approximate waveforms may be necessary and sufficient~\cite{gair2004event}. We caution, however, that inference with approximate waveforms may lead to multimodalities in the likelihood surface~\cite{katz2021fast,chua2022nonlocal}, potentially compromising robust recovery. The linear signal approximation framework adopted in our study does not capture such features, and we leave its detailed analysis to future work. Additionally, the absence of accurate models may significantly bias the inference of fully relativistic signals emitted from systems with high primary spins. The accuracy of parameter recovery also strongly depends on the frequency band of the signal, the orbital eccentricity, and the choice of the approximate model. 

\subsection{Horizon redshift of asymmetric-mass binaries with LISA}
\label{subsec:horizon_redshift}

\begin{figure*}
    \centering
    \includegraphics[width = 0.45\textwidth]{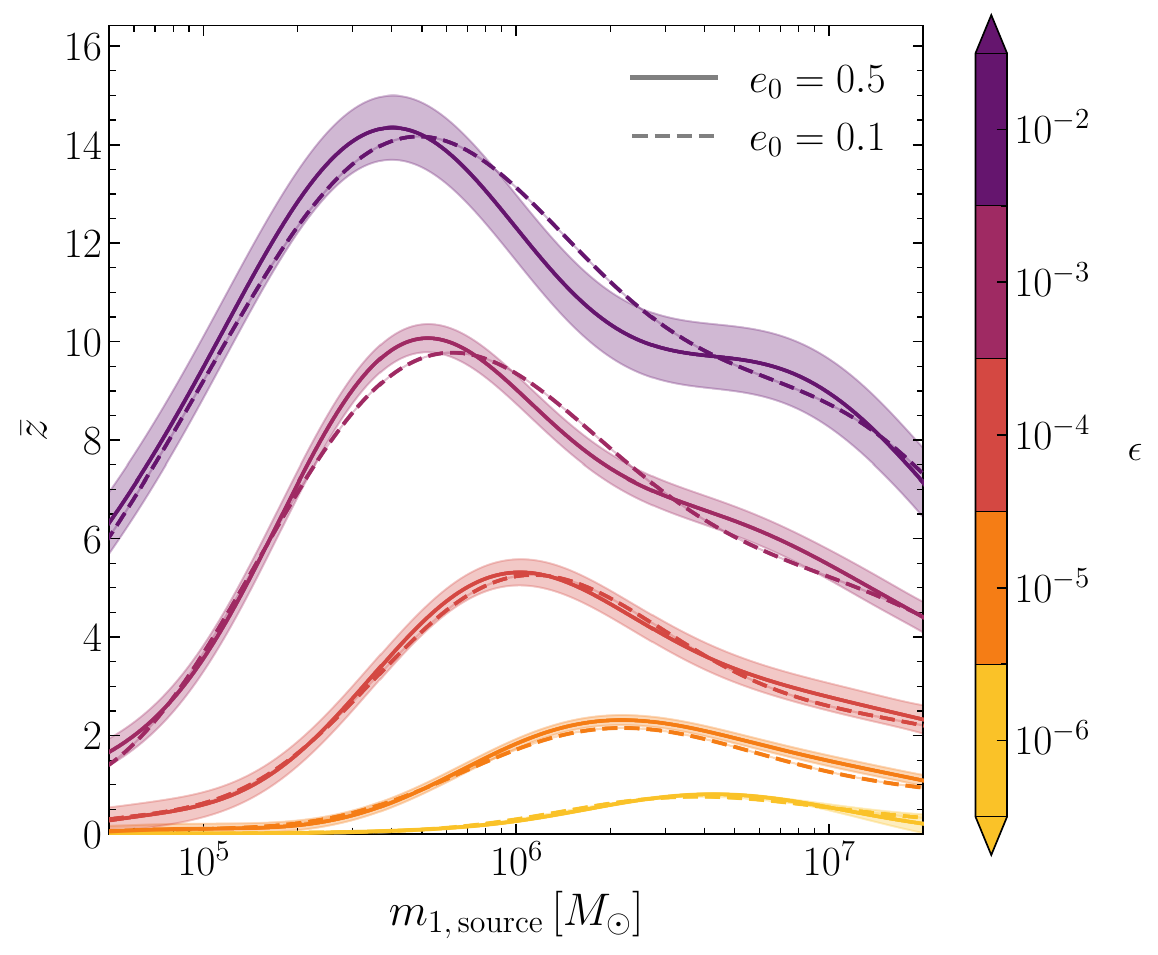}
	\includegraphics[width = 0.45\textwidth]{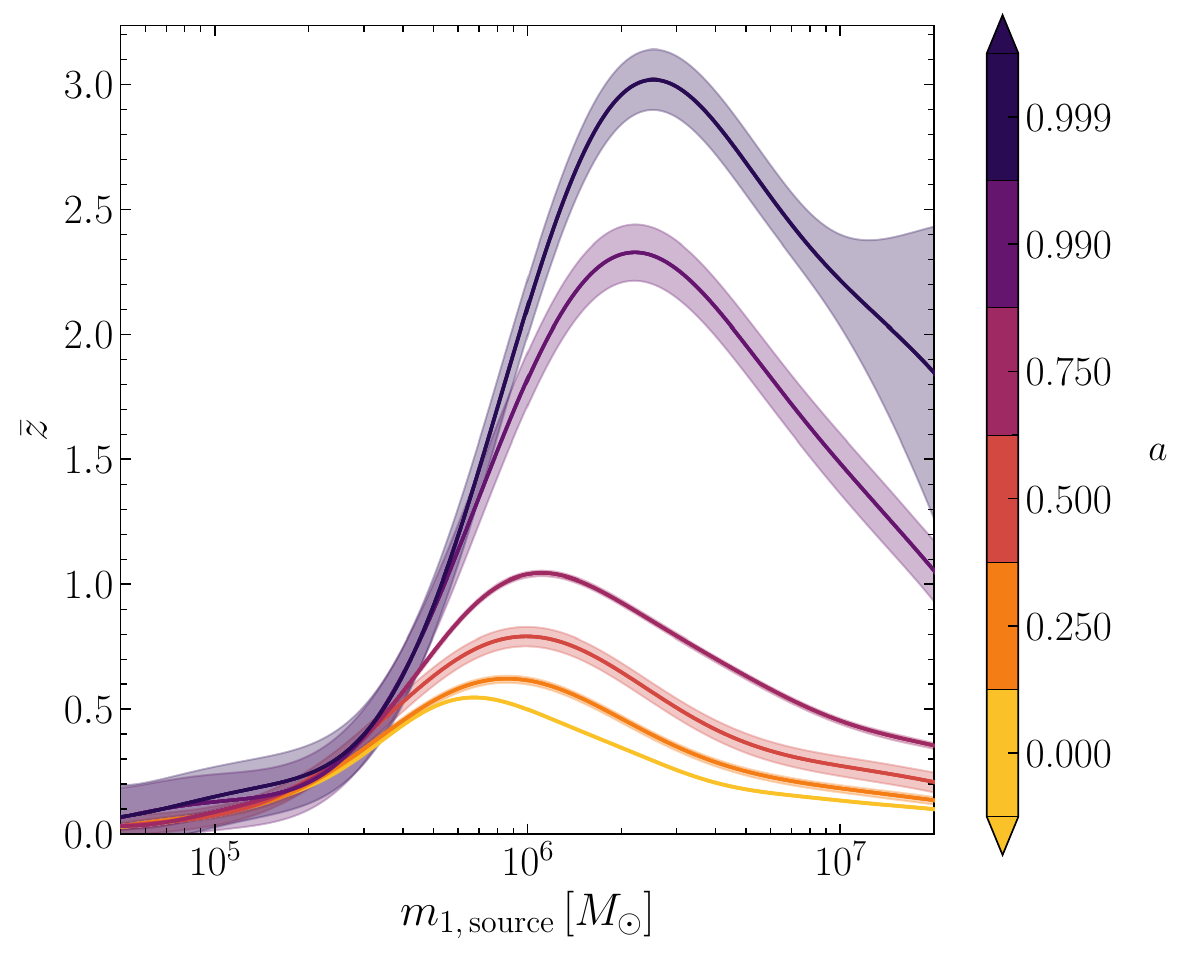}
	\caption{ 
    Here we show the evolution of the horizon redshift $\bar{z}$ at which the sky-averaged \gls{snr} of an eccentric, equatorial I/EMRI reaches a threshold \gls{snr} $\bar{\rho}=20$, for various mass ratios (\emph{left panel}) and spins (\emph{right panel}), as a function of the source-frame primary mass.
    All systems shown are two-year inspirals, and we set an initial eccentricity $e_0=0.5$ for all systems represented by solid lines. 
    Shaded regions represent the $1\,\sigma$ uncertainty region predicted by a GPR trained on the produced data.
	\emph{Left panel:} Systems with $a=0.99$ and $\epsilon \in [10^{-6}, 10^{-2}]$.
    Dashed lines represent systems with the same properties, but initial eccentricity $e_0=0.1$. 
    We do not plot the associated uncertainty intervals for visual clarity, but they are similar in scale to those of the $e_0=0.5$ case.
	\emph{Right panel:} Systems with $\epsilon=10^{-5}$ and $a \in [0.0, 0.999]$.
    }
 \label{fig:horizon_plots}
\end{figure*}

\gls{lisa} is expected to be sensitive to a wide range of \gls{gw} signals emitted by the inspiral of both stellar-mass and intermediate-mass \glspl{co} into \glspl{mbh}. 
Depending on the mass ratios of these sources, observations of \glspl{imri} and \glspl{emri} will be effective probes of the origin and evolution of \glspl{mbh} and their environments in mass ranges currently inaccessible via electromagnetic means~\cite{imbh,LISA:2024hlh}. 
In this subsection, we will investigate the range of distances (and therefore how far back in cosmic history) over which \gls{lisa} can detect these sources. 
\Cref{fig:horizon_plots} shows sky-averaged horizon redshifts $\bar{z}$ for asymmetric-mass binaries of various component masses, \gls{mbh} spins and initial eccentricities.
Each line consists of $20$ points, with detector-frame primary masses spaced uniformly in their logarithm between $5 \times 10^4$ and $5 \times 10^8 \, M_\odot$. 
We restrict the two panels in \cref{fig:horizon_plots} to the region of primary source mass $m_{1}^{\mathrm{(s)}} \in \left[10^5, 10^7 \right]\, M_\odot$.
For each combination of intrinsic parameters, we draw $100$ sets of extrinsic parameters (initial phases and orientation / sky-position angles) according to~\cref{app:monte-carlo-parameters}, and average the \glspl{snr} of the corresponding systems. 

We then compute the luminosity distance $d_\mathrm{L}$ at which each system should be placed to achieve an \gls{snr} detection threshold of 20 (a value commonly assumed in the literature~\cite{Bonetti:2020jku,Pozzoli:2023kxy,chua2017augmented}) and convert it to $\bar{z}$ assuming \textsc{Planck18} cosmology~\cite{Planck:2018vyg, Lahav:2024npe}.
Finally, to take into account the dependence on the number of generated sources during the averaging procedure and the discreteness of the mass grid used for the data generation, we fit a Gaussian Process Regressor (GPR)~\cite{gpr, scikit-learn} on the computed $(m_1^{\mathrm{(s)}}, \bar{z})$ points.
This allows us to provide an estimate of the uncertainty associated with each horizon redshift curve, and to extrapolate to regions beyond the parameter space coverage of our model for two-year inspirals.  
This extrapolation applies for large masses, large spins, and high eccentricities (which correspond to highly eccentric inspirals at low orbital separations, \AMEND{which are outside of the domain of validity of our model}), and leads to the large uncertainty associated with the high-mass tail of the $a = 0.999$ curve in the right-hand panel of~\cref{fig:horizon_plots}.

We find that the horizon redshift only mildly depends on the initial eccentricity for all values of $a$ explorable with our waveform model. 
The left panel of~\cref{fig:horizon_plots} shows the mass-ratio dependence of the horizon redshift for two different values of $e_0$, namely $0.5$ (higher eccentricity; solid lines) and $0.1$ (lower eccentricity; dashed lines). 
We find that the maximum horizon redshift of the lower-eccentricity curve is only slightly smaller than that of the higher-eccentricity curve, 
with the relative difference between the maxima of the two eccentricity curves being less than $6\%$ across all mass ratios considered.
More interestingly, the two show different correlations with the primary mass; in particular, the lower-eccentricity curves do not seem to show the \gls{snr} ``hump'' we see in the higher-eccentricity curves. 
This is especially evident in the $\epsilon \gtrsim 10^{-3}$ cases, and can be interpreted as the contribution of higher-frequency waveform modes with large $n$; at larger total masses, these modes lie in the most sensitive region of the \gls{lisa} sensitivity curve (shown in \cref{fig:tf-waveform}) and contribute significantly to the total \gls{snr} of the system. As the amplitudes of these modes scale strongly with eccentricity, this feature is more readily apparent for particularly eccentric systems.
Additionally, the value of $m_1^{\mathrm{(s)}}$ at which $\bar{z}$ is maximised varies with respect to $e_0$. 
The deviation is reasonably small, with a relative difference between the two eccentricity cases (over all primary mass values) of $\sim 18\%$ at its largest. 
This effect depends on the mass ratio: the low-eccentricity peak is shifted toward smaller masses for the $\eps=10^{-6}$ systems, while it is moved toward larger values for the remaining systems considered, with larger shifts for larger mass ratios. 
The most significant contribution to the \gls{snr} occurs as the \gls{co} enters the strong field (towards the end of inspiral); due to our initial conditions on $p_0$ in this analysis being fixed by the time-to-plunge, the eccentricity at this stage of the inspiral decreases as a function of mass ratio (as the inspiral will circularise more when starting from larger orbital separations).
This accentuates the impact of eccentricity on detectability as mass ratio increases, as depicted in the left panel of~\cref{fig:horizon_plots}.

We also find that the mass location of the detectability peak shows different trends in the two panels, moving towards lower (higher) source frame primary masses with larger mass-ratios (\gls{mbh} spins).
The relationship between mass ratio and \gls{snr} (left panel), for a given value of $m_1$, is largely driven by the proportionality of the waveform amplitudes to $\mu \sim m_2$ (\cref{eq:source-frame-waveform}).
As the rate of inspiral also increases with mass ratio, our fixed time-to-plunge leads to higher values of $p_0$; as this decreases the frequencies of strong harmonics, this in turn slightly accentuates the detectability of lower-mass systems, which manifests as a gradual shift in the location of the maximum horizon redshift in the left panel of \cref{fig:horizon_plots}.
The impact of \gls{mbh} spin (right panel) is also highly pronounced, particularly at larger values of $m_1$.
Raising \gls{mbh} spin increases the number of strong-field orbital cycles completed by the \gls{co} before it plunges, where both the amplitude and frequency of the \glspl{gw} emitted are largest.
For lower-$m_1$ systems, this high-frequency radiation sits towards the upper end of \gls{lisa}'s sensitivity and therefore does not contribute strongly to the \gls{snr} (and therefore the detectability) of these systems.
As $m_1$ increases, this strong-field emission decreases towards the frequencies where \gls{lisa} is most sensitive, significantly enhancing the overall \gls{snr}.
This trend continues until the strong-field emission shifts below the minimum in the \gls{lisa} sensitivity curve ($\sim 3\,\mathrm{mHz}$; see~\cref{fig:tf-waveform}) at which point the \gls{snr} begins to decrease.

To summarize quantitatively the prospects for probing I/EMRI sources over cosmological distances with \gls{lisa}, we find that for an \gls{snr} detection threshold of 20, we can detect \glspl{emri} with mass ratios of $10^{-5}$ at redshifts $\lesssim 3$, with the maximum horizon redshift at $m_1^\mathrm{(s)} \sim 2\times 10^6$ and $a = 0.999$. 
For systems with $a=0.99$, the horizon redshift increases to $\sim 5$ for \glspl{emri} with mass ratios of $10^{-4}$, and to $\sim 14$ for \glspl{imri} with mass ratios of $10^{-2}$.

\subsection{Distinguishing quasi-circular and mildly eccentric systems}
\label{sec:marginal-eccentricity}
The astrophysical formation scenarios of \gls{emri} systems are one of the key areas that \gls{gw} observations of these systems can address. 
One crucial parameter that can help distinguish between different formation models is eccentricity.
While event rates are largely uncertain~\cite{Babak:2017tow,Rom:2024nso,Broggi:2022udp}, 
a significant fraction of \gls{emri} systems may form in gas-dominated environments such as accretion disks~\cite{Pan:2021oob}.
The astrophysics underpinning the formation and evolution of these ``wet'' \glspl{emri} is poorly constrained at present~\cite{Pan:2021ksp}.
In general, interactions between the disk and the inspiralling object are expected to significantly dampen orbital eccentricity such that the system is quasi-circular and nearly equatorial once its \gls{gw} emission enters the \gls{lisa} band.
However, more complicated models of these disk-inspiral interactions or other environmental effects (such as interactions with other bodies in the disk) may yield \gls{lisa}-band \glspl{emri} in disks with small (and potentially measurable) eccentricities~\cite{Li:2025zgo,Spieksma:2025wex}.
Identifying these small eccentricities in \gls{emri} signals may therefore provide significant insights into the astrophysical processes that drive the formation of these systems.
To this end, in this short sub-section we will explore and identify a lower bound for measurable eccentricities in eccentric equatorial \gls{emri} systems with a rapidly-spinning \gls{mbh}.

\begin{figure}
    \centering    \includegraphics[width=\columnwidth]{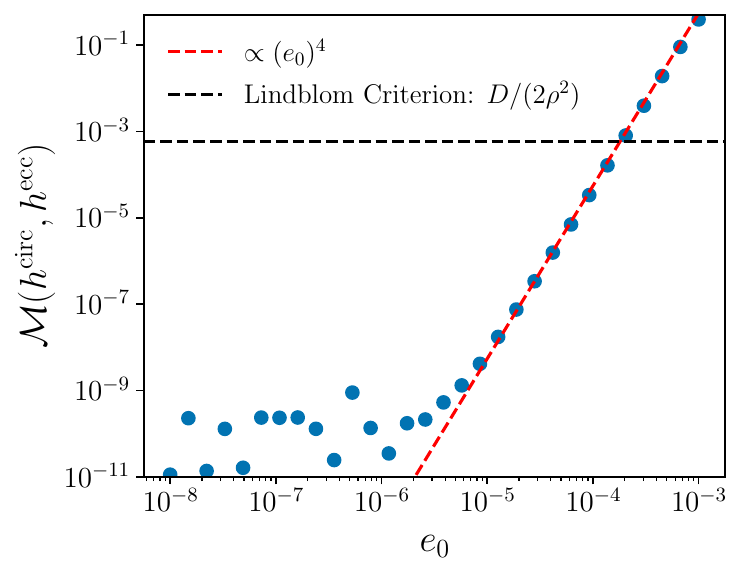}
    \caption{Mismatches (blue dots) between quasi-circular ($h^\mathrm{circ}$) and eccentric $h^\mathrm{ecc}$ waveforms as $e_0 \rightarrow0$.
    Other source parameters $\{m_1, m_2, a\} = \{10^6\,M_\odot, 25\,M_\odot, 0.998\}$, with \gls{snr} $\rho = 100$.
    The black dashed line indicates the Lindblom distinguishability criterion (\cref{eq:lindblom}) for this \gls{snr}, with $D = 12$ (12 sampled parameters).
    For $e_0 \gtrsim 10^{-5}$, mismatches grow as $(e_0)^{4}$ (red dashed line) in line with \gls{pn} scaling arguments.
    At lower initial eccentricities, other systematics in waveform generation obfuscate any physical relationship between mismatch and initial eccentricity.
    }
\label{fig:low_e0_mismatches}
\end{figure}

Intuition on measurable eccentricities can be attained through a simple mismatch case-study.
Consider an \gls{emri} system $h^\mathrm{circ} \equiv h(t;e=0)$, with masses $(m_1, m_2) = (10^{6},10)\,M_{\odot}$, dimensionless primary spin $a = 0.998$ and initial semi-latus rectum $p_0 = 10.628$ that is observed by \gls{lisa} for two years prior to the plunge of the secondary object.
When attempting to measure the system's eccentricity during data analysis, we will propose a template $h^{\rm ecc} \equiv h(t;e)$ with non-zero eccentricity and compare it with the observed signal $h^{\rm circ}$. 
To study the measurability of small eccentricities, we will examine how the mismatch $\mathcal{M}(h^{\rm circ},h^{\rm ecc})$ behaves in the limit as $e \rightarrow 0^{+}$.
A more complete study of the detectability of a small non-zero eccentricity would include comparison of Bayes' factors for the quasi-circular and eccentric hypotheses, which is beyond the scope of this paper; we simply seek to establish an order-of-magnitude estimate of what eccentricities are measurable for a typical \gls{emri} system.

We compute mismatches $\mathcal{M}(h^{\rm circ},h^{\rm ecc})$ for $e_0 \in [10^{-8},10^{-3}]$.
Our results are given in \cref{fig:low_e0_mismatches}. 
For very low eccentricities $e_{0}\in [10^{-8}, 10^{-5})$, the mismatch between circular and eccentric waveforms is dominated by \gls{ode} numerical errors.
However, for $e_{0}\gtrsim 10^{-5}$, the mismatch grows by an amount proportional to $(e_{0}^2)^2$ as given by the red dashed line. 
The black dashed line indicates the Lindblom distinguishability criterion (\cref{eq:lindblom}), where mismatch values $(<)> D/2\rho^2$ indicate that the true and approximate waveforms are (in)distinguishable. 
As in \cref{sec:waveform-validation}, we set the number of sampled model parameters $D = 12$. 
From \cref{fig:low_e0_mismatches}, we see that the two models are distinguishable for $e_{0} \gtrsim 2\cdot 10^{-4}$, highlighting that for $\rho < 100$ we cannot resolve eccentricities smaller than $e_{0} < 2\cdot 10^{-4}$ for this set of source parameters.
We have checked that our conclusions remain unchanged when decreasing the \gls{ode} integrator error $\sigma_\mathrm{tol}$.

To understand why the mismatch $\mathcal{M}(h^{\mathrm {ecc}},h^{\mathrm {circ}}) \propto (e_{0}^2)^{2}$, it helps to understand what drives the dephasing in the first place. 
If $h^\mathrm{ecc} \sim h^\mathrm{circ}e^{i\Delta\Phi}$ for $\Delta \Phi$ the dephasing between $h^\mathrm{circ}$ and $h^\mathrm{ecc}$, then $\mathcal{M} \approx 1 - \cos(\Delta \Phi) \approx \Delta \Phi^2 /2 $. 
We understand from the stability of quasi-circular inspirals~\cite{Ryan:1995zm,Kennefick:1995za,Mino:1997bx,kennefick1998stability,Fujita:2016igj} that, in the low eccentricity limit, the fluxes scale as $\sim \dot{E} \propto \dot{E}_{\mathrm{circ}} + \mathcal{O}(e^2)$.
The dephasings between $h^\mathrm{circ}$ and $h^\mathrm{ecc}$ are therefore driven by differences in the fluxes that are proportional to $e^{2}$, implying that $\Delta\Phi \sim e^{2}$. 
From these simple scaling arguments, it follows that $\mathcal{M}(h^{\mathrm{circ}},h^\mathrm{ecc})$ should grow proportionally to $\Delta \Phi^2 \sim (e_{0}^2)^2 = (e_{0})^4$, in line with our observations.

\subsection{Inference prospects for asymmetric-mass binaries with LISA}
\label{sec:inference-subsection}

\begin{table*}
    \centering
    \renewcommand{\arraystretch}{1.1}
    \setlength{\tabcolsep}{8pt}
    \begin{tabular}{c c c c c c c c c c}
        \hline\hline
        {} & Summary & $m_1 \, (M_{\odot})$ & $m_2\, (M_{\odot})$ & $a$ & $p_0$ & $e_0 \, (e_\mathrm{f})$ & $d_\mathrm{L}\, (\text{Gpc})$ & $z_\mathrm{r}$ & $\rho$ \\
        \hline
        $1$ & EMRI (Prograde) & $10^6$ & $10^1$ & $0.998$ & $7.728$ & $0.730\,(0.045)$ & $2.204$ & $0.394$ & $50$  \\
        $2$ & Strong-field EMRI & $10^7$ & $10^1$ & $0.998$ & $2.120$ & $0.425\,(0.261)$ & $3.590$ & $0.593$ & $30$ \\
        $3$ & Heavy IMRI & $10^7$ & $10^5$ & $0.950$ & $23.425$ & $0.850\,(0.023)$ &  $7.250$ & $1.058$ & $500$ \\
        $4$ & Light IMRI & $10^5$ & $10^3$ & $0.950$ & $74.383$ & $0.850\,(0.004)$ & $3.500$  & $0.581$ & $200$\\
        $5$ & EMRI (Retrograde) & $10^5$ & $10^1$ & $-0.500$ & $26.190$& $0.800\,(0.195)$ &  {$1.081$} & {$0.212$} & $30$\\
        \hline\hline
    \end{tabular}
    \caption{
    Parameters for investigated \gls{emri} and \gls{imri} sources.
    Waveforms for these systems are shown in \cref{fig:start-end-snapshots}.
    \textbf{(From left to right columns):} Detector-frame masses $m_1$ and $m_2$, dimensionless \gls{mbh} spin $a$, 
    initial semi-latus rectum $p_{0}$, initial (final) eccentricity $e_{0}$ ($e_\mathrm{f}$), luminosity distance $d_\mathrm{L}$, redshift $z_\mathrm{r}$ and the optimal \gls{snr} $\rho$. 
    Each source is observed for two years and plunges just after the end of the observation window. 
    For sources with $m_1 = 10^{5}M_{\odot}$, we use a finer sample cadence $\mathrm{d} t = 2\,\mathrm{s}$ to resolve high-frequency \glspl{gw} near plunge; all other sources use our fiducial sampling interval of $\mathrm{d} t = 5\,\mathrm{s}$. 
    In all cases, we fix the angular parameters $(\theta_{S}, \phi_{S}, \theta_{K}, \phi_{K}) = (0.8, 2.2, 1.6, 1.2)$ and initial orbital phases $(\Phi_{\phi_{0}}, \Phi_{r_0}) = (2.0, 3.0)$. 
    }
    \label{tab:emri_params}
\end{table*}

In this subsection, we investigate parameter estimation prospects for a range of \gls{imri} and \gls{emri} sources. 
As \gls{few} is capable of generating waveforms in $\sim 100\,\mathrm{ms}$ (\cref{sec:timing}), we are able to perform Bayesian inference studies on a timescale of hours, directly exploring the structure and scale of the posterior distributions of these sources. 
The motivation for performing these analyses is twofold.
First, we seek to verify that our model reproduces features expected of posterior distributions for these systems: in particular, that they are approximately Gaussian, and that parameters intrinsic to the inspiral dynamics (such as spin and eccentricity) are recovered with high precision.
Second, it was identified in Ref.~\cite{Katz:2021yft} that the mode content of \gls{emri} waveform models can be reduced significantly without significantly biasing inference results; our analyses will re-examine this conclusion with the inclusion of \gls{mbh} spin and for \gls{imri} sources.
To our knowledge, this is the first full Bayesian investigation of rapidly-spinning and highly eccentric \gls{imri} and \gls{emri} systems to appear in the literature to date (however, see Ref.~\cite{Huez:2025npe} for a parameter estimation study of mildly eccentric binaries with $q \leq 10$). 

\begin{figure*}
    \centering

    \includegraphics[width=\textwidth]{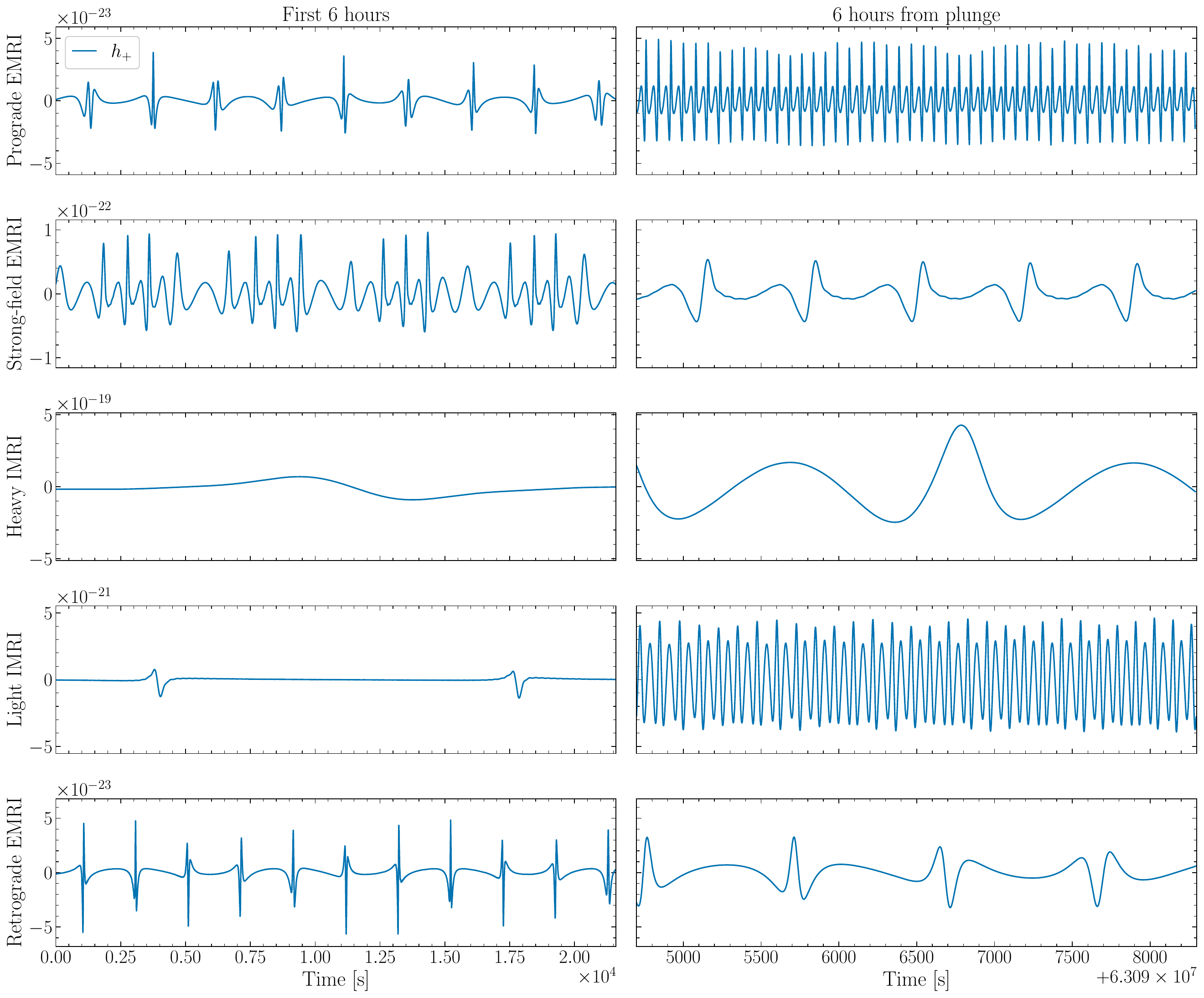}
    \caption{
    Time-domain snapshots of the science-case waveforms described in \cref{tab:emri_params} during the early (\emph{left panel}) and late (\emph{right panel}) periods of a two-year inspiral.
    This set of representative sources demonstrates the wide range of waveform morphologies captured by our model.
    In all the panels, only the plus polarization $h_+$ is shown; for the two \gls{imri} systems, initial orbital phases have also been adjusted for visual clarity. 
    }
    \label{fig:start-end-snapshots}
\end{figure*}

We perform parameter estimation on simulations of each of the sources listed in~\cref{tab:emri_params}.
The parameters of these sources were chosen to represent the wide range of waveform morphologies accessible with our waveform model~\cref{fig:start-end-snapshots}, with luminosity distances set to astrophysically-motivated values.
A notable exception to our analysis conventions is that we set a finer sampling cadence of $\mathrm{d}t=2\,\mathrm{s}$ for Sources 4 and 5 in order to better resolve their high-frequency mode content; for other sources, we use our fiducial sampling cadence of $\mathrm{d}t=5\,\mathrm{s}$.
\AMEND{While these sampling cadences are not quite high enough to resolve the highest-frequency harmonics near plunge for some of these systems (particularly Sources 1 and 4), we found that these cadences were sufficient to obtain accurate parameter constraints.}
For each simulation, we generate a waveform with mode-selection parameter $\kappa=0$ (i.e., all modes in our amplitude model are included). 
During inference, we consider $\kappa = 10^{-5}$ and $\kappa = \{10^{-3},10^{-2}\}$ (the latter value depending on the \gls{snr} of the injection), demonstrating how parameter inference results are impacted when weaker harmonics (which are more important for high eccentricities and spins; see~\cref{fig:heatmap_mode_amplitudes}) are neglected.
We do not show the (marginal) posteriors obtained from our inference runs in the main text for brevity; they can be found in~\cref{app:corner_plots}, and are referenced in the right-most column of Tab.~\ref{tab:emri_measurement_precisions}, where we summarize the relative measurement precisions for parameters of astrophysical interest. 
These include the source frame masses $m_i^{(\rm s)}$, spin $a$, final eccentricity $e_f$ and luminosity distance $d_\mathrm{L}$.
We also report the sky-localization area of each source: given the covariance matrix of the sky-angle posterior samples $\Sigma = \rm cov(\theta_S, \, \phi_S)$, and $\bar{\theta}_S= \rm{median} (\theta_S)$, we compute the $99 \%$ quantile of the sky-localization area in square degrees, given by 
\begin{equation}
    \label{eq:sky-area}
    \Delta \Omega^\mathrm{sky} = 9 \cdot 2\pi \, \sin{\bar{\theta}_S} \sqrt{\det{\Sigma}} \left(\frac{180}{\pi}\right)^2.
\end{equation}
We remind the reader that all parameter inference was conducted using second-generation \gls{tdi} variables as implemented by the \textsc{fastLISAresponse} software~\cite{katz2022assessing}, assuming constant- and equal-armlength orbits; we sampled all posterior distributions with \gls{mcmc} using the \textsc{eryn} package~\cite{Karnesis:2023ras,michael_katz_2023_7705496,Foreman-Mackey:2013} and default settings. 

\begin{table*}
    \centering
    \renewcommand{\arraystretch}{1.1}
    \setlength{\tabcolsep}{8pt}
    \begin{tabular}{c c c c c c c c}
        \hline\hline
        {} & $\delta m_1^\mathrm{(s)} \, (10^{-3})$ & $\delta m_2^\mathrm{(s)} \, (10^{-3})$ & $\delta a \, (10^{-5})$ & $\delta e_\mathrm{f} \, (10^{-4})$ & $\delta d_\mathrm{L}\, (10^{-2})$ & $\Delta \Omega^\mathrm{(sky)}\,(\mathrm{deg}^2)$ & Figure\\
        \hline
        $1$ & $9.181$ & $9.181$ & 0.023 & 0.785 & 3.802 & $1.049$ & (\ref{fig:MCMC_row_1})  \\
        $2$ & 16.506 & 16.505 & 0.040 & 0.295 & 5.326 & $3.893$ & (\ref{fig:MCMC_row_2}) \\
        $3$ & 2.459 & 2.455 & 2.600 & 2.545 & 0.776 & $17.006$ & (\ref{fig:MCMC_row_3}) \\
        $4$ & 2.650 & 2.650 & 5.234 & 69.845 & 0.870 & $1.079$ & (\ref{fig:MCMC_row_4}) \\
        $5$ & 9.200 & 9.200 & 11.830 & 0.154 & 5.847 & $1.540$ & (\ref{fig:MCMC_row_5}) \\
        \hline\hline
    \end{tabular}
    \caption{Relative precisions obtained in the parameter estimation of the sources listed in \cref{tab:emri_params} (where $\delta x$ denotes the relative precision in parameter $x$, defined as the $1\,\sigma$ width of the marginal posterior on $x$ normalised by the median value).
    The $^{(\rm s)}$ superscripts indicates that masses have been converted in the source frame. For each sample, we recover the redshift from the luminosity distance $d_\mathrm{L}$ assuming the \textsc{PLANCK18} cosmology. 
    We also report $\Delta \Omega^\mathrm{sky}$ (\cref{eq:sky-area}), which is $99\%$ of the sky-localisation area of each source in square degrees.
    In the last column, we provide the reference for the corresponding posterior distributions in \cref{app:corner_plots}. 
    Notably, $\delta m_1^\mathrm{(s)}$ and $\delta m_2^\mathrm{(s)}$ are a factor of $\sim 3$ orders of magnitude larger than the equivalent quantities computed in the detector frame, $\delta m_i^\mathrm{(d)}$. 
    This is a consequence of the mass relation $m_i^{(d)} = (1+z_\mathrm{r})\,m_i^{(s)}$, which introduces the (much larger) uncertainty in $d_\mathrm{L}$ due to its relationship with $z_\mathrm{r}$ via our assumed cosmology.
    }
    \label{tab:emri_measurement_precisions}
\end{table*}

\subsubsection{EMRIs}\label{sec:EMRI_science_cases}
We investigate three \glspl{emri} with mass ratios $\epsilon\in\{10^{-4},10^{-5},10^{-6}\}$ and parameters given by the fifth, first and second row of \cref{tab:emri_params} respectively.
As confirmed by our horizon redshift study in \cref{subsec:horizon_redshift}, these sources are indeed observable and each are unique in their own way. 
Source 1 has high spin and initial eccentricity $(a, e_0) = (0.998, 0.730)$, and represents a fairly typical \gls{emri}, assuming that many \glspl{mbh} in nature are spinning rapidly~\cite{Babak:2017tow,Barausse:2012fy}.
The more extreme mass-ratio of Source 2 causes its inspiral trajectory to evolve on a time-scale much slower than that of Source 1. 
This is reflected by its two-year inspiral starting deep in the strong field of the primary ($p_{0} \sim 2.12$) for $(a,e_0) = (0.998, 0.425)$, terminating $\sim 0.15$ away from the separatrix.
Notice that the final eccentricity $e_f \sim 0.261$; the orbit is still quite eccentric close to plunge, and therefore lies in the upper tail of the eccentricity-at-plunge population for \glspl{emri} formed via capture in Ref.~\cite{Babak:2017tow}.
The final \gls{emri} we consider, (Source 5) is a retrograde inspiral that (due to its mass ratio $\epsilon = 10^{-4}$) begins in the weak-field regime ($p_{0} =26.19$) with high $e_{0} = 0.8$ and moderate spin $a = -0.5$, terminating near the separatrix with non-trivial eccentricity $e_{f} = 0.195$. 
Based on a prior examination of \gls{emri} astrophysics in Ref.~\cite{Babak:2017tow}, we believe these are characteristic asymmetric-mass binaries that are readily observable by \gls{lisa} to plausible luminosity distances (and therefore redshifts).

Measurement precisions on the parameters of these sources are shown in the corresponding rows of~\cref{tab:emri_measurement_precisions}.
For Source 1, we see that we can constrain the intrinsic parameters to usual levels ($\Delta \theta/\theta \sim 10^{-6}$) for a source with an \gls{snr} of $30$. 
From the marginal posteriors (shown in \cref{fig:MCMC_row_1}), we find that setting $\kappa = 10^{-2}$ (retaining 235 modes) yields identical inference results to $\kappa = 10^{-5}$ (1228 modes), with any systematic biases present being too small to identify.
Similar results are obtained for Source 2 in terms of measurement precision (accounting for the lower \gls{snr} of this source), except for the improved determination of $e_\mathrm{f}$ which is expected due to the significant final eccentricity of this source.
Correlations between parameters are more pronounced for this source than the other two, which is likely a feature of the mass ratio of this source leading to slower orbital parameter evolution.
Similarly to Source 1, there is an absence of any systematic biases in \cref{fig:MCMC_row_2} between the posteriors for $\kappa=10^{-5}$ (809 modes) and $\kappa=10^{-2}$ (186 modes).
For Source \sout{3} {5}, the spin constraint are wider than for the other two sources, which is expected as the effects of spin are less pronounced for this source (as the inspiral spends a large fraction of the two-year duration in the weaker field).
The same can not be said for eccentricity effects (which are still significant in the weak field), which is reflected in the recovery of $e_\mathrm{f}$.
Once again, no systematic biases in \cref{fig:MCMC_row_5} are observable between posteriors for $\kappa=10^{-5}$ (928 modes) and $\kappa=10^{-2}$ (208 modes).

Our results confirm that \gls{emri} posterior distributions continue to be extremely narrow with the inclusion of \gls{mbh} spin.
In all cases, we recover Gaussian posteriors, which in turn implies that information matrices are an effective tool for probing the measurement precision for these sources.
These conclusions do not change when waveform mode content is reduced by $70\mathrm{-}75\%$ in all three cases, which in turn reduced inference times by more than a factor of two.
Despite the poorer measurement precisions obtained when analysing the retrograde source, luminosity distance and sky localisation area were constrained to roughly equal precision for all three analyses, which is expected due to the similar \glspl{snr} of these sources.
Notably, this leads to source-frame mass measurements of similar precision in all cases, despite the wider uncertainties in detector-frame masses for the retrograde source.
This highlights that while retrograde systems may not provide as strong a constraint on the \gls{mbh} spin, they will still be as effective as prograde \glspl{emri} for probing the \gls{mbh} mass population~\cite{Gair:2010yu,Chapman-Bird:2022tvu}.
With sky localisation areas of $\sim 1\,\mathrm{deg}^2$, both prograde and retrograde \glspl{emri} will also be valuable signals for dark-siren cosmological analyses~\cite{Gray:2023wgj,Schutz:1986gp,Laghi:2021pqk}.

\subsubsection{IMRIs}

As demonstrated in our horizon redshift study (\cref{fig:horizon_plots}), \glspl{imri} will be observable by \gls{lisa} with high \glspl{snr} to high redshifts.
To our knowledge, we have developed the first waveform model capable of modelling highly-eccentric \glspl{imri} into rapidly-spinning \glspl{mbh} (at the adiabatic order) that is sufficiently rapid for use in Bayesian inference studies.
To demonstrate these capabilities and perform an initial investigation of \gls{imri} posterior structure, we examine two \glspl{imri} with mass ratios $\epsilon = 10^{-2}$ and parameters given by row 3 and 4 in~\cref{tab:emri_params}.
Due to the large mass ratio, we see that the orbital parameters evolve far more rapidly than for the \gls{emri} sources we examined previously. 
Indeed, despite their high initial eccentricities $e_0 = 0.85$, the binaries have circularised significantly prior to plunge. 
Notably, the \glspl{snr} $\rho = \{500,200\}$ of these sources are also significantly larger, mainly due to the waveform amplitudes being proportional to $\mu \sim m_2$ (\cref{eq:source-frame-waveform}).
However, having such high \glspl{snr} may lead to more significant systematic biases when waveform models with incomplete mode content are used in parameter inference, as we explore below.

For both Source 3 and Source 4, waveforms with $\kappa = 10^{-5}$ (949 and 626 modes respectively) are sufficiently complete for performing parameter estimation with no observable biases, as shown in \cref{fig:MCMC_row_3,fig:MCMC_row_4}.
However, when attempting to infer the parameters with $\kappa = 10^{-2}$ (not shown in Figures; 223 and 156 modes respectively) we observed statistically significant biases across all the \gls{imri} parameters.
Nearly all parameters were recovered with biases of $\gtrsim 3\,\sigma$ away from the truth (with $\sigma$ computed as the standard deviation of the posterior samples) that were readily identifiable compared to the scale of the posterior. 
Even for $\kappa = 10^{-3}$ (412 and 291 modes respectively), we observe unacceptable biases in the luminosity distance and sky location on the order of $\sim 4\sigma$ and $\sim 7\sigma$ biases away from the true parameters respectively.
Biases of this scale would propagate to population / cosmological analyses and may contaminate the results of these studies, especially when considering that luminosity distance biases will affect source-frame mass estimates as well.
This behaviour is to be expected considering the higher \glspl{snr} of these sources --- from \cref{eq:lindblom}, we can expect mismatches of $10^{-3}$ (which $\kappa=10^{-3}$ roughly leads to) to be insufficient for the analysis of signals with \glspl{snr} of hundreds.
The fact that extrinsic parameters are most affected by the omission of impactful waveform modes is also expected, as this has been previously identified in the analysis of comparable-mass case binaries observed with \gls{lisa}~\cite{Yi:2025pxe}.

We report measurement precisions of parameters of astrophysical interest for these sources in \cref{tab:emri_measurement_precisions}.
Despite the higher \glspl{snr} of these signals, their spins are recovered with larger uncertainties than for the prograde \gls{emri} sources we examined.
This can be understood in terms of the rapid evolution of the \gls{imri} trajectories, which complete relatively few cycles in the strong-field regime where the effects of spin are most important.
Eccentricity-at-plunge is also recovered more poorly; this has similar justification, in addition to the \gls{imri} trajectories beginning at larger orbital separations and therefore entering the strong-field regime with lower eccentricities than the \gls{emri} sources.
Luminosity distances are recovered precisely (to better than $1\%$ precision) in both cases, which is to be expected as amplitude measurements depend strongly on the \gls{snr}.
This has the interesting consequence that despite the recovery of detector-frame masses for the \gls{imri} sources being less precise than for the \gls{emri} sources, they still provide more precise recovery of source-frame masses due to the improvement in the luminosity distance measurement.
Last, we note that the $99\%$ sky-localisation area for Source 3 ($\sim 17\,\mathrm{deg}^2$) is more than an order of magnitude larger than that of Source 4. 
This occurs because during early inspiral, the bulk of the mode spectrum for this source sits mainly at lower frequencies ($\lesssim 10^{-4}\,\mathrm{Hz}$) that are poorly measured by \gls{lisa}.
As the inspiral evolves during the observation, the mode spectrum shifts to higher frequencies (``entering'' the lower end of the \gls{lisa} band) and is measured well.
While this still enables many parameters to be recovered precisely, the sky position of the source is determined by long-timescale variations in the \gls{lisa} response function; as the waveform of this source only has significant \gls{snr} for part of the observation (such that less of these variations are measurable), this in turn leads to a poorer localisation of the source on the sky.

These analyses highlight that \gls{lisa} will be capable of measuring the parameters of \gls{imri} systems very precisely, due in part to the high \glspl{snr} of these systems.
Comparing between the extreme-mass-ratio limit (where many harmonic modes are necessary for unbiased inferences) and the comparable-mass limit (where current waveform models include only a handful of higher-order modes~\cite{Marsat:2020rtl,Toubiana:2023cwr}), it has not been explored in detail in the literature where \gls{imri} modelling lies on this scale (particularly with the inclusion of significant eccentricity).
Our results indicate definitively that \glspl{imri} (at least, of mass ratios $\sim 10^{-2}$) lie towards the \gls{emri} end of this scale --- models will need to accurately produce the harmonic mode structure of waveforms for these sources if their parameters are to be faithfully recovered.
We note that (similarly to \glspl{emri}) it is likely that models with limited mode coverage will be sufficient for detecting these sources in \gls{lisa} data, especially given the larger \glspl{snr} of these sources.
The computational benefits of doing so will be significant: for our analyses, the $\sim 55\%$ reduction in mode content achieved by setting $\kappa = 10^{-3}$ halved the parameter estimation runtime.

\section{Conclusions and future work}
\label{sec:conclusions}

\subsection{Summary of this work}
\label{sec:final_summary}

In this work, we have presented the first \gls{gsf}-based waveform model for asymmetric-mass binaries capable of efficiently generating analysis-length eccentric equatorial inspirals into rapidly-rotating black holes (at the adiabatic order). 
This model is housed in the new v2 release of the \gls{few} package, which provides the constituent components of this waveform model --- new inspiral trajectory and mode amplitude models --- as part of a modular framework that is readily adaptable to meet the requirements of the end user.
In addition to the introduction of primary spins $|a| \leq 0.999$, our model also extends semi-latus rectum and eccentricity support from $(p_0, e_0) = (\sim16, 0.7)$ to $(p_0, e_0) = (200, 0.9)$ with respect to the previous adiabatic \gls{few} model.
This expansion of parameter-space coverage extends \gls{few} to the modelling of long-duration \glspl{imri}, providing the first efficient adiabatic-order waveforms for these systems that incorporate large spins and/or eccentricities to appear in the literature.
In recent work, \gls{pn}- and Effective-One-Body-based waveform models have been developed that may be extended to the \gls{imri} regime (albeit for smaller or moderate eccentricities), e.g., Refs.~\cite{Chiaramello:2020ehz,Ramos-Buades:2021adz,Albertini:2023aol,Paul:2024ujx,Morras:2025nlp,Planas:2025feq,Gamboa:2024hli}.
Similarly to our case, these models do not have long-duration \gls{imri} simulations against which to investigate their accuracy; comparisons between these models and the one we have presented here (particularly with the inclusion of post-adiabatic effects) is therefore worthy of future investigation.

We examined significant sources of systematic error in our model --- the precise computation and interpolation of \gls{gsf} data products --- with comparisons against independent datasets and analyses of interpolation error convergence, and estimated that our model attains mismatches of $\sim 10^{-5}$ with respect to error-free adiabatic waveforms over the majority of the parameter space.
The most significant source of systematic error we identified was the interpolation of forcing functions.
Errors due to the linear interpolation of mode amplitudes with respect to spin were also found to increase significantly near the separatrix.
During this process, we identified that our model is least robust for eccentricities $e_0 \gtrsim 0.85$, and for spins $a \sim 0.998$ near the separatrix ($p \sim p_\mathrm{sep} + 1$).
Waveform generation wall-times were found to be $\sim 100\,\mathrm{ms}$, and were slightly larger than those obtained in Ref.~\cite{Katz:2021yft} for zero-spin systems.
This increase in cost is due primarily to the inclusion of larger $n$-modes in the waveform (especially for prograde inspirals with larger spins), but it also due to the implementation of the \gls{few} trajectory module entirely in \texttt{Python} (which is less efficient than the previous \texttt{C} implementation).
The latter choice is worth the performance penalty, as it significantly improves the customisability of the \gls{few} framework via modifications to the inspiral model, and access to the continuous \gls{ode} solution is an asset to waveform generation and data analysis schemes.

By comparing with both the \gls{aak} and a quadrupolar variant of our waveform model, we demonstrated the importance of relativistic amplitudes and higher-order modes when \AMEND{performing} \gls{snr} computations and parameter estimation. 
We found that \glspl{snr} computed with the \gls{aak} are more inaccurate for stronger-field high-mass systems, where the missing higher-order modes are most important and the semi-relativistic amplitudes of the \gls{aak} are least accurate.
For all component masses considered, the \gls{aak} overestimated the \gls{snr} for large spins (by up to $60\%$ in the high-mass case), and exhibited larger \gls{snr} errors for larger eccentricities.
This has ramifications for quantitative studies of the \gls{emri} detection rate in the literature (such as Ref.~\cite{Babak:2017tow}), which should be revisited with relativistic waveform models in order to obtain better rate estimates.
Exploring inference biases under the linear signal approximation, we identify biases of $\sim 1\,\sigma$ with the \gls{aak} and $\sim 0.1\,\sigma$ with the quadrupolar relativistic model.
While these biases may be identifiable in the parameter estimation of \glspl{emri}, they are sufficiently small that these approximate models may be readily applicable in the search for \gls{emri} signals in detector data (where any corresponding reductions in computational cost would be highly beneficial).
An exploration of this possibility is left to future work.

To quantify and explore \gls{lisa}'s sensitivity to \glspl{gw} from asymmetric-mass binaries, we computed sky-averaged horizon (maximum) redshifts at which \gls{lisa} can detect sources with varying component masses, spins and eccentricities.
We found that \gls{lisa} can detect \glspl{emri} of mass ratio $10^{-6}$ ($10^{-4}$) at redshifts of up to 1 (5), and \glspl{imri} of mass ratio $10^{-3}$ ($10^{-2}$) at redshifts of up to 10 (15), with only $\sim 10\%$ variations with respect to eccentricity.
In the case of \glspl{emri} with mass ratio $10^{-5}$, horizon redshifts varied from $0.5$ to $3$ as spin increased from $0$ to $0.999$.
For all configurations, we found that \gls{lisa} was most sensitive to binaries with source-frame primary masses $m_1^{\mathrm{(s)}} \sim 10^6\,M_\odot$.
Our horizon redshift results highlight that the detection and characterisation of asymmetric-mass binaries will enable \gls{lisa} to effectively probe the evolution of \gls{mbh}, intermediate-mass black hole and stellar-mass black hole populations over cosmic time, in line with \gls{lisa}'s science objectives.
The waveform model we have developed (as a component of the \gls{few} framework) is unique in the respect that it is able to accurately reproduce the physical characteristics of this wide range of \gls{gw} sources, providing the tools necessary for the exploration of \gls{lisa} science with \gls{imri} and \gls{emri} observations. 

In a similar vein, we also examined the minimum eccentricity measurable by \gls{lisa} for a representative (and rapidly-spinning) \gls{emri} system.
Understanding this limit is highly relevant to models of \gls{emri} formation in gas-dominated environments (and therefore the population of these systems as a whole).
We find that the relationship between eccentricity and mismatch (with respect to a quasi-circular waveform) scales as $e_0^4$ (in line with low-eccentricity expansions around quasi-circular inspirals), and that eccentricities less than $\sim 10^{-4}$ are unlikely to be measurable by \gls{lisa} for these systems.
This is significantly larger than the typical measurement precision achievable for \glspl{emri} with larger eccentricities, serving as a reminder that measurement precisions on inspiral parameters are not synonymous with the minimum values of these parameters that are measurable.
We only examined a single system in our analysis, as our goal was simply to obtain a rough intuition for the measurability of small eccentricities; characterising how the detectability of these small eccentricities varies across the \gls{emri} parameter space is an area worthy of future investigation.

Last, we performed Bayesian parameter estimation of several \gls{emri} and \gls{imri} sources with \gls{mcmc} methods, attaining relative precisions of $10^{-2}-10^{-3}$ for source-frame mass measurements.
The spins and final eccentricities of these systems were measured with precisions of $10^{-7}-10^{-4}$ and $10^{-5}-10^{-3}$ respectively.
In most cases, luminosity distances and $99\%$ sky-localisation areas were measured with $\sim 1\%$ precision and errors of $\sim 1\,\mathrm{deg}^2$ respectively for two-year inspirals.
We identified significant biases when performing inference on \gls{imri} observations using waveforms with reduced mode content.
This is the first demonstration of importance of higher modes in the context of \glspl{imri} for \gls{lisa}, highlighting that the analysis of these signals is more similar in principle to that of \glspl{emri} than comparable-mass binaries. 
While the systematic biases we observed when using fewer harmonic modes were significant in the context of inference, they were sufficiently small that such approximate models may be readily applicable in the detection and identification of these signals (especially given the factor of $\sim 2$ speed-up we observed in our analyses when using these models).
From our inference results, it is clear that significant effort is required to properly explore the scientific potential of \gls{imri} observations with \gls{lisa}.
As we have demonstrated, these sources are readily modelled and analysed with the \gls{few} framework, which we expect will be foundational to this future work.

\subsection{Prospects for future work}
\label{sec:future_prospects}

The implementation of this new waveform model and the general improvements to the \gls{few} framework lays the foundations for many extensions to incorporate other physical features necessary for a complete description of generic I/EMRI systems. In terms of parameter space coverage this includes extending to precessing and then generic (eccentric and precessing) orbits. Further extensions to the model include adding the merger and ringdown, resonance effects, and \gls{gw} memory. In terms of improving phase accuracy, post-adiabatic corrections from second-order self-force calculations and due to the spin on the secondary need to be incorporated. The \gls{few} framework is designed to incorporate all of these effects whilst maintaining rapid waveform generation. We now briefly discuss how each of these extensions can be made.

\textit{Extension to precessing orbits.} Precessing orbits, sometimes known as spherical orbits in the EMRI literature (e.g., Refs.~\cite{Wilkins:1972rs,Teo:2003ltt,Teo:2020sey,Lynch:2023gpu}), are orbits with $e=0$ but $|x_I| \neq 1$.
Just as with eccentric, equatorial orbits the parameter space is three dimensional consisting of $(a,p,x_I)$.
Thus all the framework in this paper for interpolation the fluxes and waveform amplitudes will carry over to the precessing case.
The main challenge is to compute the fluxes and amplitudes across the parameter space.
Codes exist to make these calculations, e.g., Refs.~\cite{Hughes:1999bq,BHPToolkit,BHPClub} and the overall computational cost will likely be less than was required for the present work as the number of polar harmonics needed to accurately compute the flux is typically less than the number of radial harmonics needed for eccentric orbits.

\textit{Extension to generic orbits.} The \gls{few} framework is designed with generic, eccentric and precessing orbits in mind but some technical limitations must be overcome before such a model can be built. 
Of particular note is the increased difficulty in interpolating data for these inspirals; the parameter-space dimensionality increases to four which greatly expands the (already large) memory requirements for mode amplitude interpolation.
Given that polar ($k \neq 0$) mode amplitudes must also be computed (increasing the total number of modes by more than an order of magnitude~\cite{Fujita:2025pc}), extrapolating from the domain of our model leads to mode amplitude grids of $\sim 500\,\mathrm{GB}$ in size or more, which is several times too large to fit in the memory of existing \gls{gpu} hardware.
The overall cost of computing the underlying data for flux and amplitude interpolation also grows significantly as the higher dimensional parameter space necessitates sampling the space at more points.
Furthermore, at each point in the parameter space the perturbation theory computation is more expensive as one must integrate over both an orbit's radial and polar motions, and one must sum over both radial and polar frequency harmonics.
This increases the computational cost of each mode which contributes to the flux by a factor of about $20$--$30$, and increases the number of modes which contribute to the flux by a similar factor.
The problem remains highly parallelizable, so will not be difficult to exploit efficient codes which can compute these data~\cite{Drasco:2005kz, Fujita:2009us,vandeMeent:2017bcc,Chen:2023lsa, Kerachian:2023oiw}.  However, the sheer quantity of data needed means that care is needed in designing the grid spacing across the parameter space before committing significant computational resources.  

\textit{Inspiral-merger-ringdown models.} For typical EMRIs, very little \gls{snr} is accumulated during the merger and ringdown portion of the waveform.
With recent work showing that \gls{gsf} results can also be used to model \glspl{imri} \cite{LeTiec:2011bk,vandeMeent:2020xgc,Wardell:2021fyy,Warburton:2021kwk},
it becomes more important to include the merger and ringdown~\cite{Apte:2019txp,Becker:2024xdi,Lhost:2024jmw,Leather:2025nhu}.
Previous approaches to computing these in perturbation theory involved computing the waveform using time-domain codes~\cite{Taracchini:2014zpa,Lim:2019xrb,Strusberg:2025qfv} and thus the results were not suitable for the \gls{few} framework.
To overcome this, recent work has cast the transition from inspiral to plunge and the plunging portions of the waveform within the multiscale framework~\cite{Kuchler:2024esj,Kuchler:2025hwx}.
Within this approach the online computation of the transition and the plunge involves solving new trajectory ODEs and the waveform is computed from a new set of amplitudes.
Structurally, this is exactly the same as the present online inspiral computation and, as such, the results of these calculations can be readily incorporated into \gls{few}.

\textit{Resonances.} Generic inspirals can experience resonant phenomena when any two of three orbital frequencies $(\hat{\Omega}_{\phi}, \hat{\Omega}_{\theta}, \hat{\Omega}_{r})$ are a low-integer multiple of each other~\cite{Mino:2005an,Tanaka:2005ue,Flanagan:2010cd, Ruangsri:2013hra, Brink:2015roa, Levati:2025ybi, Hirata:2010xn,vandeMeent:2014raa}\footnote{Similar resonances in I/EMRI systems can also occur as a consequence of the existence of a third body~\cite{Bonga:2019ycj,Yang:2019iqa,Gupta:2021cno,Gupta:2022jdt} or the presence of gravitational perturbations in the environment more generally~\cite{Apostolatos:2009vu,Pan:2023wau}. In such cases, the azimuthal frequency is implicated in the resonance.}.
In the context of the resonance caused by \gls{gsf} effects, resonances involving both the radial and polar frequencies are particularly significant. 
When resonances occur, they give a small ``kick'' to the orbital elements. 
After this kick, a phase difference with respect to an adiabatic model (that does not include resonances) accumulates over a fixed frequency window as $\nu^{-1/2}$ (e.g., Refs.~\cite{Flanagan:2010cd,vandeMeent:2013sza,Berry:2016bit,speri2021assessing, Isoyama:2021jjd, Lynch:2024ohd}); resonances therefore enter between adiabatic and post-adiabatic order, and constitute the most significant correction to adiabatic inspirals in the multiscale framework.
In order to incorporate these effects, \gls{few} will need to detect the resonant surfaces and apply the appropriate precomputed jumps to the orbital elements.
The new version of trajectory module presented in this work provides access to a high-order interpolation of the phase space trajectory and its derivatives, which will make root-finding for these resonance surfaces inexpensive and straightforward to implement.

\textit{\gls{gw} memory.} In the comparable-mass regime, it was recently shown that including \gls{gw} memory effects can help to break degeneracies between parameters~\cite{Xu:2024ybt}, and that not including memory effects in waveform models can lead to biases in parameter estimation~\cite{Rossello-Sastre:2025gtq}.
Although unlikely to be important in the EMRI regime, for IMRIs it may be important to model \gls{gw} memory effects~\cite{Islam:2021old}. 
These effects have recently been computed in the small-mass-ratio limit for quasi-circular inspirals~\cite{Cunningham:2024dog,Elhashash:2024thm}. In general, including these effects in \gls{few} is a matter of interpolating over an additional set of amplitudes.

\textit{Post-adiabatic accuracy: including the second-order \gls{gsf}.} Current second-order \gls{gsf} calculations are carried out via a multiscale framework~\cite{Hinderer:2008dm,Miller:2020bft,pound2022black,Mathews:2025nyb}.
Within this approach, the trajectory through the parameter space is computed using an extended version of~\cref{eq:elements-ode} that takes the form
\begin{equation}\label{eq:trajectory_1PA}
    \frac{\mathrm{d}\alpha}{\mathrm{d}t} = \frac{\nu}{M} \left[ \hat{f}^{(0)}_\alpha(a, p, e, x_I) + \nu \hat{f}^{(1)}_\alpha(a, p, e, x_I) + \mathcal{O}(\nu^2)\right].
\end{equation}
The post-adiabatic forcing functions $\hat{f}^{(1)}_\alpha$ are very computationally expensive to compute, but they can be pre-computed and interpolated in the same manner as the adiabatic forcing functions.
Second-order \gls{gsf} calculations also compute the second-order waveform amplitudes.
It is expected that these are not necessary to include in EMRI waveforms~\cite{Burke:2023lno}, but they may be required when modelling low-$q$ IMRIs to avoid biases in parameter estimation.
In that case, the second-order amplitudes will be interpolated along the sparse trajectory during online waveform generation alongside the adiabatic amplitudes.
This means that online waveform generation of post-adiabatic waveforms will be almost as fast as adiabatic models. 
At present, the $\hat{f}^{(1)}_\alpha$ are only known for quasi-circular inspirals into a non-rotating black hole~\cite{Wardell:2021fyy}, and these results are ready to be incorporated into \gls{few} (as was initially investigated in Ref.~\cite{Burke:2023lno}).
It is an active area of research to compute $\hat{f}^{(1)}_\alpha$ for more complex orbital configurations~\cite{Spiers:2023cip,Leather:2023dzj,Osburn:2022bby,Wei:2025lva}.
\AMEND{
In addition to post-adiabatic contributions to the forcing functions, second-order corrections to the mode amplitudes may also be important for loud \glspl{imri} (but not for \glspl{emri}, see Ref.~\cite{Burke:2023lno} for discussion).
Provided that \gls{gw} fluxes are known to second order, the extension to second-order amplitudes is straightforward, being of the form
\begin{equation}
    \mathcal{A}_{\ell m kn} = \mathcal{A}_{\ell m kn}^{(0)} + \nu \mathcal{A}_{\ell m kn}^{(1)},
\end{equation}
where $\mathcal{A}_{\ell m kn}^{(0)}$ are the adiabatic amplitudes described in \cref{sec:amplitudes}.
Post-adiabatic amplitude corrections may also include contributions from the spin of the secondary object.
}
\textit{Post-adiabatic accuracy: including the spin on the secondary.} The effect of the spin on the secondary enters the waveform phase at post-adiabatic order.
In terms of waveform generation, a spinning secondary introduces another forcing term to \cref{eq:trajectory_1PA} and the orbital frequencies are also modified.
Both of these changes can readily be incorporated into \gls{few}.
At present, codes exist that can compute the necessary forcing terms for circular~\cite{Harms:2015ixa,Akcay:2019bvk,Piovano:2020zin, Mathews:2021rod, Albertini:2024rrs}, eccentric~\cite{Skoupy:2021asz} and generic orbits~\cite{Skoupy:2023lih,Drummond:2023wqc, Piovano:2024yks}.
\AMEND{
Last, while spin precession of the secondary object enters at second post-adiabatic order, it also contributes to mode phasing and amplitude evolution differently to other terms of this order~\cite{Mathews:2025nyb}; the \gls{few} framework provides the necessary tools for the importance of secondary spin precession to be examined in future work. 
}
With the exception of \gls{gw} memory and generic orbits, work is already underway on all of these extensions (and we note that \gls{few} currently can generate generic inspirals using a $5$\gls{pn} model \cite{Isoyama:2021jjd}).
In addition to the above, \gls{few} can also be extended to include environmental and beyond-GR physics.
So long as any additional physics can be modelled as a forcing function added to \cref{eq:elements-ode} or \cref{eq:trajectory_1PA} that do not depend on the orbital phase (or if such dependence can be transformed away~\cite{VanDeMeent:2018cgn,Lynch:2021ogr,Lynch:2024hco,Lynch:2024ohd}) it can readily be incorporated into our framework.
As the \gls{few} trajectory module now exists entirely in Python, if software for computing these modifications has already been developed it can be used directly to rapidly produce a new inspiral (and hence, waveform) model suitable for subsequent data analysis investigations.
See, e.g., Refs.~\cite{Copparoni:2025jhq} and \cite{Speri:2024qak} for first efforts in this direction.

Last, in addition to these avenues for the development of more accurate and generic models, there is also significant scope to further accelerate waveform generation with the \gls{few} framework.
As we identified in \cref{sec:timing}, $\sim 90\%$ of the computational cost of the waveform derives from trajectory integration and waveform summation; there is scope to accelerate these two components in future work.
The \gls{few} integrator has been constructed entirely of vectorised array operations in preparation for batched trajectory integration on either \gls{cpu} or \gls{gpu} (depending on the batch size).
While \gls{ode} integration is inherently serial, by computing derivatives for many trajectories in parallel the time-per-trajectory can be reduced significantly (potentially by orders of magnitude with \glspl{gpu} and batch sizes $\gtrsim 10^2$).
There are also multiple potential avenues towards improving the efficiency of the mode summation.
While \gls{few} does perform online mode selection, this operation is both expensive (requiring a full set of amplitudes to be computed) and sub-optimal (as can be verified with a slow, greedy-type minimisation of waveform mismatch).
Improving this procedure may yield reduced computational cost without compromising the results of inference.
Additionally, as is highlighted by the compact nature of each waveform mode in \cref{fig:tf-waveform}, time-frequency methods may significantly improve the mode summation for I/EMRI waveforms by more than an order of magnitude.
This also presents an opportunity to integrate the application of the \gls{lisa} response --- itself an expensive operation --- with this summation operation, directly computing \gls{tdi} variables and further reducing computational costs.
Development of these techniques for implementation in \gls{few} is currently ongoing~\cite{chapman-bird2025tf}.
Bringing together the myriad of potential improvements to be made to the efficiency of \gls{few}, per-waveform wall-times of $\sim 1\,\mathrm{ms}$ are a realistic possibility, the achievement of which would greatly impact the identification and analysis of \gls{imri} and \gls{emri} signals.

\pagebreak
\newpage

\begin{acknowledgments}
We thank Ryuichi Fujita and Adam Pound for interesting discussions.
We also thank Barry Wardell for support in setting up the \gls{few} data download server, and Patrick Bourg and Christopher Whittall for improving the \gls{few} tutorials.
We are also indebted to \gls{bhpc} (in particular Ryuichi Fujita, Hiroyuki Nakano, Norichika Sago and Masaru Shibata) for providing us with independent analytical and numerical data for the forcing functions and mode amplitudes used in~\cref{sec:validation} and~\cref{app:weakfieldappendix} for the validation, upon request.
We also thank the participants of the 2025 \gls{few} Hackathon, hosted by the University of Southampton and funded by A.~Pound's Royal Society University Research Fellowship, for supporting key code developments in \gls{few} and the production of this paper.

CEAC-B acknowledges support from UKSA grant UKRI971.
LS would like to acknowledge the support of the European Space Agency through %
ESA's postdoctoral Research Fellowship programme.
ZN acknowledges support from the ERC Consolidator/UKRI Frontier Research Grant GWModels (selected by the ERC and funded by UKRI [grant number EP/Y008251/1]).
OB acknowledges financial support from the Grant UKRI972 awarded via the UK Space Agency and computational resources from the French space agency CNES in the framework of LISA. 
SK acknowledges the computing resources accessed from NUS IT Research Computing group and the support of the NUS Research Scholarship.
JM and AJKC acknowledges support from the NUS Faculty of Science, under the research grant 22-5478-A0001.
HK acknowledges the support from the Perimeter Institute for Theoretical Physics. Research at Perimeter Institute is supported in part by the Government of Canada through the Department of Innovation, Science and Economic Development and by the Province of Ontario through the Ministry of Colleges and Universities. Part of the analysis was performed on the ``Symmetry'' HPC at Perimeter Institute, using NVIDIA H200 GPU devices.
JET and AJKC acknowledge support from the NASA LISA Preparatory Science grant 20-LPS20-0005.
SI and AJKC acknowledge support from the Ministry of Education, Singapore, under the Academic Research Fund Tier 1 A-800149200-00 (FY2023).
SAH acknowledges support from NSF Grant PHY-2110384 and PHY-2409644, as well as the use of MIT Kavli Institute resources at MIT's {\tt engaging} cluster, {\tt subMIT} resources of the MIT Department of Physics, and the Open Science Grid.
NW acknowledges support from a Royal Society - Research Ireland University Research Fellowship. This publication has emanated from research conducted with the financial support of Research Ireland under Grant numbers 16/RS-URF/3428, 17/RS-URF-RG/3490 and 22/RS-URF-R/3825.
This research was also done using services provided by the OSG Consortium \cite{osg07, osg09, OSPool, OSDF}, which is supported by the National Science Foundation awards \#2030508 and \#2323298.
\end{acknowledgments}

\section*{Data Availability}

The \gls{few} software is publicly available on GitHub~\cite{chapman_bird_2025_15630565}.
Scripts and associated intermediate data products required to reproduce the Figures in this work can be found in the accompanying data release, which can be found on Zenodo~\cite{chapman_bird_2025_15631641}.
The data products required to produce waveforms with \gls{few} are also preserved on Zenodo~\cite{chapman_bird_2025_15624459} (but will be automatically downloaded from the \gls{bhpt} servers when the code is executed).

\section*{Author Contributions}
%
%

CEAC-B: Conceptualization, Data curation, Formal analysis, Investigation, Methodology, Project administration,
Resources, Software, Supervision, Validation, Visualization, Writing --- original draft, Writing --- review \& editing.

LS: Conceptualization, Formal analysis, Investigation, Methodology, Supervision, Validation, Visualization, Writing --- original draft, Writing --- review \& editing.

ZN: Conceptualization, Data curation, Formal analysis, Investigation, Methodology, Resources, Software, Validation, Visualization, Writing --- original draft, Writing --- review \& editing 

OB: Formal analysis, Investigation, Methodology, Validation, Visualization, Writing --- original draft, Writing --- review \& editing.

MLK: Conceptualization, Investigation, Methodology, Resources, Software, Supervision, Writing --- review \& editing.

AS: Formal analysis, Investigation, Methodology, Visualization, Writing --- original draft.

SK: Formal analysis, Investigation, Validation, Visualization, Software, Writing --- original draft, Writing --- review \& editing.

PL: Investigation, Validation, Visualization, Software, Writing --- original draft, Writing --- review \& editing.

JM: Formal analysis, Investigation, Validation, Visualization, Writing --- original draft, Writing --- review \& editing.

HK: Investigation, Validation, Visualization, Writing --- original draft, Writing --- review \& editing.

JET: Validation, Visualization, Writing - reviewing \& editing.

SI: Data curation, Formal analysis, Investigation, Supervision, Validation, Writing --- original draft, Writing --- review \& editing.

SAH: Conceptualization, Data curation, Formal analysis, Investigation, Methodology, Validation, Writing --- review \& editing.

NW: Conceptualization, Writing --- original draft, Writing --- review \& editing.

AJKC: Conceptualization, Supervision, Writing --- review \& editing.

MP: Software. 

\appendix

\section{Mass conventions}
\label{app:mass-convention}

The previous implementation of \gls{few} scaled quantities with respect to the primary mass $m_1$ and the small mass ratio $\epsilon$. As aforementioned, in this work we choose a symmetric mass convention, which instead scales our model with respect to the total mass $M$ and the symmetric mass ratio $\nu$. To relate these different conventions, we first re-express our equations of motion so that they are independent of mass
\begin{subequations} \label{eqn:mass-independent-eoms}
    \begin{align}
        \frac{d\alpha}{d\hat{t}} &= \hat{f}_\alpha(a,p,e,x_I) + O(\epsilon),
        \\
        \frac{d\hat{\Phi}_A}{d\hat{t}} &= \hat{\Omega}_A(a,p,e,x_I) + O(\epsilon),
    \end{align}
\end{subequations}
where 
\begin{subequations}
    \begin{align}
        \hat{t} &= \epsilon t/m_1,
        &
        \hat{\Phi} &= \epsilon \Phi,
        \\
        \hat{\Omega} &= m_1 \Omega,
        &
        \hat{f}_\alpha &= m_1 {f}_\alpha.
    \end{align}
\end{subequations}
Additionally, we take $(p,e,x_I)$ to be dimensionless and define them by their implicit relationships to the orbital frequencies $\Omega(a,p,e,x_I)$. More concretely, $(p,e,x_I)$ is related to the orbital parameters in Ref.~\cite{Fujita:2009bp}, which we denote as $(p^\mathrm{F},e^\mathrm{F},\theta_\mathrm{inc}^\mathrm{F})$, by
\begin{align}
    p &= p^\mathrm{F},
    &
    e &= e^\mathrm{F},
    &
    x_I = \cos\theta_\mathrm{inc}^\mathrm{F},
\end{align}
while it is related to those in \cite{vandeMeent:2019cam}, which we denote as $(r_1^\mathrm{V},r_2^\mathrm{V},z_1^\mathrm{V})$, by
\begin{align}
    p &= \frac{2r_1^\mathrm{V}r_2^\mathrm{V}}{m_1(r_1^\mathrm{V}+r_2^\mathrm{V})},
    &
    e &= \frac{r_1^\mathrm{V}-r_2^\mathrm{V}}{r_1^\mathrm{V}+r_2^\mathrm{V}},
    &
    x_I &= \sqrt{1-(z_1^\mathrm{V})^2},
\end{align}
and, to those in \cite{Lynch:2024hco}, which we refer to as $(p^\mathrm{LB},e^\mathrm{LB},x_I^\mathrm{LB})$, by
\begin{align}
    p &= \frac{p^\mathrm{LB}}{m_1},
    &
    e &= e^\mathrm{LB},
    &
    x_I &= x_I^\mathrm{LB}.
\end{align}

If we now consider the limit $m_1 \gg m_2$, then $\epsilon = \nu + O(\nu)$ and $m_1 = M[1 + O(\nu)]$, giving us
\begin{subequations} \label{eqn:symmetric-expansion}
    \begin{align}
        \hat{t} &= \nu t/M + O(\nu^2),
        &
        \hat{\Phi} &= \nu \Phi + O(\nu),
        \\
        \hat{\Omega} &= M \Omega + O(\nu),
        &
        \hat{f}_\alpha &= M{f}_\alpha + O(\nu). 
    \end{align}
\end{subequations}
Inserting \cref{eqn:symmetric-expansion} into \cref{eqn:mass-independent-eoms} and dropping all higher-order terms, we arrive at a new set of equations of motion, given by \cref{eq:elements-ode,eq:phases-ode}. These equations are consistent with \cref{eqn:mass-independent-eoms} at leading adiabatic order, but introduce new higher-order post-adiabatic corrections.
Note that, based on \cref{eqn:mass-independent-eoms}, if we compute an adiabatic trajectory using our symmetric mass ratio convention, we can then relate it to an adiabatic trajectory produced via \cref{eqn:mass-independent-eoms} by rescaling time by $m_1/M \times \nu/\epsilon$ and the phase by $\nu / \epsilon$. Alternatively, one can simply replace $m_1$ and $m_2$ by $M$ and $\mu$ in the small mass ratio model to produce the same trajectory as \gls{few}.

Our mass convention also affects the amplitudes of the waveform, which we scale in terms of the reduced mass $\mu = m_2[1 + O(\nu)]$. In practice, this rescaling of the strain induces a change that is on the same order as the interpolation error in our amplitudes. Therefore, it has a negligible impact when comparing to the amplitudes of other adiabatic models. Ultimately, if one wants to compare other models to \gls{few}, but these models use the small mass ratio convention, then one needs to use the masses $M$ and $\mu$ as input for the small mass ratio model and $m_1$ and $m_2$ as input in \gls{few} to produce comparable waveforms.

\section{Data grids and coordinate parameterizations}
\label{app:grid_layout_appendix}

In our model, the simulation domain is divided into two regions: an inner region (\emph{Region A}) and an outer region (\emph{Region B}), following the approach in \cite{Bardwell:2025}. Region A is compact in the $p$-dimension and designed to densely sample amplitude and flux data near the separatrix, where these quantities exhibit rapid variation. To manage computational cost, we taper the maximum eccentricity as we approach the separatrix, thereby avoiding highly eccentric orbits near plunge, which are exponentially more expensive to simulate (see~\cref{app:flux-timing} and~\cref{fig:flux-timing}). This tapering is also astrophysically motivated: gravitational radiation tends to circularize orbits, so we expect eccentricities to decrease as the separatrix is approached. Region B spans a much broader range in $p$ and $e$, and is sampled according to the post-Newtonian (\gls{pn}) scaling behaviour of fluxes and amplitudes.
In the following subsections, we detail the coordinate systems and sampling strategies used to construct our flux and amplitude datasets.

\subsection{Fluxes}

For the fluxes, the inner grid in Region A has bounds
\begin{align}
    \delta p^A_\mathrm{min} &\leq p - p_\mathrm{sep}(a,e) \leq \delta p^A_\mathrm{max},
    \\
    0 &\leq e \leq S_\mathrm{ecc}(a,p,e; e_\mathrm{sep}, e_\mathrm{max}),
    \\
    a_\mathrm{min} &\leq a \leq a_\mathrm{max},
\end{align}
with $x_I = 1$ fixed and
\begin{align} \label{eq:regionA-bounds}
    \delta p^A_\mathrm{min} &= 0.001,
    &
    \delta p^A_\mathrm{max} &= 9 + \delta p^A_\mathrm{min},
    \\
    e_\mathrm{sep} &= 0.25,
    &
    e_\mathrm{max} &= 0.9,
    \\
    a_\mathrm{min} &= -0.999,
    &
    a_\mathrm{max} &= 0.999.
\end{align}
As aforementioned, the curve $S_\mathrm{ecc}(a,p,e)$ tapers the maximum value of the eccentricity from $e=0.9$ at larger $p$-values to $e=0.25$ as we approach the separatrix. To define $S_\mathrm{ecc}$, we first transform to a new coordinate system $(u,w,y,z)$, which---for generic orbits---is related to $(a,p,e,x_I)$ via
\begin{subequations} \label{eq:uwyz}
\begin{align} 
    u &= \left[\frac{\ln(p - p_\mathrm{sep}(a,e,x_I) + C_p)  - C_\Delta}{\ln 2} \right]^\alpha,
    \\ 
    w &= \frac{e}{S_\mathrm{gen}(a,p,e,x_I)},
    \\ 
    y &= \frac{x_I - x_\mathrm{min}}{1 - x_\mathrm{min}},
    \\ 
    z &= \frac{\hat{\chi}(a) - \hat{\chi}_\mathrm{min}}{\chi_\mathrm{max} - \chi_\mathrm{min}},
\end{align}  
\end{subequations}
where 
\begin{align*}
    \hat{\chi}(a)&=(1-a)^{1/3},
    &
    \hat{\chi}_\mathrm{min} &= \hat{\chi}(a_\mathrm{max}),
    &
    \hat{\chi}_\mathrm{max} &= \hat{\chi}(a_\mathrm{min}),
\end{align*}
and $S_\mathrm{gen}(a,p,e,x_I) = \tilde{S}[u(a,p,e,x_I), z(a)]$ with
\begin{align}
    \tilde{S}(u,z) &= e_\mathrm{sep} + (e_\mathrm{max} - e_\mathrm{sep})\sqrt{z + u^\beta(1 - z)},
\end{align}
providing the eccentricity tapering mentioned at the beginning of \cref{app:grid_layout_appendix}.
Furthermore, we have introduced the constants
\begin{align*}
    C_p &= \delta p^A_\mathrm{max} - 2\delta p^A_\mathrm{min},
    &
    C_\Delta &= \ln\left(\delta p^A_\mathrm{max} - \delta p^A_\mathrm{min} \right),
    \\
    \alpha &= 1/2,
    &
    \beta &= 2,
\end{align*}
and $x_\mathrm{min}$ is a free parameter that only has to be defined for inclined orbits. Note that the coordinates are defined so that $(u, w,y,z)$ all lie in the domain $[0,1]$.

\begin{figure}[t]
    \centering
    \includegraphics[width=0.95\linewidth]{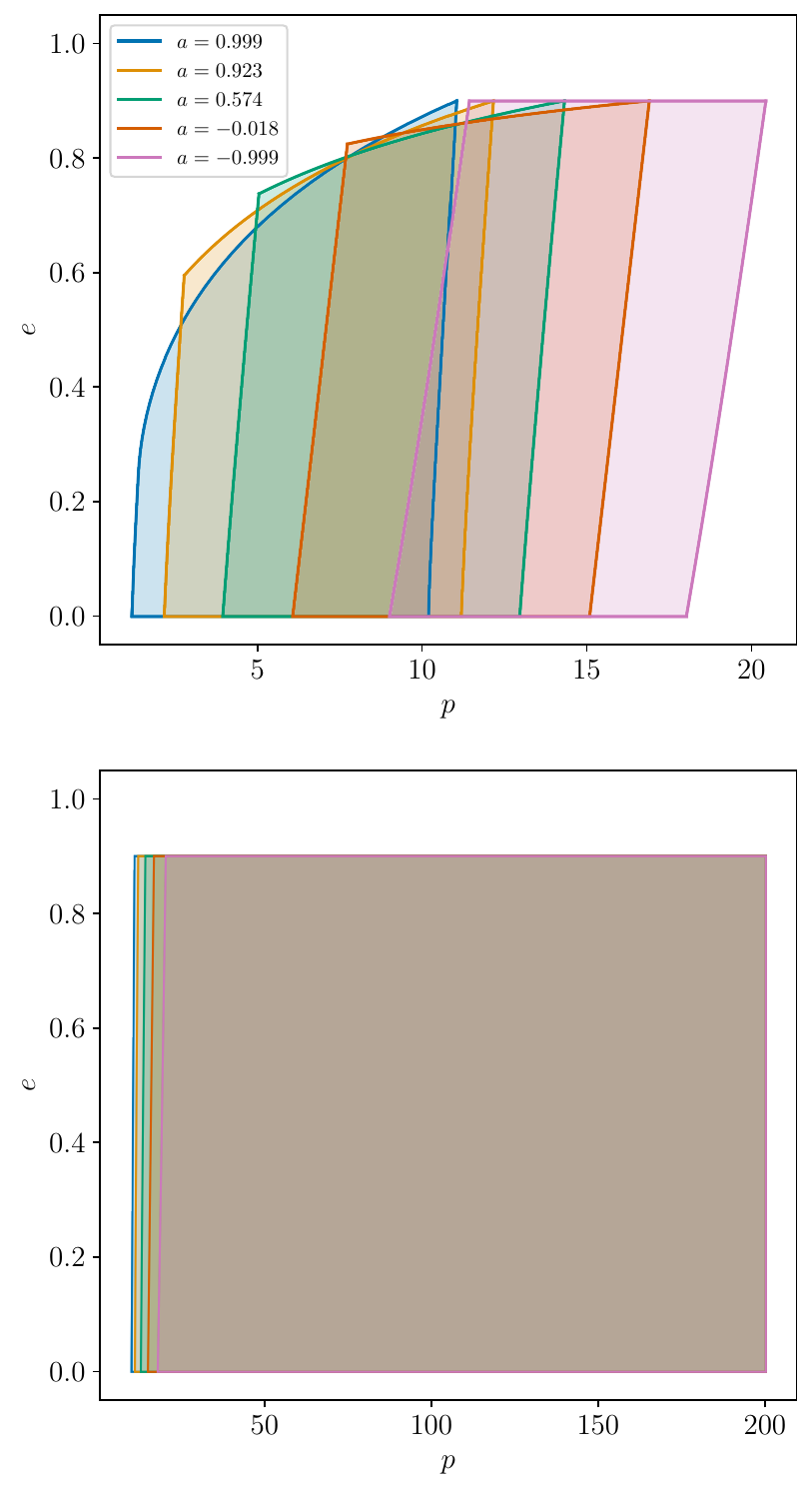}
    \caption{Visualization of the flux grid domains, plotted in the $(p, e)$ plane for various fixed spin parameters $a = \{0.999, 0.923, 0.574, -0.018, -0.999\}$. The top panel shows the inner grid (Region A), and the bottom panel shows the outer grid (Region B).}
    \label{fig:data-domains}
\end{figure}

For this work, we set $x_I = 1 = y$, and to handle retrograde orbits, we set $a<0$ rather than switching the sign of $x_I$.
Then we have $S_\mathrm{ecc}(a,p,e) = \tilde{S}_\mathrm{gen}(a,p,e,x_I=1)$.
We compute the fluxes on a three-dimensional rectilinear grid by uniformly sampling the coordinates $(u, w, z)$ over the unit cube (i.e., $[0, 1]^3$). The grid consists of $N_u \times N_w \times N_z$ points, with $(N_u, N_w, N_z) = (129, 65, 65)$.

The outer grid in {Region B} is defined by the bounds
\begin{align}
    \delta p^B_\mathrm{min} &\leq p - p_\mathrm{sep}(a,e,x_I) \leq \delta p^B_\mathrm{max},
    \\
    0 &\leq e \leq e_\mathrm{max},
    \\
    -a_\mathrm{max} &\leq a \leq a_\mathrm{max},
\end{align}
with $e_\mathrm{max}$, $a_\mathrm{max}$ being the same values used in Region A, and
\begin{align} \label{eq:regionB-bounds}
    \delta p^B_\mathrm{min} &= \delta p^A_\mathrm{max} - \delta p^A_\mathrm{min},
    \\
    \delta p^B_\mathrm{max} &= 200.
\end{align}
Like Region A, we construct a new coordinate system $(U, W, Y, Z)$, which is related to $(a,p,e,x_I)$ via
\begin{align}
    {U} &= \frac{ (\delta p^B_\mathrm{min})^{-1/2}  - \left[p-p_\mathrm{LSO}(a,e,x_I)\right]^{-1/2} }{(\delta p^B_\mathrm{min})^{-1/2}  - \left[p^B_\mathrm{max}-p_\mathrm{LSO}(a,e,x_I)\right]^{-1/2}},
    \\
    {W} &= \frac{e}{e_\mathrm{max}},
    \\
    {Y} &= \frac{x_I - x_\mathrm{min}}{1 - x_\mathrm{min}},
    \\
    {Z} &= \frac{\chi(a) - \chi_\mathrm{min}}{\chi_\mathrm{max} - \chi_\mathrm{min}}.
\end{align}
Once again, with $Y=1$ fixed, we produce a uniformly-spaced rectilinear grid with $(N_U, N_W, N_Z) = (65, 33, 33)$ points in $(u,w,z)$.

\subsection{Amplitude grid}
For the amplitudes, we use the same bounds for Region A as we did for the fluxes (which are given in \cref{eq:regionA-bounds}.
To construct the inner amplitude grid, we also use the same coordinates as the fluxes, $(u,w,y,z)$ in \cref{eq:uwyz}, but with the modified constants
\begin{align}
    \alpha & = 1/3, & \beta &= 3.
\end{align}
As before, we sample amplitudes on a uniformly-spaced rectilinear grid with $(N_u, N_w, N_z) = (33, 33, 33)$ points in $(u,w,z)$.

When constructing amplitudes in the outer domain, we use slightly different bounds for Region B as compared to the fluxes, 
\begin{align*}
    \delta p^B_\mathrm{min} &= \delta p^A_\mathrm{max} - 0.001,
    \\
    \delta p^B_\mathrm{max} &= 200 + \delta p^B_\mathrm{min}.
\end{align*}
We also use modified grid coordinates
\begin{align}
    \hat{U} &= \frac{\hat{U}_{p,\mathrm{max}}(a,e,x_I) - \hat{U}_p(p)}{\hat{U}_{p,\mathrm{max}}(a,e,x_I) - \hat{U}_{p,\mathrm{min}}(a,e,x_I)},
    \\
    \hat{W} &= \frac{e}{e_\mathrm{max}},
    \\
    \hat{Y} &= y(x_I),
    \\
    \hat{Z} &= z(a),
\end{align}
with $\hat{U}_p(p) = p^{-1/2}$, $\hat{U}_{p,\mathrm{max}}(a,e,x_I) = \hat{U}_p[\delta p^B_\mathrm{min} + p_\mathrm{sep}(a,e,x_I)]$, $\hat{U}_{p,\mathrm{min}}(a,e,x_I) = \hat{U}_p[\delta p^B_\mathrm{max} + p_{\mathrm{sep}}(a,e,x_I)]$. 
Note that $(\hat{W}, \hat{Y}, \hat{Z})$ are identical to the outer grid coordinates $(W, Y, Z)$ used for the fluxes. 
To reduce the memory footprint of our interpolant by $37.5\%$, we reduce the sampling density with respect to Region A, such that $(N_{\hat{U}}, N_{\hat{W}}, N_{\hat{Z}}) = (17, 17, 33)$ points in $(\hat{U}, \hat{W}, \hat{Z})$ --- the density with respect to $Z$ remains fixed for the purposes of vectorised linear interpolation.

There are minor differences between the flux and amplitude grids, stemming from the fact that the data were generated in two stages: amplitudes first, followed by fluxes. In the initial amplitude grid, we used $\alpha = 1/3$, which concentrated sampling density in Region A near the separatrix. This was necessary due to the sparse nature of the amplitude data in that region. However, when applying this same sampling strategy to the fluxes, which were sampled more densely overall, it led to oversampling near the separatrix with $\Delta p \lesssim 10^{-5}$ for the first several points in $u$. We found that increasing $\alpha$ to 1, while keeping the product $\alpha \beta = 1$ fixed, provided a more balanced sampling of the fluxes across all of Region A. This adjustment preserved the same eccentricity tapering used for the amplitude grid, ensuring consistency between the two datasets.

Furthermore, we set $\delta p^B_\mathrm{min} = \delta p^A_\mathrm{max}$ 
for the fluxes to create a small region of overlap between A and B, which we used for preliminary cross-validation of interpolated flux values in the shared domain. We also found that using ${U}$ for the fluxes, instead of $\hat{U}$, and truncating Region B at $\delta p^B_\mathrm{min}$ and $\delta p^B_\mathrm{max}$ in \eqref{eq:regionB-bounds} provided coordinates that much more closely align with the \gls{pn} behaviour of the fluxes at large $p$. This change resulted in a modest but noticeable improvement in flux interpolation in the outer region.

\subsection{Solving for edges of domain}

When evaluating interpolated functions of our data grids, it is useful to identify the edges of our domain. If we truncate the evaluation of our amplitude grid at the same maximum value of $p$ as the fluxes, then we have
\begin{align}
    p_\mathrm{lower}(a,e) &= p_\mathrm{sep}(a,e,x_I=1) + \delta p^A_\mathrm{min},
    \\
    p_\mathrm{upper}(a,e) &= 200,
    \\
    a_\mathrm{lower}(p,e) &= a_\mathrm{min},
    \\
    a_\mathrm{upper}(p,e) &= a_\mathrm{max},
    \\
    e_\mathrm{lower}(a,p) &= 0,
\end{align}
where the ``lower'' and ``upper'' subscripts refer to the minimum and maximum values of $a$, $p$, or $e$, respectively, on our grid when the two other orbital parameters are fixed.

The process for finding maximum or ``upper'' $e$ given a constant $(a,p)$ is slightly more involved due to the tapered eccentricity bound used in Region A. To identify $e_\mathrm{max}(a,p)$, we use the following method:
\begin{enumerate}
    \item if $p \geq p_\mathrm{sep} + \delta p^B_\mathrm{min}$, then $e_\mathrm{upper} = 0.9$, otherwise
    \item we solve for the root $e^0_\mathrm{lower}$ that satisfies \begin{align*}w(a,p,e^0_\mathrm{lower})-1 &= 0
    \\ &= e^0_\mathrm{lower}/\tilde{S}[u(a,p,e^0_\mathrm{lower},1),z(a)],\end{align*}
    in the interval $e^0_\mathrm{lower}\in (0, e_\mathrm{max})$;
    \item if $p_\mathrm{sep}(a,e^0_\mathrm{lower},1) < p $, then $e_\mathrm{lower} = e^0_\mathrm{lower}$;
    \item else, this is a false root, and $e_\mathrm{lower}$ is instead the root of the equation $p-p_\mathrm{sep}(a,e_\mathrm{lower},1)=0$ in the interval $e_\mathrm{lower} \in (0, e^0_\mathrm{max})$.
\end{enumerate}

\subsection{Timing of flux calculations}
\label{app:flux-timing}

\begin{figure}
    \centering
    \includegraphics[width=0.88\linewidth]{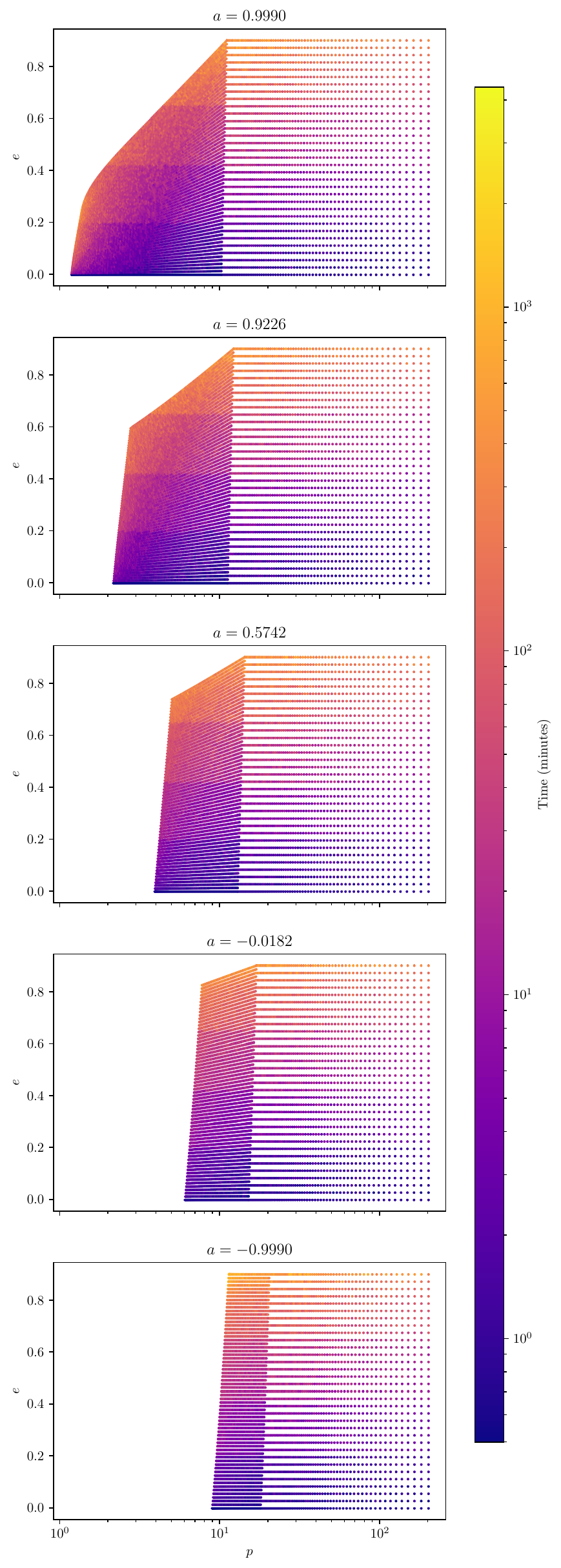}
    \caption{The number of CPU minutes required to calculate the fluxes at each grid point for the same slices of $a$ visualized in~\cref{fig:data-domains}.
    }
    \label{fig:flux-timing}
\end{figure}

Based on the grids and sampling rates described above, we compute the fluxes at $615,810$ points in the $(a,p,e)$ parameter space. We perform these calculations on the Open Science Grid (OSG), a free open-access high throughput computing infrastructure \cite{osg07, osg09, OSPool, OSDF}. In~\cref{fig:flux-timing} we plot the computational cost of each flux calculation for constant values of $a$. The computational cost is most significantly impacted by the value of the eccentricity and grows exponentially with $e$. However, this growth is not smooth due to the nature of the flux calculations performed by \textsc{pybhpt}. As the eccentricity grows, \textsc{pybhpt} needs to increase the numerical resolution of various computations, which it does by increasing the sampling rate by factors of two. Thus, at certain regions in parameter space, the calculation effectively doubles in computational cost because the calculation is hard coded to run at double the numerical resolution to meet precision requirements. This leads to the sharp horizontal bands in~\cref{fig:flux-timing}. Furthermore, particularly for large prograde spins, we observe a rapid growth in computational cost as we approach the separatrix. All together, the flux grid took just under 605,000 CPU hours to calculate.

\section{Setup for Monte-Carlo studies of waveform characteristics}
\label{app:monte-carlo-parameters}

At multiple points throughout this work, we examine the accuracy and performance of our waveform model by performing Monte-Carlo draws of source parameters.
For consistency, we randomly sample these parameters from the same distributions in each case; these are summarized in \cref{tab:monte-carlo-param-ranges}.
Given a set of these drawn parameters, $p_0$ is chosen according to the desired time-to-plunge of the inspiral in question (typically either two or four years).
For parameter sets where no such point can be found in our domain of validity (e.g., highly eccentric systems with $\epsilon \lesssim 10^{-6}$), 
a new point is drawn to replace it.
This rejection procedure yields the more complicated distribution that appears in the top panel of \cref{fig:downsample-mismatch-scatters}.

For waveform-level comparisons, angular parameters are distributed in the typical manner: azimuthal and polar angles are distributed uniformly on the surface of the unit sphere, and phase parameters are uniformly distributed in the range $[0, 2\pi)$.
As none of our Monte-Carlo analyses entail random sampling of $d_\mathrm{L}$, we set this parameter to a fiducial value of $1\,\mathrm{Gpc}$ in all cases.

\begin{table*}
    \centering
    \renewcommand{\arraystretch}{1.1}
    \setlength{\tabcolsep}{8pt}
    \begin{tabular}{c c c c c}
        \hline\hline
        Parameter & Unit & Min & Max & Distribution \\
        \hline
        $m_1$ & $M_\odot$ & $10^5$ & $10^7$ & Uniform in $\log m_1$ \\
        $m_2$ & $M_\odot$ & $10^5$ & $10^7$ & Uniform in $\log m_2$ \\
        $a$ & - & $0$ & $0.999$ & Uniform \\
        $e_0$ & - & $0$ & $0.9$ & Uniform \\
        \hline\hline
    \end{tabular}
    \caption{Distributions used for the Monte-Carlo studies of waveform characteristics performed in this work.
    Given a set of these drawn parameters, $p_0$ is chosen according to the desired time-to-plunge of the inspiral in question; see text of \cref{app:monte-carlo-parameters} for discussion.
    }
    \label{tab:monte-carlo-param-ranges}
\end{table*}

Notably, \cref{tab:monte-carlo-param-ranges} only covers prograde inspirals.
The retrograde half of the parameter space is significantly less challenging to model accurately at all stages of waveform generation.
We therefore exclude this region from many of our Monte-Carlo studies and focus primarily on prograde inspirals.
None of the conclusions presented in this work change significantly when considering the retrograde case (to the contrary, the median performance of our model over the parameter space improves with such an extension).

\section{GW data analysis framework}
\label{app:data_analysis_fundamentals}

The strain data-stream observed by \gls{lisa} will be a superposition of multiple \gls{gw} signals of various source types that overlap in both time and frequency \cite{Speri:2022kaq}.
These signals will be buried in a mixture of both instrumental noise and astrophysical confusion noise (particularly the Galatic foreground~\cite{LISA:2022yao}). 
For simplicity, we assume that the data stream $d(t)$ contains a single \gls{emri} signal in additive noise, such that
\begin{equation}\label{eq:data_stream}
d(t) = h(t;\boldsymbol{\theta}_\mathrm{tr}) + n(t)\,,
\end{equation}
where $h(t;\boldsymbol{\theta}_\mathrm{tr})$ is the \gls{emri} signal described by the parameters $\boldsymbol{\theta} = \boldsymbol{\theta}_\mathrm{tr}$ (which we wish to infer in data analysis). 
This assumption is well-motivated by the negligible overlap between \glspl{emri} and other sources in the data stream~\cite{chua2022nonlocal}.

The probabilistic noise process $n(t)$ determines the form of the likelihood function used in the inference of $\boldsymbol{\theta}$. 
In our analysis, we assume that $n(t)$ is Gaussian, (weakly) stationary and circulant. 
While the cyclostationarity of the Galactic foreground signal invalidates this assumption~\cite{Pozzoli:2024wfe,Digman:2022jmp}, including this feature in our noise model will significantly increase the computational cost of our analyses without greatly affecting any of our conclusions, so we do not consider this noise feature in this work.
Our assumptions regarding the noise process leads to a diagonal noise-covariance matrix in the Fourier domain~\cite{whittle:1957},
\begin{equation}\label{eq:diagonal_noise_cov_FD}
\Sigma(f,f') = \langle \tilde{n}(f)\tilde{n}^{\star}(f')\rangle = \frac{1}{2}\delta(f - f')S_{n}(f')\,\text{d}f'\,.
\end{equation}
with Fourier-domain convention 
\begin{equation}
\tilde{a}(f) = \int_{-\infty}^{\infty}a(t)\exp(-2\pi i f t)\,\text{d}t\,.
\end{equation}
The \gls{psd} $S_{n}(f)$ describes the average mean-square fluctuations of $\tilde{n}(f)$ as a function of $f$. 
The diagonal noise covariance matrix defined in \cref{eq:diagonal_noise_cov_FD} gives rise to the Whittle likelihood function that is ubiquitous in the field of \gls{gw} data analysis.
The log of this likelihood function is
\begin{equation}\label{eq:Whittle_likelihood}
\log\mathcal{L}(d \mid \boldsymbol{\theta}) \propto -\frac{1}{2}(d - h_{\text{m}} \mid d - h_{\text{m}})\,,
\end{equation}
with the noise-weighted inner product
\begin{equation}\label{eq:Inner_Product}
(a\mid b) = 4\,\text{Re}\int_{0}^{\infty}\frac{\tilde{a}^{\star}(f)\tilde{b}(f)}{S_{n}(f)}\,\text{d}f\, ,
\end{equation}
where $\star$ denotes complex conjugation.
The quantity $h_m (t;\boldsymbol{\theta}) \equiv h_{m}$ represents a template from the waveform model used to perform inference on the signal $h$ in the data stream. 
Deviations between models for the signal and template waveforms yield systematic biases: for $h_{\text{m}}(t;\boldsymbol{\theta}) \neq h(t;\boldsymbol{\theta})$, a parameter estimation analysis would recover biased parameters $\boldsymbol{\theta}_{\text{bf}}$. 
This will be discussed later. 
The difference between two waveforms $h_1$ and $h_2$ can be expressed in terms of the normalised overlap and mismatch; respectively,
\begin{align}
    O(h_1, h_2) &= \frac{(h_1\mid h_2)}{\sqrt{(h_1 \mid h_1) (h_2 \mid h_2)}}\,, \\
    \mathcal{M}(h_1,h_2) &= 1 - O(h_{1},h_2)\,.\label{eq:mismatch}
\end{align}
The mismatch $\mathcal{M}$ is equal to zero for $h_1 \propto h_2$ and equal to one if the two waveforms $h_1$ and $h_2$ are orthogonal. We will use~\cref{eq:mismatch} extensively in \cref{sec:waveform-validation} as a figure of merit for the faithfulness of our waveform model.

A useful quantity that determines the strength of an \gls{emri} signal $h$ with respect to the noise is the optimal matched-filter \gls{snr} 
\begin{equation}\label{eq:SNR_eq}
\rho = \sqrt{(h\mid h)}\,.
\end{equation}
\Cref{eq:SNR_eq} is optimal in the sense that it is the mean \gls{snr} (averaging over all noise realisations) under the assumption that the template waveform model is correct.
Statistical biases in measured parameters due to noise fluctuations scale inversely with $\rho$, whereas systematic biases due to modelling errors do not; considering the range of expected \glspl{snr} for typical \glspl{emri} is therefore an important facet of our analysis of waveform systematics in \cref{sec:waveform-validation}.

Data analysis in GW astronomy is typically performed with Bayesian inference techniques, at the heart of which lies Bayes' theorem:
\begin{equation}
\label{eq:bayes}
p(\boldsymbol{\theta} \mid d) = \frac{\pi(\boldsymbol{\theta}) \mathcal{L}(\boldsymbol{\theta} \mid d)}{\mathcal{Z}(d)}\,,
\end{equation}
where $\pi(\boldsymbol{\theta})$ is the prior probability, and the normalising constant $\mathcal{Z}$ is usually denoted the evidence (and is independent of $\boldsymbol{\theta}$).
The posterior distribution $p(\boldsymbol{\theta} \mid d)$ encodes all information regarding the values of $\boldsymbol{\theta}$ that are best-supported by the observational data, given our prior expectations; insight regarding the form of $p(\boldsymbol{\theta} \mid d)$ is attained via parameter estimation.
In \cref{sec:EMRI_science_cases}, we perform \gls{mcmc} analyses (as implemented by the \textsc{eryn} package~\cite{Karnesis:2023ras,michael_katz_2023_7705496,Foreman-Mackey:2013}, with default settings) to sample \gls{emri} posterior distributions, from the results of which we compute parameter variances as summary statistics.
We assume wide uniform priors on all sampled parameters (such that they do not truncate the posterior bulk).
Initial \gls{mcmc} walker positions are chosen to be near the injected \gls{emri} parameters, as our goal is not to perform search, only to probe the form of the posterior bulk. 
To incorporate the response function of \gls{lisa} in our analyses, we use the \textsc{fastlisaresponse} package~\cite{katz2022assessing} to generate second-generation $\{A, E, T\}$ \gls{tdi} variables~\cite{Prince:2002hp} from our \gls{emri} waveforms, assuming spacecraft orbits with constant and equal arm-lengths.
Our model for $S_n(f)$ is the \textsc{SciRDv1} sensitivity curve~\cite{babak2021lisa} for this \gls{tdi} variable convention, including the effects of the Galactic confusion noise~\cite{PhysRevD.104.043019} as a stationary noise process~\cite{robson2019construction}.

Despite the computational efficiency of our model on \glspl{gpu}, sampling \cref{eq:bayes} still typically takes $\sim 12\,\mathrm{hr}$ per \gls{emri} source.
This is prohibitively expensive for more exhaustive studies of parametric biases due to waveform model differences, such as those performed in \cref{sec:higher-mode-importance}, in which a signal with parameters $\boldsymbol{\theta}_{\rm tr}$ is injected with one model and analysed with another, returning (potentially biased) best-fit parameters $\boldsymbol{\theta}_{\rm bf}$.
To facilitate such an analysis, we estimate these biases via the linear signal approximation given by Cutler and Vallisneri~\cite{Cutler:2007mi},
\begin{align}
\Delta\boldsymbol{\theta} \approx \left(\Gamma^{-1}(\boldsymbol{\theta}_{\rm tr})\right)^{ij}\left(\left. \partial_jh_{\rm ap}(\boldsymbol{\theta}_{\rm tr})\right|h_{\rm tr}(\boldsymbol{\theta}_{\rm tr}) - h_{\rm ap}(\boldsymbol{\theta}_{\rm tr})\right)\,, \label{eq:CVbias}
\end{align}
where $\Delta\boldsymbol{\theta} := \boldsymbol{\theta}_{\rm bf} - \boldsymbol{\theta}_{\rm tr}$ and $\partial_jh := \partial h/\partial\theta_j$, and the $\approx$ indicates we have ignored higher-order terms. 
The information matrix $\Gamma$ has the elements
\begin{align}\label{eq_app:FM}
    \Gamma_{ij} := \left(\left.\partial_ih_{\rm ap}\right|\partial_jh_{\rm ap}\right)
\end{align}
evaluated in \cref{eq:CVbias} at $\boldsymbol{\theta}_{\rm tr}$.

Parameters estimated from noisy observations will be modified by a probabilistic shift away from the true parameter values.
In a similar vein to \cref{eq:CVbias}, at leading order this shift is approximately
\begin{equation}\label{eq:noise_bias_CV}
\Delta \boldsymbol{\theta}_{\text{noise}}   \approx \left(\Gamma^{-1}(\boldsymbol{\theta}_{\rm tr})\right)^{ij}\left(\left. \partial_jh_{\rm ap}(\boldsymbol{\theta}_{\rm tr})\right|n_\mathrm{nr}\right)\,,
\end{equation}
where $n_\mathrm{nr}$ is a specific noise realisation. 
Since $n_\mathrm{nr}$ has zero mean, $\Delta \boldsymbol{\theta}_{\rm noise}$ is unbiased (i.e., $\mathbb{E}[\Delta\boldsymbol{\theta}_{\rm noise}] = 0$). 
For all of the parameter estimation analyses we perform in this work --- both in the form of Cutler-Vallisneri bias estimates (\cref{sec:higher-mode-importance}) and \gls{mcmc} explorations of posterior distributions (\cref{sec:inference-subsection}) --- we assume a ``zero-noise'' realisation, such that $n_\mathrm{nr}$ is the zero-vector.
We make this choice to eliminate the statistical parameter shifts due to noise, leaving only the systematic shifts due to modelling errors (the characterisation of which is the goal of these analyses).
Our results should therefore be interpreted as the expectation over all noise realisations.
A pertinent question is why the posteriors recovered in \cref{sec:inference-subsection} have non-zero widths, despite the apparent absence of noise. It is important to emphasise that the ``zero-noise'' realisation is a perfectly valid draw from the distribution of all possible noise realisations, and that we should therefore expect our parameter estimates to be uncertain for zero noise according to our noise model (given by the \gls{psd}), just as we would expect to occur for any other realisation.

The linear signal approximation is typically valid for \glspl{emri} with detectable \glspl{snr} $\rho \gtrsim 20$ (which gives rise to the Gaussian posterior distributions obtained in \cref{sec:inference-subsection}).
We use the \textsc{StableEMRIFisher (SEF)} package~\cite{kejriwal_2024_sef} to obtain robust information matrices given a set of \gls{emri} parameters. 
Once the best-fit parameter point is obtained, we quantify its bias with respect to $\boldsymbol{\theta}_{\rm tr}$ by its corresponding sigma contour level. 
It is given by the Mahalanobis distance $D_{\rm Maha}$~\cite{mahalanobis}, i.e. the multivariate generalization of the z-score, and is defined as
\begin{align}
    D_{\rm Maha}^2 := \frac{\Delta\boldsymbol{\theta}^T\cdot\Gamma\cdot\Delta\boldsymbol{\theta}}{D}.\label{eq:Mahalanobis}
\end{align}
We scale by $1/D$ here compared to the original definition since the unscaled Mahalanobis distance increases with dimensionality, which can lead to deceptively large sigma-contour levels. 
Consider, e.g., $\Gamma$ to be a $D\times D$ identity matrix and $\Delta\boldsymbol{\theta}$ a length $D$ vector of $1$'s, representing a $1\,\sigma$-biased point from the peak of a standard normal in $D$-dimensions. 
The numerator of \cref{eq:Mahalanobis} is $D$ in this case, justifying the scaling by $1/D$ to recover $D_{\rm Maha} = 1$. 
Finally, note that parameter space correlations also impact the Mahalanobis distance through $\Gamma$.

\section{Comparison against other implementations in limiting cases}
\label{app:model-comparison}

In addition to the self-contained examination of our \gls{few} v2 carried out in~\cref{sec:validation}, we can also verify that it correctly matches other models in limiting cases. 
Specifically, we compare against accurate models for quasi-circular inspirals into spinning \glspl{mbh}~\cite{Khalvati:2024tzz,Nasipak:2023kuf} (\cref{app:quasicircular}), and verify that elements of our model approach \gls{pn}-\gls{gsf} based results in the weak-field limit ($p \gg p_\mathrm{sep}$) with moderate eccentricities (\cref{app:weakfieldappendix}).

The results of the comparisons conducted in this Appendix, when combined with the systematic validation performed throughout the main body of text, serve to confirm that our model produces accurate waveforms for spinning and/or equatorial \gls{imri} and \gls{emri} across its domain of validity.

\subsection{Asymmetric-mass, quasi-circular inspirals into rapidly-spinning black holes}
\label{app:quasicircular}
In order to compare the model developed in this work (\gls{few} v2) against other implementations that are expected to be faithful in the strong-field regime, in this appendix we benchmark its performance in the non-eccentric limit against two existing Kerr quasi-circular equatorial waveform models that are similarly accurate to adiabatic order. 
The first reference model \textsc{KerrCirc}, detailed in \cite{Khalvati:2024tzz}, is an independent extension to earlier versions of \gls{few} but using flux and amplitude data generated by a distinct Teukolsky solver. 
The second reference model is \textsc{BHPWave} \cite{Nasipak:2023kuf}, which provides waveforms from a completely independent generative framework. 
Our comparison focuses on evaluating the orbital phase evolution by comparing trajectories, examining interpolated waveform amplitudes, and analyzing the mismatch between the full waveforms as a function of the primary black hole's spin parameter.

\begin{figure}[t]
    \centering
    \includegraphics[width=0.98\linewidth]{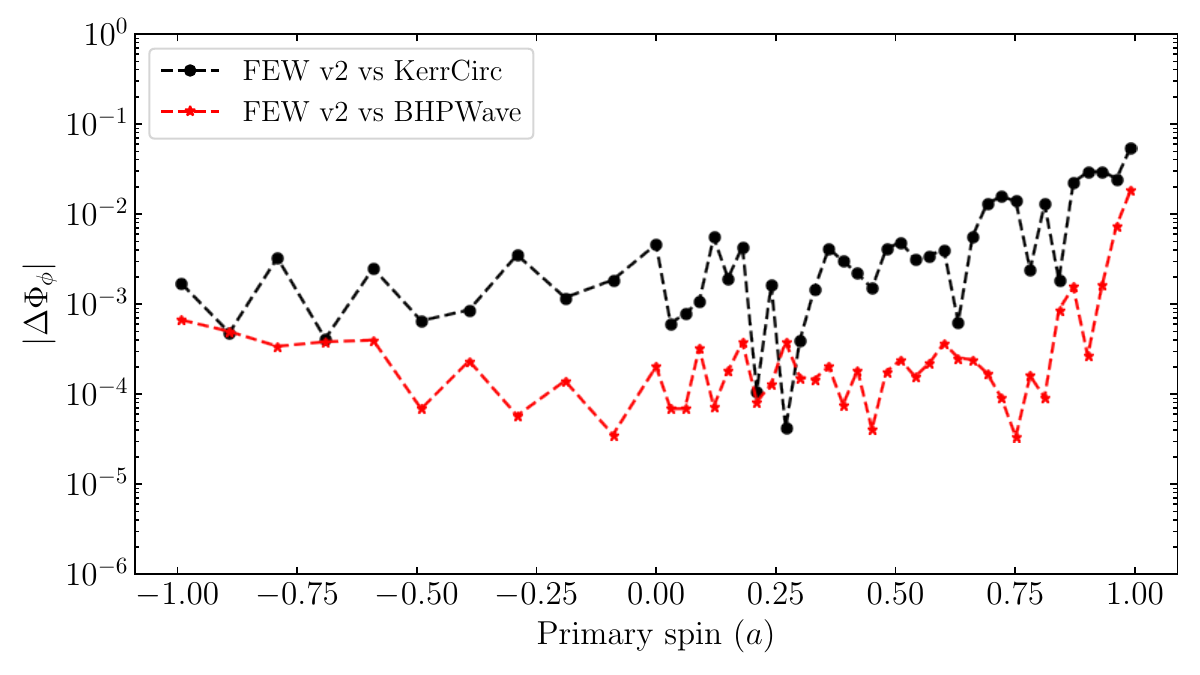}
    \caption{ Absolute value of the final orbital phase shift between our model and two other models, \textsc{KerrCirc} (black dashed line) and \textsc{BHPWave} (red dashed line), in the circular orbit limit for different values of primary spin, $a \in [-0.99, +0.99]$. The phase shifts are for four-year inspirals and the masses are fixed to $(m_1, m_2) = (10^6, 10)\,M_\odot$.}
\label{fig:phase_shift_model_comparison}
\end{figure}

Before computing the phase shifts, we rescaled the masses as discussed in \cref{app:mass-convention} to have a similar mass convention across all three models. Moreover, since the overall scaling of the phases, and phase errors are known with the mass ratio and with the primary mass, we only report the comparison as a function of the primary spin. We chose masses of $(m_1, m_2) = (10^6, 10)M_\odot$. Throughout this section, for each value of the primary spin parameter $a$, the inspiral is started at a corresponding $p_0$ such that the evolution lasts four years before reaching the plunge.

The absolute value of the final phase shift as a function of primary spin compared to both models is shown in \cref{fig:phase_shift_model_comparison}. The phase shift $\Delta\Phi_{\phi} \lesssim 1$ for all spin values; we observe some oscillations alongside the overall increase for higher spins in prograde orbits. The primary source of these phase shifts lies in differences in flux and interpolation accuracy. These discrepancies arise from several factors, such as the accuracy of the flux data, choice of parameterization, the scaling applied to the flux data prior to interpolation, the type of interpolant used, and the interpolation grid itself, specifically, the density of said grid. 

The \textsc{KerrCirc} model employs a uniform grid in spin with $\Delta a = 0.01$, whereas \textsc{BHPWave} uses a non-uniform grid with a higher density of points concentrated at larger spin values. The latter setup is more similar to ours, in which the grid resolution is denser at higher spin and sparser at lower spin. This grid structure likely explains why \textsc{BHPWave} shows better agreement with our model.  A detailed study of grid-induced and interpolation-induced errors as a function of spin is 
\AMEND{reported in ~\cite{Khalvati:2025znb}}.

\begin{figure}[b]
    \centering
    \includegraphics[width=0.98\linewidth]{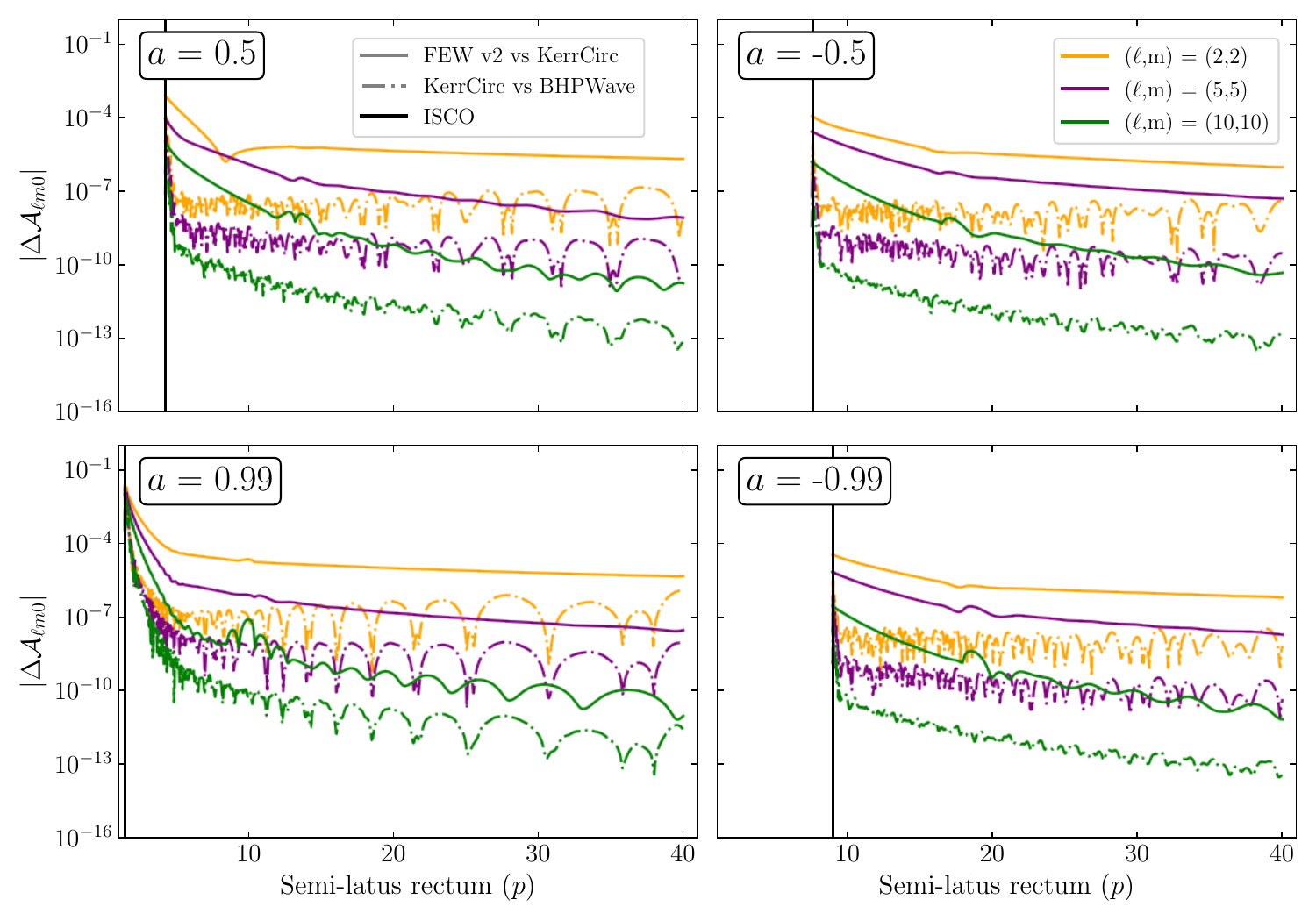}
    \caption{Comparison of the real part of the mode amplitude differences, $|\Delta \mathcal{A}_{\ell m 0}|$, as a function of orbital separation $p$. Each panel corresponds to a fixed primary spin value, $a \in {0.99, 0.5, -0.5, -0.99}$, and shows three different modes, $(\ell, m) \in {(2,2), (5,5), (10,10)}$. The solid lines compare our model to \textsc{KerrCirc}, while the dashed lines compare \textsc{KerrCirc} to \textsc{BHPWave}. The innermost stable circular orbit is indicated by a vertical marker.
    }
    \label{fig:amplitude_comparison}
\end{figure}

Furthermore, we compared the interpolated amplitudes in our model for four different primary spin values, $a \in \{0.99, 0.5, -0.5, -0.99\}$ and three different modes, $(\ell, m) \in \{(2,2), (5,5), (10,10)\}$ as a function of orbital separation, $p$, with both \textsc{KerrCirc}, and \textsc{BHPWave} in \cref{fig:amplitude_comparison}. We dropped the index $n$ here because in the circular orbit limit, there are no radial harmonics and $n = 0$. We plot $|\Delta \mathcal{A}_{\ell m 0}|$ instead of the fractional error, as the latter can appear misleadingly large for higher modes due to their smaller amplitudes. We observe that the mode amplitudes agree more closely between \textsc{KerrCirc} and \textsc{BHPWave} than with our model. This is primarily because we do not employ a very dense grid in $p$ for computing the amplitudes (\cref{app:grid_layout_appendix}). We remind the reader that amplitude errors do not accumulate over the inspiral and therefore do not lead to significant deviations in the total mismatch between waveforms (this is observed during amplitude validation in \cref{sec:amplitude-validation}).

\begin{figure}
    \centering
    \includegraphics[width=0.98\linewidth]{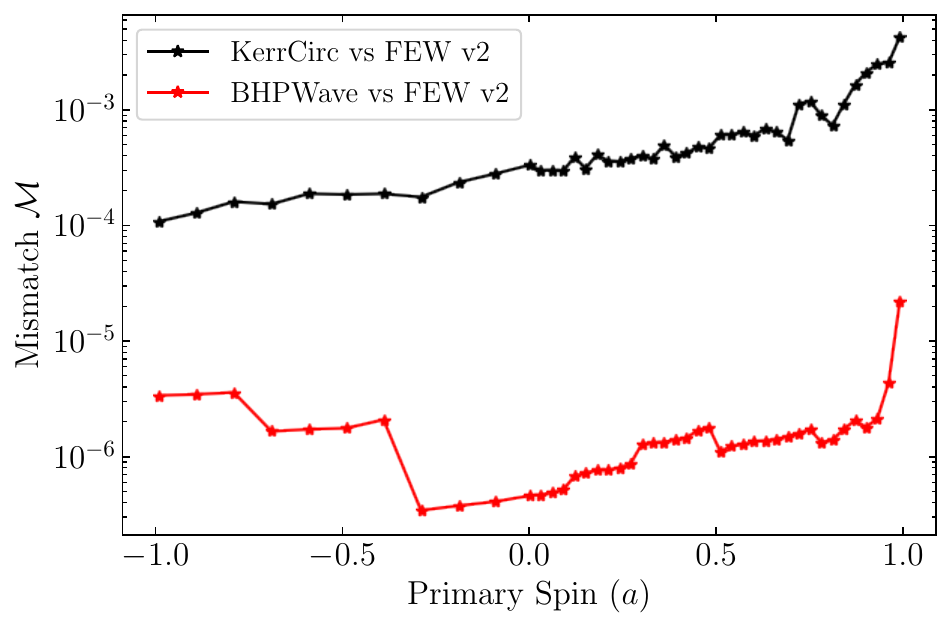}
    \caption{
    Mismatch between our waveform and \textsc{KerrCirc} (solid line) or \textsc{BHPWave} (dashed line) as a function of the primary spin $a$, in the circular orbit limit. Mismatches were computed using a flat \gls{psd}, with a four-year observation time before plunge (i.e. reaching the innermost stable circular orbit). The source parameters were fixed at $(M, \mu) = (10^6, 10),M_\odot$, and an arbitrary viewing configuration with $(\theta, \phi) = (\pi/5, \pi/3)$ 
    }
    \label{fig:mismatch_across_models}
\end{figure}

Finally, to provide an overall comparison of the full waveforms, we computed the mismatch (see \cref{app:data_analysis_fundamentals} for definitions) between our waveform and both \textsc{KerrCirc} and \textsc{BHPWave} as a function of the primary spin for $a \in [-0.99,0.99]$ in the circular orbit limit. 
The mismatches were calculated using a flat \gls{psd}, such that differences between models are captured uniformly across all frequencies.  As for the phase shifts, we fixed the source masses to $(m_1, m_2) = (10^6, 10)M_\odot$ and viewing angles are fixed to typical value for the viewing angles $(\theta, \phi) = (\pi/5, \pi/3)$.
The results are shown in \cref{fig:mismatch_across_models}. The mismatch between our waveform and \textsc{BHPWave} remains below $10^{-5}$ for almost all spin values, in both prograde and retrograde orbits, except for a sharp increase as $a \rightarrow 0.99$. 
This increase coincides with the region where the phase shifts and amplitude differences are largest, as previously discussed. Typically, for larger spin values, as the system enters deeper into the strong-field regime, any discrepancies sourced from differences in amplitudes, fluxes, grid structures, or interpolation schemes become more pronounced.

In comparison, the mismatch with \textsc{KerrCirc} is systematically larger, which is consistent with the phase shift trends observed in~\cref{fig:phase_shift_model_comparison}. 
It is worth mentioning that these mismatch values do not necessarily reflect the absolute accuracy of any of the models considered here, but rather indicate the level of agreement between them. In most cases, the discrepancies are small enough to be below the resolvable threshold of the detector and thus unlikely to impact observational outcomes. 
A detailed investigation into the sources of these differences is beyond the scope of this article and is presented separately in a study focusing on the systematics errors in circular equatorial waveforms~\cite{Khalvati:2025znb}.

\subsection{Weak-field behaviour against PN-GSF approaches}
\label{app:weakfieldappendix}

The combined \gls{pn}-\gls{gsf} approach of black hole perturbation theory yields closed-form analytic solutions to the (frequency-domain) Teukolsky equation and thus the inspiral forcing functions in \cref{eq:elements-ode} and the \gls{gw} strain amplitudes in \cref{eq:Hlmn}
in turn; we refer a reader to Refs.~\cite{Mino:1997bx,Sasaki:2003xr} for reviews. 
In the context of the \gls{emri} and \gls{imri} modelling, this approach is dependent upon order-by-order expansions in powers of the (dimensionless) velocity parameter: $v = \sqrt{1 / p}$
(or similar parameters) and often eccentricity: $e$ (while keeping the inclination and primary spin arbitrary), limiting the accuracy of the analytic expressions in the stronger field ($v \simeq 1$) and at larger eccentricities ($e \simeq 1$). At the same time, this method is particularly advantageous to account for a vast region of the weak-field parameter space ($v \ll 1$) that would be expensive to cover with numerical \gls{gsf} computations. 
%
%

Building upon the results reported in~\cite{Sago:2005fn,Ganz:2007rf,Sago:2015rpa}, Ref.~\cite{Fujita:2020zxe} derived the analytic expressions for the inspiral forcing functions equivalent to \cref{eq:elements-ode} for generic (eccentric and precessing) adiabatic inspirals, valid to $\mathcal{O}(v^{11}, e^{11})$ {relative} to the leading ``Newtonian-circular'' terms. 
Shortly afterwards, Ref.~\cite{Isoyama:2021jjd} combined that result with the analytic GW strain amplitudes (derived to the same \gls{pn} accuracy) to produce a generic adiabatic waveform model.
In this work, we label these analytical results simply ``PN5'' for brevity but we highlight the additional eccentricity expansion employed in their derivation. The PN5 expressions for both the spheroidal harmonic modes of the Teukolsky amplitudes and the inspiral forcing functions are publicly available from the \gls{bhpc} website~\cite{BHPClub}\footnote{There are a number of other \gls{pn}-\gls{gsf} calculations that explicitly consider the eccentric equatorial inspirals in Kerr spacetimes~\cite{Tagoshi:1995sh,Bini:2016dvs,Bini:2019lcd,Munna:2023wce}. However, the results obtained are lower orders in $p$ and $e$ expansion (sometimes with slow primary spin $a \ll 1$ assumed), or have yet to be implemented in \gls{few}. For this reason, we have not used them as a weak-field benchmark for the results in this work.}.

\begin{figure}[t]
    \centering
    \includegraphics[width=0.95\columnwidth]{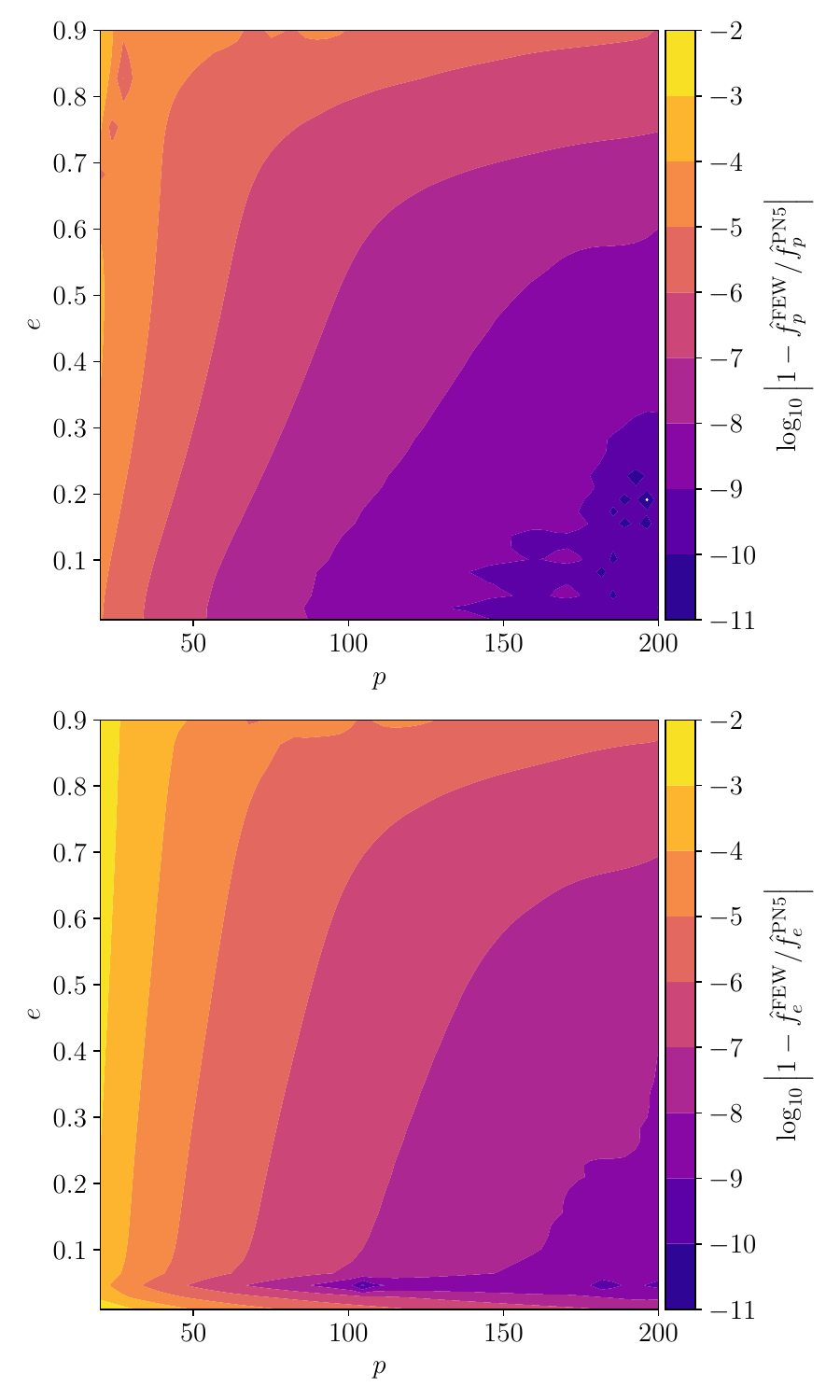}
    \caption{The relative error in $\hat{f}^{(0)}_{p, e}$ between the interpolated forcing function in this work (\gls{few}, $\hat{f}^{\mathrm {FEW}}_{p,e}$) and the \gls{pn} forcing function in \cref{app:weakfieldappendix} (PN5, $\hat{f}^{\mathrm {PN5}}_{p, e}$) in the weak-field for $a = 0.998$ (we find similar behaviour for other values of $a$).}
    \label{fig:PN5FluxErrors}
\end{figure}


\begin{figure}[b]
    \centering
    \includegraphics[width=0.98\columnwidth]{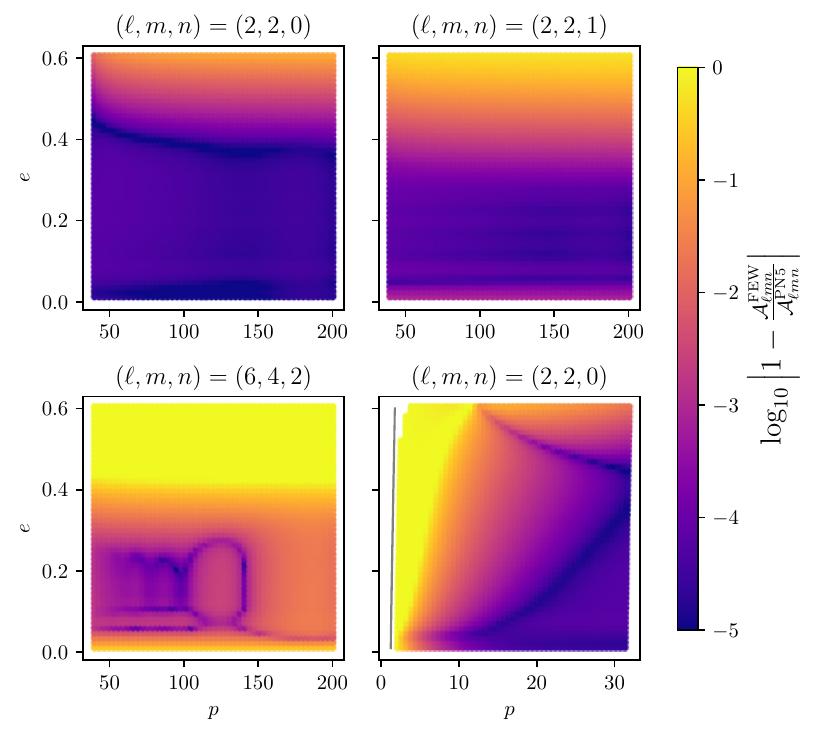}
    \caption{
    Absolute relative error for three selected modes of the interpolated (complex) mode amplitude in this work (\gls{few}, ${\mathcal A}^{\mathrm {FEW}}_{\ell m n}$) and the PN5 waveform mode amplitude in \cref{app:weakfieldappendix} (PN5, ${\mathcal A}^{\mathrm {PN5}}_{\ell m n}$) with $a = 0.998$ (prograde); note the difference in the range of $p$ in the right-bottom panel, which depicts the error in the strong field region. 
    We find similar levels of agreement across other values of $a$.}
\label{fig:PN5AmpMagErrors}
\end{figure}


\begin{figure*}
    \centering
    \includegraphics[width = 0.95\textwidth]{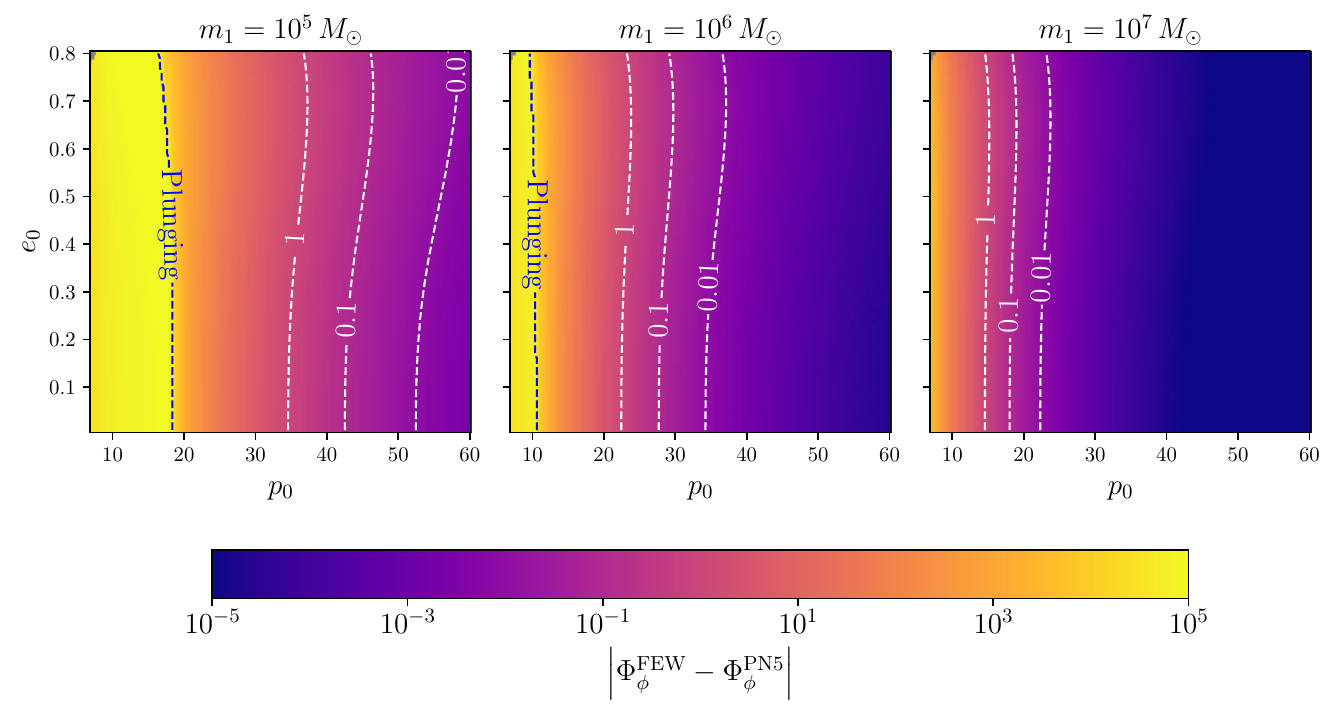}
    \caption{\textbf{(From left to right):} Orbital dephasings between the trajectory model presented in this work (Kerr) and the \gls{pn} trajectory discussed in \cref{app:weakfieldappendix} (PN5) for $m_1 \in \{10^5, 10^6, 10^7\}\,M_\odot$.
    All inspirals are evolved for $4$ years, with $a = 0.998$ and $\epsilon = 10^{-5}$.
    The white dashed contours indicate regions where the dephasing crosses $\{0.01,0.1,1\}\,\mathrm{rad}$. 
    The blue dashed line indicates the outer boundary in $(p_{0},e_{0})$ where the body plunges within four years in the PN5 model.
    }
    \label{fig:traj_comparison_PN_Kerr_dephasing}
\end{figure*}

Some of these PN5 results have already been integrated into \gls{few}. For instance, the PN5-AAK waveform model combines the PN5 inspiral trajectory with the semi-relativisitic (``kludge'') amplitudes~\cite{chua2017augmented} as described in Ref.~\cite{Katz:2021yft}, providing a semi-relativistic waveform model defined over the generic inspiral parameter space: 
the \gls{few} implementation of the PN5 mode amplitudes is also under development (cf.~\cref{sec:future_prospects}). 

We have used the equatorial limit of PN5 results to validate our model (\gls{few} v2) in the weak field by: 
\begin{enumerate}
    \item Validating the interpolations of the forcing functions $\hat{f}^{(0)}_{p,e}$ in the trajectory module against their PN5 equivalents (\cref{fig:PN5FluxErrors}). 
    \item Validating the interpolations of the mode amplitudes $\mathcal{A}_{\ell m n}$ against their PN5 equivalents for sample harmonic modes; the PN5 spherical harmonic waveform mode amplitudes were obtained from the PN5 Teukolsky spheroidal harmonic mode amplitudes via~\cref{eq:amps} and projected onto spherical harmonics as described in~\cref{eq:sphericalproj} (\cref{fig:PN5AmpMagErrors}).
    \item Assessing the orbital dephasing (over four years) between the (fully-relativistic) trajectory model in \gls{few} v2 and the PN5 inspiral model scales(\cref{fig:traj_comparison_PN_Kerr_dephasing}). 
    %
    %
\end{enumerate}
To do this, we have since implemented two minor improvements to the PN5 trajectory module. Firstly, we have corrected minor typos in the (adiabatic) PN5 fluxes of the angular momentum and Carter constant and the associated PN5 forcing functions $\hat{f}^{\mathrm {PN5}}_{p,e}$. Secondly, unlike the public database at BHPC~\cite{BHPClub}, $\hat{f}^{\mathrm {PN5}}_{p,e}$ here are computed from the PN5 fluxes of energy and angular momentum (in the equatorial limit) \emph{without} expanding the Jacobian elements in~\cref{eq:forcesJac} in powers of $v$ or $e$, as doing so enhances the accuracy of $\hat{f}^{\mathrm {PN5}}_{p,e}$ in comparison to their numerical counterparts (cf. Ref.~\cite{Isoyama:2021jjd}).

%
We find in \cref{fig:PN5FluxErrors} that the agreement between the interpolations of the forcing functions $\hat{f}^{\mathrm {FEW}}_{p,e}$ and the PN5 forcing functions $\hat{f}^{\mathrm {PN5}}_{p,e}$ improves as $p$ increases and $e$ decreases. This agreement in ${\hat f}^{(0)}_{e}$ breaks down for small $e$, since both magnitudes of $\hat{f}^{\mathrm {PN5, Few}}_{e}$ become very small. 
A similar trend can be observed in the amplitude comparisons displayed in the top two panels of \cref{fig:PN5AmpMagErrors}, aside from the bottom left panel where the magnitude of the higher multipole amplitude varies rapidly and become small, and the bottom right panel where the accuracy of the PN5 amplitudes decrease closer to the separatrix. However, we find the interpolation error is predominant at smaller values of $e$ in this case. 
The deterioration of the agreement in ${\mathcal {A}}_{\ell m n}$ as the eccentricity becomes larger (typically $e \gtrsim 0.4$) is a result of the eccentricity expansion in the PN5 amplitudes, which become significantly less accurate in that region. This observation has been corroborated through a comparison of the PN5 amplitudes with numerical amplitude data from both the Black Hole Perturbation Toolkit (BHPT)~\cite{BHPToolkit} and the \gls{bhpc}~\cite{Isoyama:2021jjd,BHPClub}.
For weak-field trajectories with initial separation $p_{0} \gg p_{\text{sep}}$ in \cref{fig:traj_comparison_PN_Kerr_dephasing}, we observe that the dephasing between both models tends to zero (independent of the initial eccentricity $e_0$), indicating that the two models converge in the weak field limit, as expected.

Moving forwards, in addition to validating the accuracy of the interpolated amplitudes and forcing functions, these comparisons underpin the integration of \gls{pn}-\gls{gsf} results into future \gls{few} frameworks. The current model could be extended much deeper into the weak field and the accuracy of the amplitude and forcing functions may be improved by switching from interpolated data to their \gls{pn}-\gls{gsf} counterparts at sufficiently large $p$ (and lower values of $e$). This motivates further developments in the \gls{pn}-\gls{gsf} framework such as extending the analytic expressions to higher \gls{pn} orders and in turn improving their accuracy at smaller separations. See Refs.~\cite{Munna:2023wce,Skoupy:2024jsi,Sago:2024mgh,Castillo:2024isq} for recent efforts in this direction. Furthermore, deriving the \gls{pn} expressions without expanding in powers of the eccentricity might be a step-function improvement for the use of \gls{pn}-\gls{gsf} information when $e\gtrsim 0.4$~\cite{Isoyama:2021jjd,Fujita:2025pc,David:2025pc}. We leave these improvements to future work.

\section{Marginal posterior distributions for science-case sources}
\label{app:corner_plots}

\begin{figure}[p]
    \includegraphics[width=0.9\linewidth]{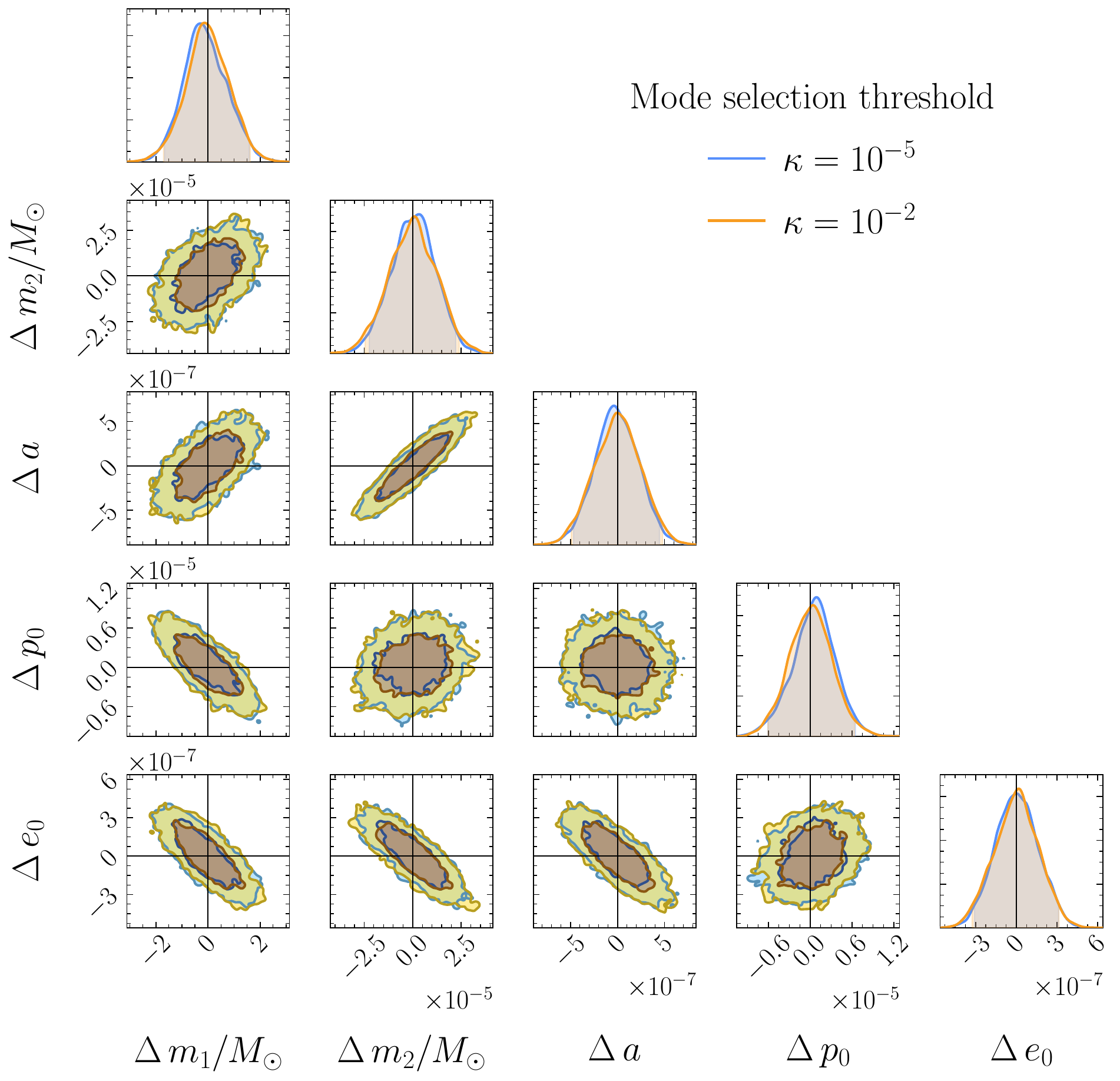}
    \caption{Posterior distributions of the intrinsic parameters for an \gls{emri} with parameters given in the first row of~\cref{tab:emri_params} for two values of the mode selection threshold. The blue (orange) posteriors refer to a threshold set to $\kappa=10^{-5}$ ($10^{-2}$). 2D (1D) shaded areas refer to 1- and 2-sigma (95\%) credible regions. For the intrinsic parameters, we plot the quantities $\Delta X = X - X_{\rm true}$ for visual clarity.
    }\label{fig:MCMC_row_1}
\end{figure}

\begin{figure}[p]
    \includegraphics[width=0.9\linewidth]{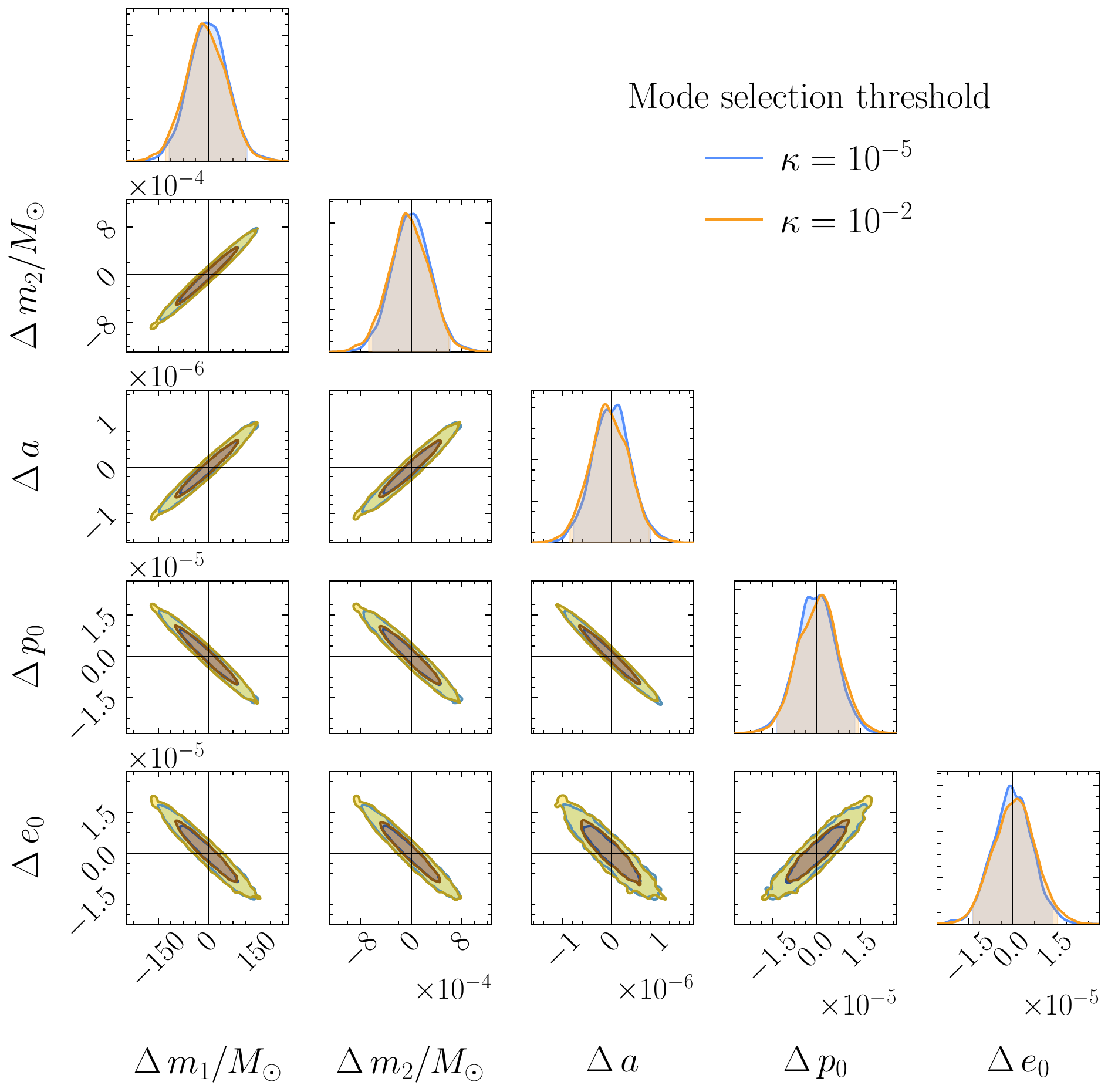}
    \caption{Posterior distributions of the intrinsic parameters for an \gls{emri} with parameters given in the second row of~\cref{tab:emri_params} for two values of the mode selection threshold. The blue (orange) posteriors refer to a threshold set to $\kappa=10^{-5}$ ($10^{-2}$). 2D (1D) shaded areas refer to 1- and 2-sigma (95\%) credible regions. For the intrinsic parameters, we plot the quantities $\Delta X = X - X_{\rm true}$ for visual clarity.
    }\label{fig:MCMC_row_2}
\end{figure}

\begin{figure}[p]
    \includegraphics[width=0.9\linewidth]{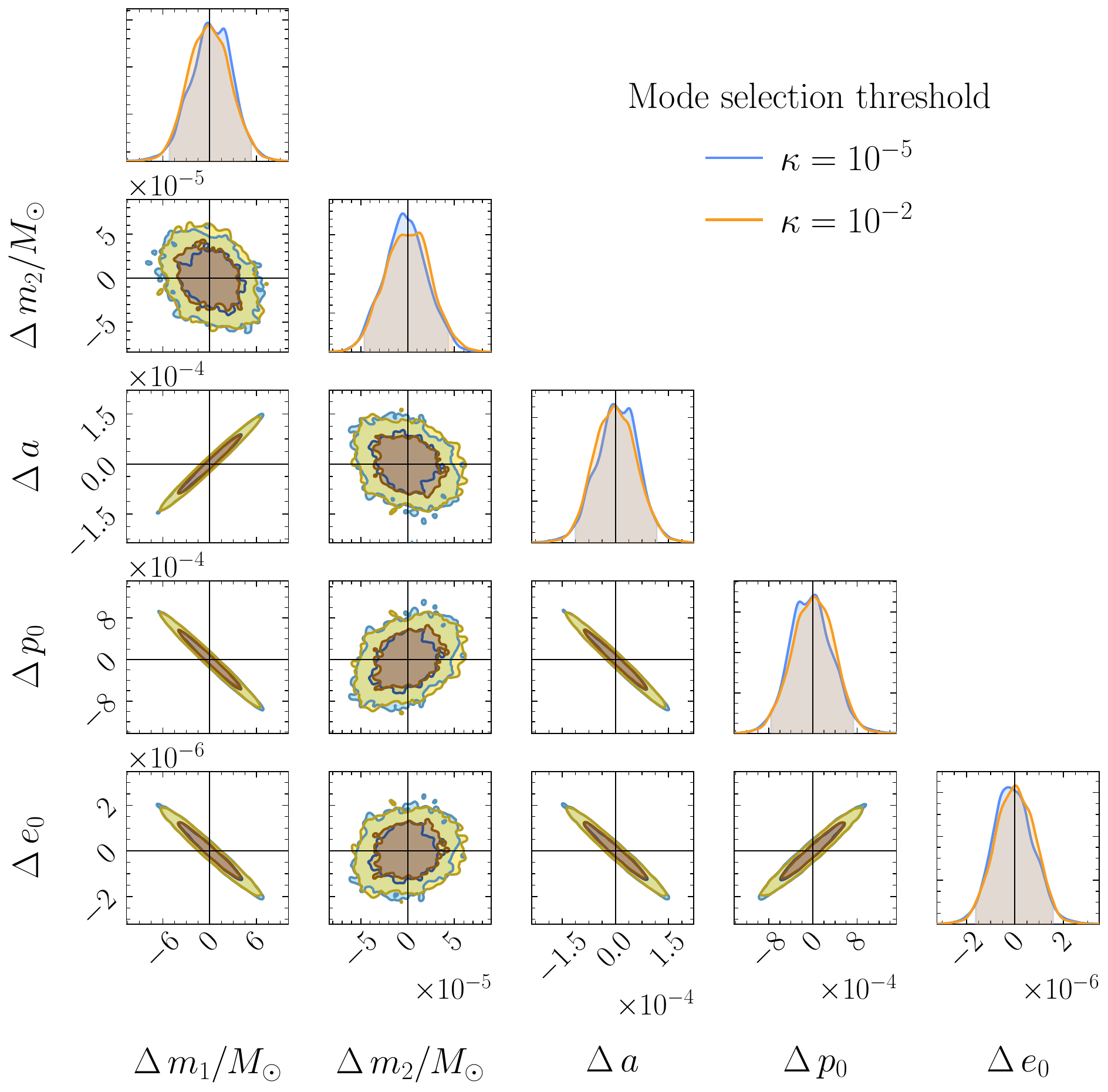}
    \caption{Posterior distributions of the intrinsic parameters for an \gls{emri} with parameters given in the fifth row of~\cref{tab:emri_params} for two values of the mode selection threshold. The blue (orange) posteriors refer to a threshold set to $\kappa=10^{-5}$ ($10^{-2}$). 2D (1D) shaded areas refer to 1- and 2-sigma (95\%) credible regions. For the intrinsic parameters, we plot the quantities $\Delta X = X - X_{\rm true}$ for visual clarity.
    }\label{fig:MCMC_row_5}
\end{figure}

In this Appendix, we show corner plots of the marginal posterior distributions obtained from the parameter estimation analyses performed in \cref{sec:inference-subsection}.
\Cref{fig:MCMC_row_1,fig:MCMC_row_2,fig:MCMC_row_5} correspond to the analysis of \glspl{emri} with parameters given by rows 1, 2 and 5 of \cref{tab:emri_params} respectively.
We do not observe any significant biases in the intrinsic parameters of these sources for $\kappa=10^{-2}$, which is expected (based on \cref{eq:lindblom}) given that these sources have \glspl{snr} $\leq 50$.

\begin{figure}[p]
    \includegraphics[width=0.9\linewidth]{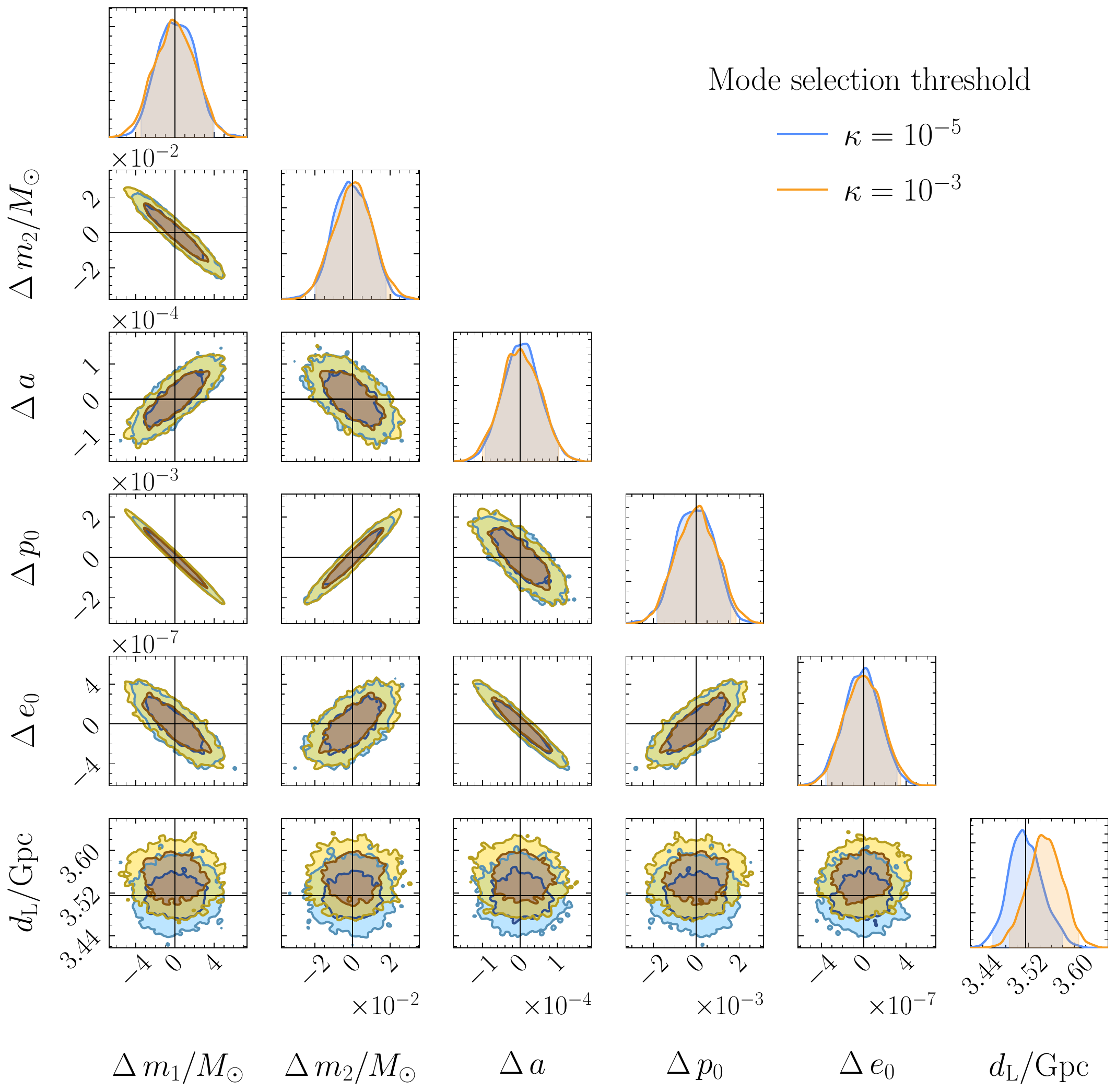}
    \caption{Posterior distributions of the intrinsic parameters for an \glspl{imri} with parameters given in the fourth row of~\cref{tab:emri_params} for two values of the mode selection threshold. The blue (orange) posteriors refer to a threshold set to $\kappa=10^{-5}$ ($10^{-3}$). 2D (1D) shaded areas refer to 1- and 2-sigma (95\%) credible regions. For the intrinsic parameters, we plot the quantities $\Delta X = X - X_{\rm true}$ for visual clarity.
    We also include the luminosity distance $d_\mathrm{L}$ to show the bias induced by the mode selection threshold. 
    }\label{fig:MCMC_row_4}
\end{figure}

\begin{figure*}[p]
    \includegraphics[width=0.95\linewidth]{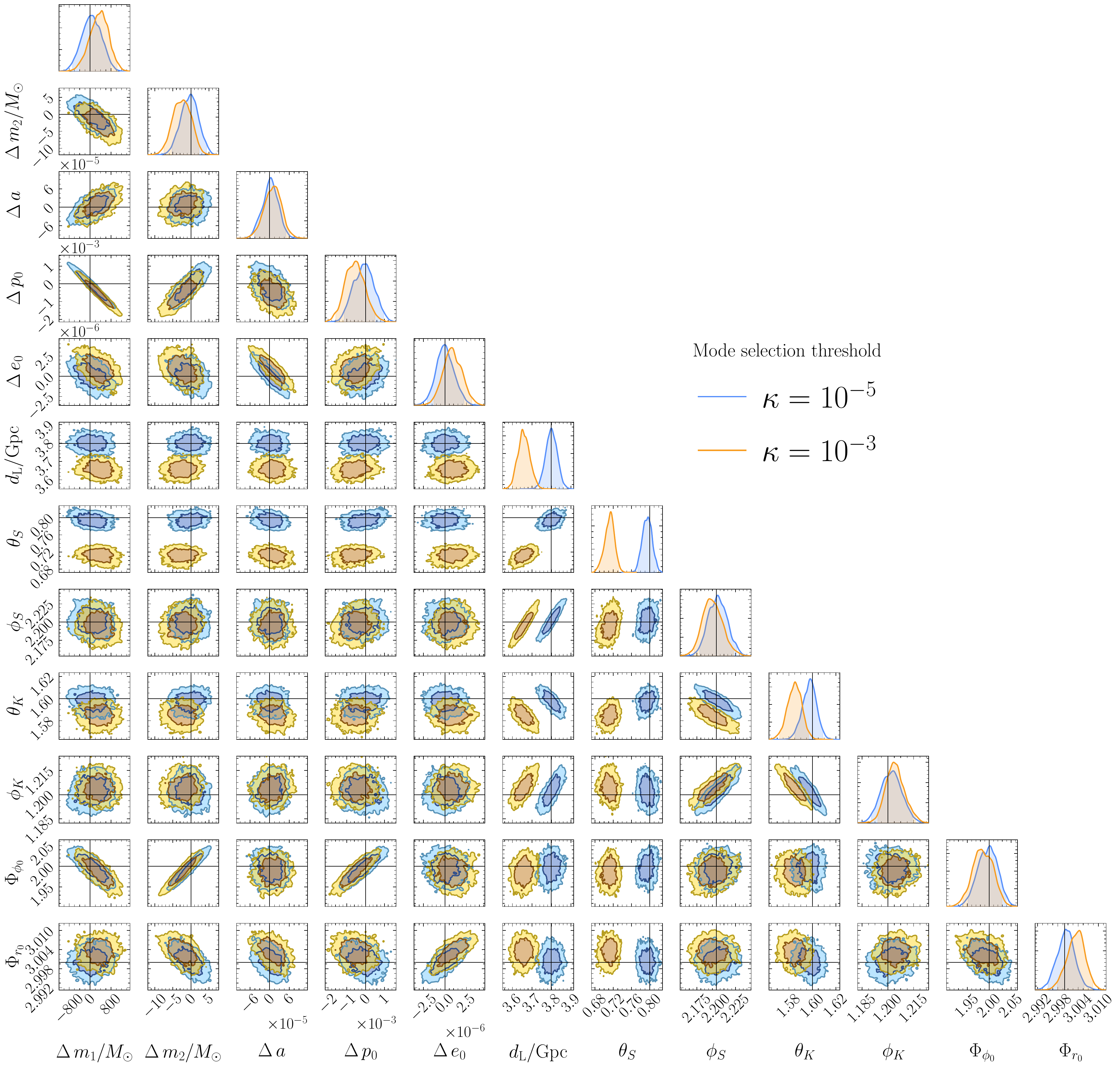}
    \caption{Posterior distributions for an \gls{imri} with parameters given in the third row of~\cref{tab:emri_params} for two values of the mode selection threshold. The blue (orange) posteriors refer to a threshold set to $\kappa=10^{-5}$ ($10^{-3}$). 2D (1D) shaded areas refer to 1- and 2-sigma (95\%) credible regions. For the intrinsic parameters, we plot the quantities $\Delta X = X - X_{\rm true}$ for visual clarity. The $\kappa=10^{-3}$ posteriors yield significant biases in the extrinsic parameters due to the high \gls{snr} of this source (500).}
    \label{fig:MCMC_row_3}
\end{figure*}

Similarly, \cref{fig:MCMC_row_3,fig:MCMC_row_4} correspond to the analysis of \glspl{imri} with parameters given by rows 3 and 4 of \cref{tab:emri_params} respectively.
As the \glspl{snr} of these sources is higher (500 and 300 respectively), analyses with $\kappa=10^{-2}$ failed to converge due to large mismodelling errors shifting the posterior bulk far from the parameters of the injections.
For $\kappa=10^{-3}$, biases in intrinsic parameters are generally below $1\,\sigma$, but more significant biases are observed in the recovery of extrinsic parameters.

\clearpage 

\clearpage

%
%
\bibliographystyle{IEEEtran}
\bibliography{Kerr_FEW}
\newpage\hbox{}\thispagestyle{empty}{\newpage}

\end{document}